\documentclass[a4paper]{emulateapj}

\usepackage{newtxtext,newtxmath}
\shorttitle{Optical Variability of Ten Blazars}
\usepackage{graphicx}
\usepackage{threeparttable}
\usepackage{amsmath}

\begin{document}
\makeatletter
\newcommand*{\rom}[1]{\expandafter\@slowromancap\romannumeral #1@}
\makeatother

\title{Multi-color optical monitoring of ten blazars from 2005 to 2011}
\author{Nankun Meng\altaffilmark{1},
Xiaoyuan Zhang\altaffilmark{1},
Jianghua Wu$^{1}$,
Jun Ma$^{2}$,
Xu Zhou$^{2}$}
\affil{\altaffilmark{1}Department of Astronomy, Beijing Normal University, Beijing 100875, China; E-mail: jhwu@bnu.edu.cn\\
        \altaffilmark{2}National Astronomical Observatories, Chinese Academy of Sciences, 20A Datun Road, Chaoyang District, Beijing 100012, China}

\begin{abstract}

We carried out multi-color optical monitoring of a sample of ten blazars from 2005 to 2011.
The sample contains 3 LBLs, 2 IBLs, 4 HBLs, and 1 FSRQ.
Our monitoring focused on the long-term variability and the sample included nine BL Lac objects and one flat-spectrum radio quasar.
A total number of 14799 data points were collected.
This is one of the largest optical database for a sample of ten blazars.
All objects showed significant variability except OT 546. 
Because of the low sampling on each single night, only BL Lacertae was observed to have intra-day variability on 2006 November 6. 
Most BL Lac objects showed a bluer-when-brighter chromatism, while the flat-spectrum radio quasar, 3C 454.3, displayed a redder-when-brighter trend.
The BWB color behaviors of most BL Lacs can be at least partly interpreted by the fact of increasing variation amplitude with increasing frequency observed in these objects. 
The average spectral index of LBLs is around 1.5, as expected from 
the model dominated by Synchrotron Self-Compton (SSC) loss. 
The optical emission of HBL is probably contaminated by the thermal emission from the host galaxies.
Correlation analysis did not reveal any time delay between variations at different wavelengths.

\end{abstract}

\section{introduction}

Blazars are the most violently variable class of active galactic nuclei (AGNs), whose relativistic jets are close to our line of sight. They show characteristics of strong variability of flux and polarization of non-thermal radiation across the electromagnetic spectrum. 
Blazars can be divided into flat-spectrum radio quasars (FSRQs) and BL Lac objects.
The former show broad emission lines (equivalent width, EW $> 5 \AA$) in their spectra, while the latter have weak (EW $< 5 \AA $) or absent emission lines.
The broadband spectral energy distribution (SED) of blazars has two characteristic humps. The low-frequency hump extends from radio to UV or X-ray and is dominated by synchrotron radiation. The high-frequency hump is located from X-ray to $\gamma$-ray, and is believed to be due to the inverse Compton scattering emission \citep[e.g.][]{1997ARA&A..35..445U,2007Ap&SS.309...95B}. The exact frequency ranges of the two humps may vary from object to object. BL Lac objects can be subdivided into high-, intermediate-, and low-frequency-peaked objects (HBLs, IBLs, and LBLs, respectively), depending on the positions of the peak frequencies of their synchrotron emission \citep[e.g.][]{1995ApJ...444..567P,1998MNRAS.299..433F}.

The variability of blazar does not exhibit periodic behavior over a large range of timescales from years down to minutes \citep[e.g.][]{1995ARA&A..33..163W,2004AJ....128...47D,2011ApJS..192...12I,2013ApJS..206...11S,2013ApJS..204...22D,2015ApJS..218...18D,2018PASP..130b4102Z}. 
Optical variability properties have served as a tool to investigate the emission processes.
Variability timescale is important to constrain the size and location of the emitting region. Previous observations at multiple wavelengths revealed complex color behaviors of blazars. Some authors found a bluer-when-brighter (BWB) chromatism \citep[e.g.][]{2003ApJ...590..123V,2012AJ....143..108W}, while some others claimed the opposite, the redder-when-brighter (RWB) trend \citep[e.g.][]{2012ApJ...756...13B}, or no clear tendency \citep[e.g.][]{2009ApJ...694..174B}. It appears that BL Lac objects are BWB while FSRQs are RWB \citep[e.g.][]{2006A&A...450...39G,2011MNRAS.418.1640W,2012MNRAS.425.3002G,2017ApJ...844..107I}.

We carried out multi-color optical monitoring of a sample of ten blazars from 2005 to 2011. Our monitoring focused on the long-term variability and the sample included nine BL Lac objects and one FSRQ.
Here we present the observational and analysis results.
Section 2 describes our observations and photometries. 
Section 3 gives a brief introduction of the data analyses.
Section 4 interprets the results of individual targets. The conclusions and discussions are given in Section 5.

\section{Observations and Photometries}

Our monitoring program was performed with a 60/90 cm Schmidt telescope located at the Xinglong Station of the National Astronomical Observatories, Chinese Academy of Sciences (NAOC).
Before 2006, the telescope was equipped with a Ford Aerospace 2048$\times$2048 CCD camera. The CCD had a pixel size of 15 $\mu m$ and a field of view (FOV) of $58' \times 58'$, resulting in a spatial resolution of $1."$7 pixel$^{-1}$. After 2006, a new 4096$\times$4096 E2V CCD took place of the old one. It has a FOV of $96' \times 96'$ and a spatial resolution of $1.''$3 pixel$^{-1}$.
This telescope is equipped with 15 intermediate-band filters, covering the wavelengths from 300 to 1000 nm \citep{1996AJ....112..628F,2000PASP..112..691Y,2003A&A...397..361Z}. 
This photometric system can sample a 15-color SED for the targets.
Five of the 15 filters, $c$, $e$, $i$, $m$, and $o$ were adopted for our multi-color quasi-simultaneous observation. Their central wavelengths and bandwidths are listed in Table~\ref{filters}.
The central wavelength of the $i$ band is similar to that of the $R$ band. The magnitudes in these two bands can be transformed with a formula of $R=0.897 \times i+1.127 $ for AGNs \citep{2013ApJS..204...22D}.

In our program, ten blazars were monitored from 2005 January 29 to 2011 June 13.
In order to reduce the readout time and to increase the sampling rate, only the central 512$\times$512 pixels are read out as a frame. Each frame has a FOV of about $14.5' \times 14.5'$ and $11' \times 11'$ for the old and new CCDs, respectively. 
The observations were mostly made in the $c$, $i$, and $o$ bands. 
Some objects were monitored in the $e$, $i$, and $m$ bands at the beginning.
The observation details of our targets are listed in Table~\ref{sources}.
The exposure time was determined by the brightness of the source, the moon phase, and weather conditions.

The data reduction procedures include positional calibration, bias subtraction, flat-fielding, extraction of instrumental aperture magnitude, and flux calibration.
The comparison and check stars are selected in the finding charts on the website of Landessternwarte
K$\ddot{\mathrm{o}}$nigstuhl\footnote{https://www.lsw.uni-heidelberg.de/projects/extragalactic/charts/}.
Their $c$, $e$, $i$, $m$, and $o$ magnitudes were obtained by observing them and the standard star HD 19945 on photometric nights.
The $c$, $e$, $i$, $m$, and $o$ magnitudes of each blazar were calibrated relative to those magnitudes of comparison stars.
Example results are
given in Table~\ref{example}. The full results for all 34 observations of the blazars are available in electronic form in the online Journal.
The columns are the observation date and time in universal date and time, Julian date, exposure time in seconds, magnitude and error, and differential magnitude (its nightly average values were set to zero) of the check star.

\section{Data Analyses}

The long-term light curves are displayed in Figure~\ref{fig:all}. 
Since we had low sampling for a single object on each single night, the intranight magnitude changes for a certain object are all very small (as will be manifested later), so we plotted for each object only the median measurements on all single nights in Figure~\ref{fig:all}. 
Unless specifically mentioned, the following analyses are all based on the median light curves.

The spectral indices were calculated by assuming a power-law for the spectrum of each blazar and by using measurements in three passbands. Before calculating the spectral indices, we corrected the Galactic interstellar reddening. Based on the model of \citet{2011ApJ...737..103S}, the extinctions in our passbands were obtained by cubical interpolation from those of 88 passbands provided by NED\footnote{The NASA/IPAC Extragalactic Database (NED) is operated by the Jet Propulsion Laboratory, California Institute of Technology, under contract with the National Aeronautics and Space Administration.}. The extinctions of all sources are listed in Table~\ref{extinction}. The spectral index curves are also plotted in Figure~\ref{fig:all}.

The statistical results of the variability behaviors are summarized in Table~\ref{observations}.
The columns are the maximum and minimum magnitudes, average magnitudes, and amplitudes of the whole observations.
The long-term color-magnitude diagrams are presented for all ten blazars in Figure~\ref{fig:allcolor}.
Based on the original complete data, we searched for the intraday variability (IDV) for each target and used four statistical tests, the C, $\chi^2$, F, and ANOVA tests \citep{1997AJ....114..565J,1999A&AS..135..477R,1983ApJ...272...11P,1998ApJ...501...69D,2010AJ....139.1269D}, to make a quantitative assessment on whether there is IDV. 
The correlation analyses were performed on the variations at different wavelengths for those objects with relatively intensive monitoring in order to search for the possible inter-band time lags. The results are listed in Table~\ref{alldelay}.
We now discuss the results of each blazar in details.

\section{Individual Comments on Each Blazar}

\subsection{3C 66A}

The blazar 3C 66A is classified as an IBL, because its synchrotron peak is located between $10^{15}$ and $10^{16}$ Hz \citep{2003A&A...407..453P,2010ApJ...716...30A}. The redshift was determined to be z = 0.444 by independent authors \citep{1978bllo.conf..176M,1993ApJS...84..109L}. Ever since its optical counterpart was identified by \citet{1974ApJ...190L..97W}, 3C 66A was studied over a wide frequency range, from radio to $\gamma$-ray. 
Very Long Baseline Interferometry (VLBI) images from 2.3 to 43 GHz revealed a typical core-jet structure on parsec scales \citep[e.g.][]{2001ApJS..134..181J,2002ApJ...577...85M}.
It is one of the sources detected in TeV \citep[e.g.][]{2009ApJ...692L..29A,2009ApJ...693L.104A}.

Its optical variability has been studied on different timescales by a lot of investigators \citep[e.g.][]{1978ApJ...220...19M,1991BAAS...23.1420C,1992ApJS...80..683X,1996A&AS..120..313T,2000ApJ...537..101F,2011Ap&SS.333..213H}. 
The study of \citet{2000ApJ...537..101F} showed that the variability amplitude increased with increasing frequency. 
IDV was reported by \citet{1991BAAS...23.1420C}, but some observations did not reveal IDV \citep{1978ApJ...220...19M,1992ApJS...80..683X,2011Ap&SS.333..213H}. 
3C 66A has been intensively monitored by two Whole Earth Blazar Telescope (WEBT) campaigns \citep{2005ApJ...631..169B,2009ApJ...694..174B}. 
The first WEBT campaign from 2003 July through 2004 April revealed IDV within a 2-hour timescale \citep{2005ApJ...631..169B}.
Time lag of shorter wavelength preceding longer wavelength was shown by correlation analysis \citep{2011Ap&SS.333..213H}. 
Optical spectrum turns bluer/flatter when the source brightens, and this tendency was confirmed by many other authors \citep[e.g.][]{2003ApJ...590..123V,2006A&A...450...39G,2010MNRAS.404.1992R,2011Ap&SS.333..213H}.

Our optical observations of 3C 66A in the $c$, $i$, and $o$ bands were carried out for 237, 248, and 235 nights, respectively, between 2005 January and 2011 February (Table~\ref{sources}).
Part of these data have been already published in \citet{2009ApJ...694..174B} and \citet{2011Ap&SS.333..213H}. 
\citet{2009ApJ...694..174B} included our data from 2007 September to December.
\citet{2011Ap&SS.333..213H} covered the period from 2005 January 26 to 2008 October 16.

3C 66A remained active throughout the whole period and there are significant correlations between the light curves in different passbands.
One can see from Figure~\ref{fig:all} that the overall brightness of 3C 66A showed a wave pattern with two peaks at around JDs 2454400 and 2455200 and three troughs at around JDs 2454100, 2454800, and 2455600. 
A number of short-term oscillations overlapped on the overall wave pattern. 
On the other hand, we did not detect any IDV because of the low sampling on each single night. 
The variation amplitude tends to decrease with decreasing frequency (see Table~\ref{observations}). This is consistent with the result of \citet{2000ApJ...537..101F}.
The correlation analysis did not find any inter-band time lag at long and short timescales (see Table~\ref{alldelay}).

No systematic spectral variability was detected in the second WEBT campaign covered the autumn and winter of 2007-2008 \citep{2009ApJ...694..174B}.
It can be seen from Figure~\ref{fig:all}, the spectral index became smaller (the spectrum became flatter) when the brightness of the source increased.
One can see from Figure~\ref{fig:allcolor} that 3C 66A shows significant BWB trends (p $\leq$ 0.05) in all three colors (see Table~\ref{color}), which is in accord with the previous results mentioned above.

\subsection{PKS 0735+178}

PKS 0735+178 was first classified as a BL Lac object by \citet{1974ApJ...190L.101C}. The broadband spectrum suggested that it is a LBL \citep{1996ApJ...463..424U}. \citet{2012A&A...547A...1N} derived a redshift of $0.45\pm0.06$ for this object by using the host galaxy as a standard candle.

Historical optical data over a century show its violent long-term variations with an amplitude of 3-4 mags \citep{1979AJ.....84.1658P,1997A&AS..125..525F}.
It probably has periods of 4.89 and 14.2 years \citep{1988AJ.....95..374W,1997A&AS..125..525F}. \citet{2007A&A...467..465C} found 3 characteristic timescales of about 4.5, 8.5 and 11-13 years. For short timescales, some IDV and inter-day variations were reported in its optical history. However, \citet{2009MNRAS.399.1622G} reported an unusual quiescence of IDV between 1998 and 2008.
Optical variations of this source often show larger amplitudes than those in the infrared \citep{2000ApJ...537..101F}, which indicates a BWB trend.

We observed this source on 13, 15, and 10 nights in the $c$, $i$, and $o$ bands, respectively, between 2009 December 18 and 2010 November 30.
The recorded maximum brightness of 15.819 mags in the $i$ band occurred on 2010 January 10 (JD 2455207).
The brightness dropped by 0.58 mags at the end.
The variation amplitude decreases in the wavelength sequence of $c$, $i$, and $o$ (see Table~\ref{observations}).
No IDV was detected because of the low sampling.

As can be seen in Figure~\ref{fig:all}, the spectral index varied slightly between 1.1 and 1.8. 
One can see from Figure~\ref{fig:allcolor}, PKS 0735+178 exhibits different levels of positive correlations for all its observed colors.
Significant BWB trends were found for the $c-i$ and $c-o$ colors, while the $i-o$ color had significance less than 95\%.
These BWB behaviors are consistent with decreasing amplitude at decreasing frequency observed in this object, as mentioned above.
\citet{2000ApJ...537..101F} reported the similar color trends.
On the other hand, this type of correlation appears to be weak \citep{2006A&A...450...39G} or opposite \citep{2000ApJS..127...11G} in other observations.

\subsection{OI 090.4}

OI 090.4 was first identified as a BL Lac object by \citet{1977ApJ...215L..71T}. Cross identification of optical and radio source was accomplished by \citet{1977AJ.....82..692C}. A tentative redshift z = 0.66 was proposed by \citet{1986ASSL..121..675P} and was then questioned by \citet{1991MNRAS.252..482A}. 
\citet{1996MNRAS.283..241F} suggested z $\sim$ 0.3 on the basis of the host galaxy properties. \citet{2003A&A...412..651C} identified [OII] $\lambda$3727 and [OIII] $\lambda$5007 indicating redshift z = 0.266$\pm$0.001.
\citet{1995ApJS..100...37G} suggested it was a LBL by its multi-frequency spectrum. \citet{2009A&A...507L..33P} reported its viewing angle of jet at about 6.9 degs.

Its optical variability was studied by a number of authors \citep[e.g.][]{1980PASP...92..156B,1981A&A...103..342Z,1988AJ.....96.1215P,1991A&AS...88..225S,1991A&AS...87..461X,1994A&AS..106..361X,2000A&AS..143..357K}. 
\citet{1981A&A...103..342Z} reported 2 mags fluctuations in the $B$ band. \citet{1988AJ.....96.1215P} observed 0.8 mags changes in the $B$ band on a 2 year timescale. \citet{1991A&AS...88..225S}  detected 0.9 mags variations in the $V$ band from 1984 to 1989. \citet{1994A&AS..106..361X} claimed 1.04 mags variations both in the $V$ and $B$ bands from 1985 to 1991, and a rapid optical flare of 0.56 mags in the $B$ band within 80 mins on 1988 December 11 was also reported by \citet{1991A&AS...87..461X}. 
There were no systematic color changes with brightness in the results of \citet{1997AJ....113.1995N}.
And they argued the underlying emission mechanism of this source was of a non-thermal nature.

We observed this source on 19, 21, and 18 nights in the $c$, $i$, and $o$ bands, respectively, between 2006 March 14 and 2011 March 30 (Table~\ref{sources}). 
Due to the few data points, brief statistical results are listed for this source. No IDV was found because of the low sampling.
The amplitude of the variations is systematically larger at higher frequencies (see Table~\ref{observations}).

As can be seen in Figure~\ref{fig:all}, the average spectral index was 1.46.
One can see from Figure~\ref{fig:allcolor}, OI 090.4 shows a stronger BWB trend with $i-o$ color than with the $c-i$ and $c-o$ colors.
The low confidence levels (p $>$ 0.1) for the $c-i$ and $c-o$ colors may relate to the few data points and big errors.
The observed BWB trend is inconsistent with the results from \citet{1997AJ....113.1995N}.

\subsection{Mrk 421}

Mrk 421 is one of the nearest blazars and it has a redshift of z = 0.03 \citep{2005ApJ...635..173S}. It was first reported as an elliptical galaxy with a highly condensed nucleus emitting non-thermal continuum radiation but no emission lines in the optical regime \citep{1975ApJ...198..261U}. It is classified as a HBL due to its synchrotron peak higher than 0.1 keV \citep{2012AJ....143...23G}. 

In optical region, \citet{1975ApJ...201L.109M} found variations of $\ge$ 4.7 mags from 1899 to 1975. 
Observations on 17 nights between 2009 November 15 and 2011 May 2  were made by \citet{2012MNRAS.425.3002G}, which showed a maximum brightness of $\sim$ 12.46 in the $V$ band on JD 2455684.
\citet{1999ApJS..121..131F}  found that the variations in the $J$, $H$, and $K$ bands were much smaller than those in the optical.
\citet{2004A&A...422..505G} reported typical brightness variations of $\sim$ 0.4 mags and a maximum variation of 0.89 mags in the $J$ band from February to March, 2003.
A BWB behavior was claimed for this source \citep{1988A&AS...72..347S,1997A&AT...14..119F,1998A&A...339...41T,2014Ap&SS.349..909C}.
\citet{1997A&AT...14..119F} and \citet{1998A&A...339...41T} claimed that this BWB chromatism may be due to the thermal contribution of the host galaxy.
But \citet{2014Ap&SS.349..909C} argued that only a host galaxy contribution was not sufficient, and some intrinsic mechanism was also needed.

We observed this source on 15, 5, and 16 nights in the $c$, $m$, and $o$ bands, respectively, between 2005 January 30 and 2008 May 05.
The observed maximum brightness of 13.159 mags in the $c$ band occurred on 2008 March 2 (JD=2454528).
As can be seen from Table~\ref{observations}, the amplitude in this band was larger than the one in the $o$ band for the same period.

We only observed Mrk 421 simultaneously in the $c$ and $o$ bands for several days.
So we didn't calculate its spectral index. 
One can see from Figure~\ref{fig:allcolor}, the positive correlation derives that Mrk 421 exhibits a strong BWB trend with the $c-o$ color, which is in accordance with the previous results mentioned above.

\subsection{ON 231}

ON 231 was discovered as a variable star by \citet{1916AN....202..415W}, and was reported as a new quasar by \citet{1971Natur.231..515B}. The detection of a few weak emission lines in its spectrum made it possible to determine a redshift of z = 0.102 \citep{1985ApJ...292..614W}. The synchrotron peak frequency log $\nu_{peak}$ = 14.84 was calculated by \citet{2006A&A...445..441N} and it is classified as an IBL.

In the optical regime, the optical outbursts in 1940, 1953, and 1968 were observed by \citet{1998A&AS..130..109T}. A series of 2 mags outbursts were claimed in 1975-1976, and this type of rapid activity continued through 1986 and early 1987 \citep{1988AJ.....95..374W}. A variation with a timescale of 6 days was detected in 1983 \citep{1987A&AS...67...17X}. \citet{1991A&AS...87..461X} reported three outbursts with similar profiles from 1987 to 1989.
\citet{1999A&A...342L..49M} observed an extraordinary outburst from April to May, 1988, and the magnitude in the $R$ band reached a maximum brightness of 12.2.
In 1997, the $R$ magnitude showed an oscillating behavior between 14 and 14.5 for about two months \citep{2002A&A...395...11T}. 
\citet{2013MNRAS.429.2773C} claimed the larger variation amplitude at higher frequency and the BWB chromatism.
They also found that the variability at the $R$ band lagged that at the $B$ band by $\sim$1200 s on March 17.

Our optical $c$, $i$, and $o$ bands observations of ON 231 were carried out for 76, 80, and 78 nights, respectively, between 2009 May 23 and 2011 May 15.
At the begin, the object was at a relatively high state. Then it got fainter and showed some oscillations at the second and third stages. 
At the last stage, the brightness dropped gradually from JDs 2455572 to 2455680 and raised again after that.
One can see from Table~\ref{observations} that the variation amplitude decreases towards longer wavelengths, which is in accord with \citet{2013MNRAS.429.2773C}.
We did not find any IDV because of the low sampling on each single night.
No inter-band time lags were found at long and short timescales for this source (see Table~\ref{alldelay}).

One can see from Figure~\ref{fig:all}, the spectral indices decreased with increasing luminosity. However, \citet{2006MNRAS.371.1243H} reported the spectral index $\alpha$ decreased with increasing luminosity.
ON 231 shows strong BWB behavior in all three color-magnitude diagrams (Figure~\ref{fig:allcolor}).
This is in agreement with the results of \citet{1998A&AS..130..109T} and \citet{2014A&A...562A..79S}.

\subsection{PG 1553+113}

Located in the Serpens Caput constellation, PG 1553+113 was classified as a BL Lac object based on its featureless spectrum \citep{1990PASP..102.1120F} and was classified as a HBL blazar \citep{1994ApJS...93..125F,1995A&AS..109..267G,2002A&A...383..410B}. Evidence of very-high-energy $\gamma$-ray emission from this source was first reported by H.E.S.S. in 2005 \citep{2006A&A...448L..19A} and was later confirmed by observations above 200 GeV with the MAGIC telescope at a significance level of 8.8 $\sigma$ \citep{2007ApJ...654L.119A}. Due to its featureless optical spectrum, the redshift of PG 1553+113 remains highly uncertain. Measurements using the Hubble Space Telescope (HST) Cosmic Origins Spectrograph (COS) yielded a lower limit of 0.395 \citep{2010ApJ...720..976D}. The small statistical uncertainties of the VERITAS energy spectrum provided a robust upper limit of 0.62 \citep{2015ApJ...799....7A}. 

\citet{2011MNRAS.416..101G} claimed the detection of its IDV in the optical region.
A general BWB trend was found in the optical observations by a WEBT campaign from April to August of 2013 \citep{2015MNRAS.454..353R}. However, \citet{2015A&A...573A..69W} did not find a definite long-term trend when analyzing the color behavior in 2007-2012.

We observed this source on 28 nights in the $c$, $i$, and $o$ bands between 2009 May 26 and 2010 March 28. 
One can see from Figure~\ref{fig:all} that there is an outburst of about 0.2 mags on a timescale of 23 days (from JD 2454988 to JD 2455011).
The source got fainter at the second stage and changed into a relatively high state at the last.
Since we only got the differential magnitudes, we did not calculate the spectral indices.
The overall variation amplitudes were $\Delta c$=$0^{m}.475$, $\Delta i$=$0^{m}.47$, and $\Delta o$=$0^{m}.462$, which showed a decrease trend with decreasing frequency.
No IDV was observed on each single night.
Although we did not collect much data for this object, it showed a complete outburst at the beginning. So we made a correlation analysis for it.
As can be seen from Table~\ref{alldelay}, however, no inter-band time lags were found for this source.

\subsection{OT 546}

OT 546  was first noticed by \citet{1966ApJ...143..192Z} and was classified as a BL Lac object by \citet{1980ARA&A..18..321A}. The redshift of z = 0.0554 $\pm$ 0.0003 was measured by \citet{1978ApJ...219L..97O}. It was a nearby HBL \citep{2015ApJ...808..110A} and was discovered as a very high energy (VHE) source by MAGIC \citep{2014A&A...563A..90A}.
OT 546 was also found to have a massive black hole $\sim$ $5.4 \times 10^8 M_{\odot}$ at its center \citep{1995Ap&SS.229..157F}.

Its optical variations were discovered by \citet{1967ApJ...150L.177S}. 
From 1975 to 1987, the lightcurves from Rosemary Hill Observatory observations showed variations with an overall amplitude of 0.8 mags and an average magnitude of $\sim$ 16.7 in the $B$ band \citep{1988AJ.....96.1215P}. The monitoring from 1996 to 1997 in Tuorla Observatory displayed that the brightness in the $V$ band varied between 15.76 and 16.12 \citep{2000A&AS..143..357K}.

The monitoring of OT 546 can be divided into two phases. 
Phase 1 is from JDs 2453444 to 2453919. 
The object was observed on 55, 54, and 52 nights in the $e$, $i$, and $m$ bands, respectively. 
Phase 2 is from JDs 2454134 to 2455713. 
The object was observed on 113, 121, and 110 nights respectively in the $c$, $i$, and $o$ bands. 
The brightness of this object was quite stable during our monitoring period. 
There are four flares at around JDs 2453650, 2454800, 2454960, and 2455330. 
The last flare has an amplitude of about 0.2 mags in the $i$ band.
As can be seen from Table~\ref{observations}, the variation amplitude decreases towards longer wavelengths in both phases.
The overall amplitudes are less than 1 mags and this is in agreement with the results of \citet{1988AJ.....96.1215P} and \citet{2000A&AS..143..357K}.
We did not find any IDV on each single night because of the low sampling.

One can see from Figure~\ref{fig:all} that the spectral indices fluctuated slightly around 1.8.
From the results in Figure~\ref{fig:allcolor} and Table~\ref{color},
strong BWB chromatisms were found for the $c-i$, $c-o$, $e-i$, and $e-m$ color behaviors, at $\gg$ 99.99\% confidence.
Meanwhile, BWB trends with the $i-o$ and $i-m$ colors were detected at a relatively low confidence level.
This may be because of the closely separated frequencies and the differences between the variation amplitudes at different wavelengths.
On the other side, \citet{1999ApJS..121..131F} claimed a RWB trend from their observations.

\subsection{1ES 1959+650}

1ES 1959+650, with a redshift of z = 0.047 \citep{1993ApJ...412..541S}, was classified as a HBL because its synchrotron bump peaked at UV--X-ray frequencies \citep{2010ApJ...719L.162B}. 
\citet{2003ApJ...583L...9H} found it to be a TeV emission source. 
The mass of its black hole was estimated to be $\sim 1.5 \times 10^8 M_{\odot}$ \citep{2002ApJ...569L..35F}.

Previous studies suggested that 1ES 1959+650 was highly variable in the optical wavelengths \citep[e.g.][]{2000A&AS..144..481V,2004ApJ...601..151K,2009ASPC..408..266K,2013ApJS..206...11S}. \citet{2000A&AS..144..481V} obtained a decrease of 0.28 mags within 4 days. 
\citet{2009ASPC..408..266K} reported a maximum variability of $\Delta R$=$1.07$ mags within 358 nights. 
Sorcia et al. (2013) found that the maximum and minimum $R$ magnitudes were 14.08 and 15.20, respectively, from 2009 to 2010.
\citet{2012MNRAS.420.3147G} did not detect the genuine IDV, whereas IDV was observed on two nights in 2009 and no quasi-periodicity was found in the long-term lightcurves of \citet{2015AJ....150...67Y}.
\citet{2017MNRAS.469.1682Z} also claimed no detection of IDV for 38 intra-day light curves and they argued that 
the larger variation amplitude in the higher frequency can be demonstrated by the BWB trend.

We observed this source in the $c$, $i$, and $o$ bands on 81, 81, and 71 nights, respectively, between 2009 July 03 and 2011 June 13.
There was one outburst at the first stage. After that, the light curves were characterized by continuous increase in brightness and the $i$ magnitude ran up to 14.760 on 2010 September 13 (JD 2455453). Several data points were recorded at the last stage of our observation and the object turned faint.
During the whole period of our monitoring, the light curves are quite "clean" and no fast flares were overlapped on long-term variability.
The variation amplitude tends to decrease with decreasing frequency (see Table~\ref{observations}).
No IDV was claimed because of the low sampling for this source and actually this is agree with no appearance of fast flares for our observations.
We did not detect any inter-band time lags (see Table~\ref{alldelay}).

As can be seen from Figure~\ref{fig:all}, the spectral indices become smaller when the brightness of the source increases.
One can see from Figure~\ref{fig:allcolor} that this source shows very strong BWB chromatisms in all three color-magnitude diagrams, at $\gg$ 99.99\% confidence.
This is in agreement with the result obtained by \citet{2015AJ....150...67Y,2017MNRAS.469.1682Z}.
The BWB color behaviors can be at least partly interpreted by the fact of increasing variation amplitude with increasing frequency, in agreement with \citet{2017MNRAS.469.1682Z}.

\subsection{BL Lacertae}

BL Lacertae, with a redshift of z = 0.069 \citep{1977ApJ...212L..47M}, is the prototype of BL Lac objects. 
The first bump of its SED peaks at optical/IR bands and it was classified as a LBL object \citep{1998MNRAS.299..433F}. 
\citet{2002ApJ...579..530W} found that the black hole mass is $1.7  \times 10^8 M_{\odot}$.
It is highly variable in all wavelengths ranging from radio to $\gamma$-ray.
\citet{2013MNRAS.436.1530R} suggested that the optical and $\gamma$-ray emitting regions were located upstream from the zone producing the mm radiation in the jet.

Within the optical bands, numerous investigations have been carried out to search for its flux variability and spectral behaviors \citep[e.g.][]{1992AJ....104...15C,2002A&A...390..407V,2006MNRAS.373..209H,2012MNRAS.425.3002G,2015A&A...573A..69W,2015MNRAS.450..541A,2017MNRAS.469.3588M}.
Most observations detected its IDV and revealed larger IDV amplitudes at higher frequencies. They also confirmed the presence of BWB trend, but yielded no evidence for periodicities.
\citet{1998A&A...332L...1N} did not find any significant time lags between different optical bands during a single night.
However, a possible time lag between the $e$ and $m$ bands was reported as $\sim$11.6 mins by \citet{2006MNRAS.373..209H}.
\citet{2017MNRAS.469.3588M} revealed possible time delay of $\sim$ 10 mins between the variations in the $V$ and $R$ bands.
\citet{2015A&A...573A..69W} observed a positive correlation in their entire dataset, albeit with a large scatter, and they argued  
the observed BWB behavior was intrinsic to the jet emission regions.

The observation period of BL Lacertae can be divided into two phases. 
In phase 1, the object was monitored on 10, 11, and 10 nights in the $e$, $i$, and $m$ bands, respectively, from JDs 2453918 to 2454063.  
In phase 2, the object was observed on 155, 162, and 161 nights respectively in the $c$, $i$, and $o$ bands from JDs 2454074 to 2455559.
The object kept fluctuating through the whole period. It showed two overall brightening trends in phase 1 and in the first stage of phase 2 and an overall darkening behavior in the last stage of phase 2. 
The light curves demonstrated several fluctuations before JD 2455200. Afterwards, this object dropped its brightness to a minimum of 15.042 in the $i$ band on 2010 November 12 (JD 2455513). With a turnover, the object brightened again.
As can be seen from Table~\ref{observations}, the variation amplitude decreases towards longer wavelengths in phase 2.


We detected the IDV of BL Lacertae in the $e$, $i$, and $m$ bands on 2006 November 6 by visual inspection and by using C, $\chi^2$, and F statistical tests. The presence of IDV in the $i$ and $m$ bands are confirmed by the ANOVA test. Previous work also claimed significant IDVs \citep[e.g.][]{1998AJ....115.2244W,1998A&A...332L...1N,2017MNRAS.469.3588M}.
The IDV light curves are displayed in Figure~\ref{fig:idv}.
The correlation analyses were performed for both the overall and IDV light curves and no time lags were found for both long- and short-term variability (see Table~\ref{alldelay}).

As can be seen from Figure~\ref{fig:all}, the spectral indices fluctuated slightly around 1.8.
On the color-magnitude diagram, BL Lacertae exhibits strong BWB trends with $c-i$, $c-o$, and $i-o$ colors at a high confidence level, while, no clear correlations were claimed between the $e-i$, $e-m$, and $i-m$ colors and their corresponding magnitudes due to very low confidence.
The observed BWB trends are in agreement with the previous results \citep{2002A&A...390..407V,2004A&A...421..103V,2006A&A...450...39G,2011PASJ...63..639I,2015A&A...582A.103G,2015A&A...573A..69W}.

\subsection{3C 454.3}

3C 454.3 is a highly variable blazar with a redshift of 0.859 and was classified as a FSRQ. 
Its black hole mass was 4.4$\times 10^9 M_{\odot}$ \citep{2001MNRAS.327.1111G}. It is one of the best studied blazars at all wavelengths.
In 2005 May, a dramatic outburst was recorded from the radio to X-ray energy \citep{2010ApJ...712..405V}.

In the optical region, its variability has been studied on different timescales by a lot of authors \citep[e.g.][]{2006A&A...453..817V,2007A&A...473..819R,2008ApJ...676L..13V,2011A&A...531A..90Z,2015MNRAS.451.3882A}.
In 2005 May, it reached a historical maximum with $R$ = 12.0 and then decayed to $R$ = 15.8 \citep{2006A&A...453..817V}. 
Based on the continuous monitoring of the object with the WEBT, a quiescent state from spring of 2006 to 2007 was reported by \citet{2007A&A...473..819R}.
After this quiescent state, the object again underwent multi-frequency flaring activity in 2007 July \citep{2008ApJ...676L..13V}.
\citet{2011A&A...531A..90Z} claimed it reached to a brightness of $R$ = 12.69. 
\citet{2015MNRAS.451.3882A} detected IDV for this target only in the $B$ and $R$ bands, and not in both $V$ and $I$ bands.

We observed this source on 39, 42, and 40 nights respectively in the $c$, $i$, and $o$ bands between 2007 July 19 and 2010 December 14. 
The object was at first at a high state and then dropped dramatically to 15.3 mags in the $i$ band.
At the second stage, the object kept getting faint and reached a minimum of 16.096 mags, which correspond to $R \sim 15.565$.
A dramatic 1.17 mags decline was detected from the second half of 2008 to early 2009 (around JD 2454800).
And then it brightened again and showed a strong flare with an amplitude of about 2 mags at the end.
One more data point at the beginning was obtained in the $i$ band when compared with the other two wavelength observations.
The overall variation amplitude was $\Delta i$=$2^{m}.688$ after removing that point.
The variation amplitude tends to increase with decreasing frequency.
On the other hand, we did not detect any IDV because of the low sampling on each single night.

One can see from Figure~\ref{fig:all} that the spectral indices became smaller when the brightness decreased.
In the color-magnitude diagram, as the only FSRQ in our sample, 3C 454.3 showed different levels of negative correlations, which were obviously different from other objects.
Significant RWB trends were found for the $c-i$ and $c-o$ colors, while the $i-o$ color had the same chromatism at $<$ 95\% confidence.
\citet{2011A&A...531A..90Z} found RWB chromatism in its faint states.

It has been suggested that a few components in the spectrum of 3C 454.3 may be responsible for the RWB phenomenon \citep{2008A&A...491..755R}. 
These components are the blue bumps between $\sim$ 2000 and 4000 $\AA$ in the rest frame. It is the combination of many emission features produced in the broad line region (BLR), in particular a forest of the Fe $\rm \rom{2}$ emission lines, the Mg $\rm \rom{2}$ lines, and the Balmer continuum. 
In the case of 3C 454.3, which is at a redshift of 0.859, its $c$-band flux is enhanced by the fluxes of the Fe $\rm \rom{2}$ lines.
The Fe $\rm \rom{2}$ lines are evident when the synchrotron radiation from jet is faint. When 3C 454.3 is in the low state, these less variable emission components dominate the fluxes and dilute the variability at high frequencies. So the variation amplitudes at higher frequencies are smaller than those at lower frequencies, as can be seen in Table~\ref{observations} for this object, which will lead to the RWB behaviors \citep{2007A&A...473..819R,2008A&A...491..755R,2011A&A...534A..87R,2010PASJ...62..645S,2011MNRAS.418.1640W,2011MNRAS.410..368B}.

\section{Conclusions and Discussions}

We performed long-term multi-color observations of ten blazars in the $c$, $e$, $i$, $m$, and $o$ bands from 2005 January 29 to 2011 June 13 by using the 60/90 cm Schmidt telescope located at the Xinglong Station of the NAOC. 
We collected 14799 data points. This is one of the largest optical database for a sample of blazars.
They can not only be used to study long-term flux and spectral variabilities, but also can be correlated with data in radio, X-ray, or gamma-ray wavelengths to investigate the broadband behaviors. 

Nine of our targets are BL Lac objects. Only 3C 454.3 belongs to FSRQs. 
Most of them were active and displayed small amplitude oscillating behavior superposed on the long-term variability, with the exception of OT 546. 
A total range of 0.5 mags in the $i$ band were attained in our 6-year time span for OT 546.
As a contrast, 3C 454.3 demonstrated a total amplitude of 2.93 mags in the $i$ band.

This project focused on the long-term variability of blazars. We collected only several data points within a short period for individual target on each single night, so we did not detect IDV for most objects. Only BL Lacertae was captured to show IDV on 2006 November 6.
We adopt the z-transformed discrete correlation functions (ZDCF; in autocorrelation mode) method to get a quantitative calculation IDV timescale \citep{1997ASSL..218..163A}.
We choose the minimum zero-crossing time of the DCF as the correlation timescale and get the timescale of variability of 2.13 hrs in the $i$ band (see Figure~\ref{fig:BHmass}). 
The minimum timescale $\Delta t_{\rm obs}$ will provide an upper limit to the mass of super-massive black hole (BH).
According to \citet{2012NewA...17....8G}, the mass of BH can be estimated by
\begin{equation}
   M_{\rm BH}=\frac{c^3\Delta t_{\rm obs}}{10G(1+z)}.
\end{equation}
For BL Lacertae, its $ M_{\rm BH}$ is calculated to be 1.47$\times 10^8  \rm M_{\odot}$.
This result is consistent with the BH mass of 0.1-6 $\times 10^8  \rm M_{\odot}$ obtained by \citet{1999A&AS..136...13F,2002ApJ...579..530W,2010MNRAS.402..497G,2010A&A...516A..59C,2012NewA...17....8G,2017MNRAS.469.3588M}.

In our observations, most BL Lac objects showed strong BWB trends except for PKS 0735+178, OI 090.4, and BL Lacertae in phase 1. 
PKS 0735+178 displayed BWB, RWB, or weak trends in the literatures \citep{2000ApJ...537..101F,2000ApJS..127...11G,2006A&A...450...39G}.
Our result on its color behavior was hindered by the few data points and relatively large measurement errors (see Figure~\ref{fig:allcolor}). 
OI 090.4 may be intrinsically achromatic when varying its brightness \citep{1997AJ....113.1995N}. 
Our few data points and relatively large measurement errors can hardly give any new result for this object. 
BL Lacertae illustrated strong BWB chromatism in phase 2 of our observations and in the literatures \citep[e.g.][]{2006A&A...450...39G,2011PASJ...63..639I}. 
The weak BWB in phase 1 should be resulted from the few data points.
The BWB color behaviors of most BL Lacs can be at least partly interpreted by the fact of increasing variation amplitude with increasing frequency observed in these objects.
A possible physical explanation for the observed BWB chromatism is that more than one mechanism are at work, as proposed by \citet{2004A&A...421..103V}.
The optical variability can be interpreted in terms of two components:  
"strong-chromatic" fast flares and "mild-chromatic" long-term variations.
The former are likely due to intrinsic shock-in-jet model \citep[e.g.][]{2002PASA...19..138M}.
The flattening of spectrum can be caused by increasing luminosity due to injection of instantaneous renewed electrons with the energy distribution harder than that of the previously cooled ones.
As the shock propagates down the jet, it strikes a region where the local plasma density is enhanced.
The radiations at different frequencies are then produced at different distances behind the shock.
Higher frequency radiations emerge sooner and closer to the shock front
than the lower frequency radiations, thus causing color variations.
\citet{2007A&A...470..857P} claimed the BWB "mild-chromatic" long-term variations can be explained in terms of
a variable Doppler factor $\delta$ due to changes in the view angle of a curved and inhomogeneous jet.
Most objects in our sample showed fast flares overlapped on long-term variability (see Figure~\ref{fig:all}), 
which is consistent with the suggestions of \citet{2004A&A...421..103V} and \citet{2007A&A...470..857P}. 
On the other hand, the FSRQ, 3C 454.3, displayed clear RWB behavior. 
Several less variable components, such as the Fe $\rm \rom{2}$ lines and Balmer continuum at short wavelengths, may be responsible for this kind of color behavior, as discussed by many authors \citep{2007A&A...473..819R,2008A&A...491..755R,2011A&A...534A..87R,2010PASJ...62..645S,2011MNRAS.418.1640W,2011MNRAS.410..368B}.
When 3C 454.3 is in the low state, these less variable components dominate the fluxes and dilute the variability at short wavelengths. Thus, the variation amplitudes at short wavelengths are smaller than those at long wavelengths, leading to a RWR behavior.

Our sample contains 3 LBLs, 2 IBLs, 4 HBLs, and 1 FSRQ. For 2 HBLs (Mrk 421 and PG 1553+113), we didn't get their spectral indices.
In Figure~\ref{fig:spectra1} we plotted the spectral index against the synchrotron peak frequency for eight objects except Mrk 421 and PG 1553+113.
The synchrotron peak frequencies of these eight objects are taken from \citet{2016ApJS..226...20F} and \citet{2010ApJ...716...30A}.
\citet{2010ApJ...716...30A} took three new acronyms (low synchrotron peaked blazars or LSP, intermediate synchrotron peaked blazars or ISP, and high synchrotron peaked blazars or HSP) to redefine all types of non-thermal-dominated AGNs depending on their synchrotron hump frequency, $\nu_{peak}$. They suggested the following classification: LSP with $\nu_{peak} \le 10^{14} $ Hz; ISP with $10^{14} < \nu_{peak} < 10^{15} $ Hz and HSP with $\nu_{peak} \ge 10^{15}$ Hz. The large majority of FSRQs are of the LSP type.

As we can see from Figure~\ref{fig:spectra1}, our 3 LBLs and 1 FSRQ, 3C 454.3, belong to LSP.
Thus we expect their optical spectral index should be greater than 1 due to the fact that they are on the descending part of the synchrotron peak.
Except for BL Lacertae, the other two LBLs have spectral indices around 1.5.
In the case of radiative losses dominated by Synchrotron Self-Compton (SSC) emission, the time-averaged synchrotron spectrum should be 
   $J_{syn}  \propto \nu^{-(2+p)/4}$.
For a particle distribution in the power-law form $N(E)=N_0E^{-p}$, the synchrotron emission above the break energy $(Be\hbar \gamma_{max}^2)/(m_ec^3)$ will have a spectral shape $F_{\nu} \propto \nu^{-3/2}$.
\citet{2002ApJ...564...92C} further demonstrated that the SSC-loss-dominated synchrotron emission exhibits a distribution of spectral indices around 1.5 in the optical regime as the particle injection index varies.
Then, the fact that 2 LBLs have spectral indices of about 1.5 can be interpreted with the SSC model.
For 3C 454.3, even though its optical emission at short wavelengths is contaminated by the blue bumps as mentioned above, the average spectral index is 1.486.
This indicates that the optical continuum of 3C 454.3 is likely dominated by the synchrotron component.
This interpretation is also supported by the fact that 3C 454.3 shows strong variability in our observations.

Our 2 IBLs show intermediate synchrotron hump frequencies, belong to ISP. Their average spectral indices are 1.058 and 1.384, respectively, which are in agreement with the expected value of $\alpha \ge 1$. 
This also indicates that the synchrotron process dominates their optical emission.

For HSP, the predicted spectral index under simple synchrotron emission model should be less than or equal to 0.80, corresponding to the optically thin synchrotron emission \citep{1995PASP..107..803U}. 
However the spectral indices of our 2 HBLs, OT 546 and 1ES 1959+650, are about 1.816 and 1.08, respectively.
This indicates that the optical emission of these two HSP is contaminated by other components such as the thermal contribution from the host galaxies.
For the nearby blazars, the host galaxy contribution is relatively important \citep{1994ApJ...432..547P,1996MNRAS.283..241F}.
For example, \citet{2012MNRAS.425.3002G} determined the spectral index of 1ES 1959+650 to be 0.67 after the subtraction of the host galaxy contribution.
In addition, the Galactic absorption has a great influence on 1ES 1959+650.
Before the dereddening, the average spectral index was 1.59. It is necessary for this target to correct the Galactic absorption.

It will be interesting to check the changes of the SEDs in response to the brightness variations for these blazars.
In Figure~\ref{fig:sed}, the SEDs at the highest, intermediate, and lowest states are displayed for those eight objects with spectral index calculation in Figure~\ref{fig:all}.
The SEDs are compiled by using the fluxes at the $c$, $i$, and $o$ wavelengths. The observations in the $e$ and $m$ bands were made only for OT 546 and BL Lacertae. They have either much fewer data points (for BL Lacertae) or much larger measurement errors (for OT 546) than in the $c$ and $o$ bands.
So we did not use them in the SEDs.
The slopes of the SEDs changed clearly between different states in all eight objects. 
This is consistent with the color behavior demonstrated in Figure~\ref{fig:allcolor}. 
More than half SEDs can be well fitted with a power-law, especially for 3C 66A, ON 231, and BL Lacertae, whose SEDs at three states are all power-laws. 
The lowest SEDs of PKS 0735+178 and OT 546 and the intermediate SED of OI 090.4 show concave shapes. 
However, because of the relatively large errors in the fluxes, the concave shapes are of relatively low confidence level and the precise shapes are unknown. 
The highest SED of OT 546 has a significant concave shape. This is quite unexpected even the fluxes may be polluted by the contribution of its host galaxy. We then plotted its second highest SED, which is clearly a power-law. Given the small variation amplitudes of OT 546 in all five passbands, such a significant change in its SED is unexpected and spurious measurement might occur in one or more passbands in the highest state. 
The lowest SED of 3C 454.3 has also a concave shape. The excess in the $c$ band should be the result of the contribution of the less variable blue bump. 
The three SEDs of 1ES 1959+650 all have convex shapes. As a low redshift object, its host galaxy is expected to contribute more flux in the $i$ band than in the $c$ and $o$ bands, which leads to the observed convex SEDs. 
\citet{2009ApJ...694..174B} compiled six convex SEDs for 3C 66A in the optical-near-IR domain. This is not confirmed in our work due partly to our less sampling in frequency. 
\citet{2014A&A...562A..79S} presented the optical-IR SEDs for seven blazars. Their SEDs of PKS 0735+178 and ON 231 are basically power-laws and are consistent with our results. 
The SEDs with a complete coverage in the UV-optical-IR will help reveal the possible shift of the synchrotron peak with varying flux for at least the LBL/IBL objects \citep[e.g.][]{2000A&A...363..108V,2007ApJ...657L..81F}.
However, this is beyond the scope of this work.

Depending on the injection and acceleration of relativistic electrons and subsequent radiative cooling and escape process, the observed inter-band time lags are expected \citep[e.g.][]{1998A&A...333..452K,2004ApJ...613..725S}.
In the optical regime, a few lags have been reported in several blazars \citep{2000AJ....120.1192R,2000PASJ...52.1075Q,2003A&A...397..565P,2006MNRAS.366.1337S,2009ApJS..185..511P,2012AJ....143..108W,2017MNRAS.469.3588M}.
For example, \citet{2009ApJS..185..511P} found a possible lag of about 11 mins between $B$ and $I$ band variations.
\citet{2000A&A...363..108V} presented an upper limit to the possible 10 mins delay between $B$ and $I$ variations.
\citet{2012AJ....143..108W} reported the variability at the $B'$ and $V'$ bands leads that at the $R'$ band by 30 mins.
\citet{2017MNRAS.469.3588M} derived a $\sim$10 mins lag between variations in the $V$ and $R$ bands.
We did the inter-band correlation analysis for 3C 66A, ON 231, PG 1553+113, 1ES 1959+650, and BL Lacertae.
No time delay was identified for these targets at long and short timescales.
The previous lags are mainly focus on a short period.
Actually, most authors claimed no detection of inter-band delays in the optical regime, especially for long-term variability \citep[e.g.][]{2010RAA....10..125H,2013MNRAS.431.2481D,2015ApJS..218...18D}.
Only several data points were collected for our observations for individual target on each single night 
and were not enough for the detection of time lags.

\section*{Acknowledgements}

We would like to thank the anonymous referee for insightful comments
and suggestions to improve this manuscript. This work has been supported by the Chinese National Natural Science Foundation grants U1531242.

\bibliography{2017apjs}

\begin{thebibliography}{}
\expandafter\ifx\csname natexlab\endcsname\relax\def\natexlab#1{#1}\fi

\bibitem[{{Abdo} {et~al.}(2010){Abdo}, {Ackermann}, {Agudo}, {Ajello}, {Aller},
  {Aller}, {Angelakis}, {Arkharov}, {Axelsson}, {Bach}, \&
  et~al.}]{2010ApJ...716...30A}
{Abdo}, A.~A., {Ackermann}, M., {Agudo}, I., {et~al.} 2010, \apj, 716, 30

\bibitem[{{Abraham} {et~al.}(1991){Abraham}, {Crawford}, \&
  {McHardy}}]{1991MNRAS.252..482A}
{Abraham}, R.~G., {Crawford}, C.~S., \& {McHardy}, I.~M. 1991, \mnras, 252, 482

\bibitem[{{Acciari} {et~al.}(2009){Acciari}, {Aliu}, {Arlen}, {Beilicke},
  {Benbow}, {B{\"o}ttcher}, {Bradbury}, {Buckley}, {Bugaev}, {Butt}, {Byrum},
  {Cannon}, {Celik}, {Cesarini}, {Chow}, {Ciupik}, {Cogan}, {Cui}, {Daniel},
  {Dickherber}, {Ergin}, {Falcone}, {Fegan}, {Finley}, {Fortin}, {Fortson},
  {Furniss}, {Gall}, {Gibbs}, {Gillanders}, {Godambe}, {Grube}, {Guenette},
  {Gyuk}, {Hanna}, {Hays}, {Holder}, {Horan}, {Hui}, {Humensky}, {Imran},
  {Kaaret}, {Karlsson}, {Kertzman}, {Kieda}, {Kildea}, {Konopelko},
  {Krawczynski}, {Krennrich}, {Lang}, {LeBohec}, {Maier}, {McCann},
  {McCutcheon}, {Millis}, {Moriarty}, {Mukherjee}, {Nagai}, {Ong}, {Otte},
  {Pandel}, {Perkins}, {Petry}, {Pizlo}, {Pohl}, {Quinn}, {Ragan}, {Reyes},
  {Reynolds}, {Roache}, {Rose}, {Schroedter}, {Sembroski}, {Smith}, {Steele},
  {Swordy}, {Theiling}, {Toner}, {Varlotta}, {Vassiliev}, {Wagner}, {Wakely},
  {Ward}, {Weekes}, {Weinstein}, {Williams}, {Wissel}, {Wood}, \&
  {Zitzer}}]{2009ApJ...693L.104A}
{Acciari}, V.~A., {Aliu}, E., {Arlen}, T., {et~al.} 2009, \apjl, 693, L104

\bibitem[{{Agarwal} \& {Gupta}(2015)}]{2015MNRAS.450..541A}
{Agarwal}, A., \& {Gupta}, A.~C. 2015, \mnras, 450, 541

\bibitem[{{Agarwal} {et~al.}(2015){Agarwal}, {Gupta}, {Bachev}, {Strigachev},
  {Semkov}, {Wiita}, {B{\"o}ttcher}, {Boeva}, {Gaur}, {Gu}, {Peneva},
  {Ibryamov}, \& {Pandey}}]{2015MNRAS.451.3882A}
{Agarwal}, A., {Gupta}, A.~C., {Bachev}, R., {et~al.} 2015, \mnras, 451, 3882

\bibitem[{{Aharonian} {et~al.}(2006){Aharonian}, {Akhperjanian}, {Bazer-Bachi},
  {Beilicke}, {Benbow}, {Berge}, {Bernl{\"o}hr}, {Boisson}, {Bolz}, {Borrel},
  {Braun}, {Breitling}, {Brown}, {B{\"u}hler}, {Carrigan}, {Chadwick},
  {Chounet}, {Cornils}, {Costamante}, {Degrange}, {Dickinson},
  {Djannati-Ata{\"i}}, {O'C.~Drury}, {Dubus}, {Egberts}, {Emmanoulopoulos},
  {Espigat}, {Feinstein}, {Fontaine}, {Funk}, {Gallant}, {Giebels},
  {Glicenstein}, {Goret}, {Hadjichristidis}, {Hauser}, {Hauser}, {Heinzelmann},
  {Henri}, {Hermann}, {Hinton}, {Hofmann}, {Holleran}, {Horns}, {Jacholkowska},
  {de Jager}, {Kh{\'e}lifi}, {Komin}, {Konopelko}, {Latham}, {Le Gallou},
  {Lemi{\`e}re}, {Lemoine-Goumard}, {Lohse}, {Martin}, {Martineau-Huynh},
  {Marcowith}, {Masterson}, {McComb}, {de Naurois}, {Nedbal}, {Nolan},
  {Noutsos}, {Orford}, {Osborne}, {Ouchrif}, {Panter}, {Pelletier}, {Pita},
  {P{\"u}hlhofer}, {Punch}, {Raubenheimer}, {Raue}, {Rayner}, {Reimer},
  {Reimer}, {Ripken}, {Rob}, {Rolland}, {Rowell}, {Sahakian}, {Saug{\'e}},
  {Schlenker}, {Schlickeiser}, {Schuster}, {Schwanke}, {Siewert}, {Sol},
  {Spangler}, {Steenkamp}, {Stegmann}, {Superina}, {Tavernet}, {Terrier},
  {Th{\'e}oret}, {Tluczykont}, {van Eldik}, {Vasileiadis}, {Venter}, {Vincent},
  {V{\"o}lk}, {Wagner}, \& {Ward}}]{2006A&A...448L..19A}
{Aharonian}, F., {Akhperjanian}, A.~G., {Bazer-Bachi}, A.~R., {et~al.} 2006,
  \aap, 448, L19

\bibitem[{{Albert} {et~al.}(2007){Albert}, {Aliu}, {Anderhub}, {Antoranz},
  {Armada}, {Baixeras}, {Barrio}, {Bartko}, {Bastieri}, {Becker}, {Bednarek},
  {Berger}, {Bigongiari}, {Biland}, {Bock}, {Bordas}, {Bosch-Ramon}, {Bretz},
  {Britvitch}, {Camara}, {Carmona}, {Chilingarian}, {Ciprini}, {Coarasa},
  {Commichau}, {Contreras}, {Cortina}, {Curtef}, {Danielyan}, {Dazzi}, {De
  Angelis}, {de los Reyes}, {De Lotto}, {Domingo-Santamar{\'{\i}}a}, {Dorner},
  {Doro}, {Errando}, {Fagiolini}, {Ferenc}, {Fern{\'a}ndez}, {Firpo}, {Flix},
  {Fonseca}, {Font}, {Fuchs}, {Galante}, {Garczarczyk}, {Gaug}, {Giller},
  {Goebel}, {Hakobyan}, {Hayashida}, {Hengstebeck}, {H{\"o}hne}, {Hose}, {Hsu},
  {Jacon}, {Jogler}, {Kalekin}, {Kosyra}, {Kranich}, {Kritzer}, {Laille},
  {Liebing}, {Lindfors}, {Lombardi}, {Longo}, {L{\'o}pez}, {L{\'o}pez},
  {Lorenz}, {Majumdar}, {Maneva}, {Mannheim}, {Mansutti}, {Mariotti},
  {Mart{\'{\i}}nez}, {Mazin}, {Merck}, {Meucci}, {Meyer}, {Miranda},
  {Mirzoyan}, {Mizobuchi}, {Moralejo}, {Nilsson}, {Ninkovic},
  {O{\~n}a-Wilhelmi}, {Otte}, {Oya}, {Paneque}, {Paoletti}, {Paredes},
  {Pasanen}, {Pascoli}, {Pauss}, {Pegna}, {Persic}, {Peruzzo}, {Piccioli},
  {Poller}, {Puchades}, {Prandini}, {Raymers}, {Rhode}, {Rib{\'o}}, {Rico},
  {Rissi}, {Robert}, {R{\"u}gamer}, {Saggion}, {S{\'a}nchez}, {Sartori},
  {Scalzotto}, {Scapin}, {Schmitt}, {Schweizer}, {Shayduk}, {Shinozaki},
  {Sidro}, {Sillanp{\"a}{\"a}}, {Sobczynska}, {Stamerra}, {Stark}, {Takalo},
  {Temnikov}, {Tescaro}, {Teshima}, {Tonello}, {Torres}, {Turini}, {Vankov},
  {Vitale}, {Wagner}, {Wibig}, {Wittek}, {Zanin}, \&
  {Zapatero}}]{2007ApJ...654L.119A}
{Albert}, J., {Aliu}, E., {Anderhub}, H., {et~al.} 2007, \apjl, 654, L119

\bibitem[{{Aleksi{\'c}} {et~al.}(2014){Aleksi{\'c}}, {Antonelli}, {Antoranz},
  {Asensio}, {Backes}, {Barres de Almeida}, {Barrio}, {Becerra Gonz{\'a}lez},
  {Bednarek}, {Berger}, {Bernardini}, {Biland}, {Blanch}, {Bock}, {Boller},
  {Bonnefoy}, {Bonnoli}, {Borla Tridon}, {Borracci}, {Bretz}, {Carmona},
  {Carosi}, {Carreto Fidalgo}, {Colin}, {Colombo}, {Contreras}, {Cortina},
  {Cossio}, {Covino}, {da Vela}, {Dazzi}, {de Angelis}, {de Caneva}, {de
  Lotto}, {Delgado Mendez}, {Doert}, {Dom{\'{\i}}nguez}, {Dominis Prester},
  {Dorner}, {Doro}, {Eisenacher}, {Elsaesser}, {Farina}, {Ferenc}, {Fonseca},
  {Font}, {Fruck}, {Garc{\'{\i}}a L{\'o}pez}, {Garczarczyk}, {Garrido Terrats},
  {Gaug}, {Giavitto}, {Godinovi{\'c}}, {Gonz{\'a}lez Mu{\~n}oz}, {Gozzini},
  {Hadamek}, {Hadasch}, {H{\"a}fner}, {Herrero}, {Hose}, {Hrupec}, {Idec},
  {Jankowski}, {Kadenius}, {Klepser}, {Knoetig}, {Kr{\"a}henb{\"u}hl},
  {Krause}, {Kushida}, {La Barbera}, {Lelas}, {Lewandowska}, {Lindfors},
  {Lombardi}, {L{\'o}pez}, {L{\'o}pez-Coto}, {L{\'o}pez-Oramas}, {Lorenz},
  {Lozano}, {Makariev}, {Mallot}, {Maneva}, {Mankuzhiyil}, {Mannheim},
  {Maraschi}, {Marcote}, {Mariotti}, {Mart{\'{\i}}nez}, {Masbou}, {Mazin},
  {Meucci}, {Miranda}, {Mirzoyan}, {Mold{\'o}n}, {Moralejo}, {Munar-Adrover},
  {Nakajima}, {Niedzwiecki}, {Nilsson}, {Nowak}, {Orito}, {Paiano},
  {Palatiello}, {Paneque}, {Paoletti}, {Paredes}, {Partini}, {Persic}, {Prada},
  {Prada Moroni}, {Prandini}, {Puljak}, {Reichardt}, {Reinthal}, {Rhode},
  {Rib{\'o}}, {Rico}, {R{\"u}gamer}, {Saggion}, {Saito}, {Saito}, {Salvati},
  {Satalecka}, {Scalzotto}, {Scapin}, {Schultz}, {Schweizer}, {Shore},
  {Sillanp{\"a}{\"a}}, {Sitarek}, {Snidaric}, {Sobczynska}, {Spanier}, {Spiro},
  {Stamatescu}, {Stamerra}, {Steinke}, {Storz}, {Sun}, {Suri{\'c}}, {Takalo},
  {Takami}, {Tavecchio}, {Temnikov}, {Terzi{\'c}}, {Tescaro}, {Teshima},
  {Tibolla}, {Torres}, {Toyama}, {Treves}, {Uellenbeck}, {Vogler}, {Wagner},
  {Weitzel}, {Zandanel}, {Zanin}, \& {MAGIC
  Collaboration}}]{2014A&A...563A..90A}
{Aleksi{\'c}}, J., {Antonelli}, L.~A., {Antoranz}, P., {et~al.} 2014, \aap,
  563, A90

\bibitem[{{Alexander}(1997)}]{1997ASSL..218..163A}
{Alexander}, T. 1997, in Astrophysics and Space Science Library, Vol. 218,
  Astronomical Time Series, ed. D.~{Maoz}, A.~{Sternberg}, \& E.~M.
  {Leibowitz}, 163

\bibitem[{{Aliu} {et~al.}(2009){Aliu}, {Anderhub}, {Antonelli}, {Antoranz},
  {Backes}, {Baixeras}, {Balestra}, {Barrio}, {Bartko}, {Bastieri}, {Becerra
  Gonz{\'a}lez}, {Becker}, {Bednarek}, {Berger}, {Bernardini}, {Biland},
  {Bock}, {Bonnoli}, {Bordas}, {Borla Tridon}, {Bosch-Ramon}, {Bretz},
  {Britvitch}, {Camara}, {Carmona}, {Chilingarian}, {Commichau}, {Contreras},
  {Cortina}, {Costado}, {Covino}, {Curtef}, {Dazzi}, {DeAngelis}, {DeCea del
  Pozo}, {de los Reyes}, {DeLotto}, {DeMaria}, {DeSabata}, {Delgado Mendez},
  {Dominguez}, {Dorner}, {Doro}, {Elsaesser}, {Errando}, {Ferenc},
  {Fern{\'a}ndez}, {Firpo}, {Fonseca}, {Font}, {Galante}, {Garc{\'{\i}}a
  L{\'o}pez}, {Garczarczyk}, {Gaug}, {Goebel}, {Hadasch}, {Hayashida},
  {Herrero}, {H{\"o}hne-M{\"o}nch}, {Hose}, {Hsu}, {Huber}, {Jogler},
  {Kranich}, {La Barbera}, {Laille}, {Leonardo}, {Lindfors}, {Lombardi},
  {Longo}, {L{\'o}pez}, {Lorenz}, {Majumdar}, {Maneva}, {Mankuzhiyil},
  {Mannheim}, {Maraschi}, {Mariotti}, {Mart{\'{\i}}nez}, {Mazin}, {Meucci},
  {Meyer}, {Miranda}, {Mirzoyan}, {Mold{\'o}n}, {Moles}, {Moralejo}, {Nieto},
  {Nilsson}, {Ninkovic}, {Otte}, {Oya}, {Paoletti}, {Paredes}, {Pasanen},
  {Pascoli}, {Pauss}, {Pegna}, {Perez-Torres}, {Persic}, {Peruzzo}, {Prada},
  {Prandini}, {Puchades}, {Raymers}, {Rhode}, {Rib{\'o}}, {Rico}, {Rissi},
  {Robert}, {R{\"u}gamer}, {Saggion}, {Saito}, {Salvati}, {Sanchez-Conde},
  {Sartori}, {Satalecka}, {Scalzotto}, {Scapin}, {Schweizer}, {Shayduk},
  {Shinozaki}, {Shore}, {Sidro}, {Sierpowska-Bartosik}, {Sillanp{\"a}{\"a}},
  {Sitarek}, {Sobczynska}, {Spanier}, {Stamerra}, {Stark}, {Takalo},
  {Tavecchio}, {Temnikov}, {Tescaro}, {Teshima}, {Tluczykont}, {Torres},
  {Turini}, {Vankov}, {Venturini}, {Vitale}, {Wagner}, {Wittek}, {Zabalza},
  {Zandanel}, {Zanin}, \& {Zapatero}}]{2009ApJ...692L..29A}
{Aliu}, E., {Anderhub}, H., {Antonelli}, L.~A., {et~al.} 2009, \apjl, 692, L29

\bibitem[{{Aliu} {et~al.}(2015){Aliu}, {Archer}, {Aune}, {Barnacka}, {Behera},
  {Beilicke}, {Benbow}, {Berger}, {Bird}, {Buckley}, {Bugaev}, {Byrum},
  {Cardenzana}, {Cerruti}, {Chen}, {Ciupik}, {Connolly}, {Cui}, {Dickinson},
  {Dumm}, {Eisch}, {Errando}, {Falcone}, {Federici}, {Feng}, {Finley},
  {Fortin}, {Fortson}, {Furniss}, {Galante}, {Gillanders}, {Griffin},
  {Griffiths}, {Grube}, {Gyuk}, {H{\aa}kansson}, {Hanna}, {Holder}, {Hughes},
  {Humensky}, {Johnson}, {Kaaret}, {Kar}, {Kertzman}, {Khassen}, {Kieda},
  {Krawczynski}, {Krennrich}, {Kumar}, {Lang}, {Madhavan}, {McArthur},
  {McCann}, {Meagher}, {Millis}, {Moriarty}, {Nieto}, {O'Faol{\'a}in de
  Bhr{\'o}ithe}, {Ong}, {Orr}, {Otte}, {Park}, {Perkins}, {Pohl}, {Popkow},
  {Prokoph}, {Pueschel}, {Quinn}, {Ragan}, {Rajotte}, {Reyes}, {Reynolds},
  {Richards}, {Roache}, {Sembroski}, {Shahinyan}, {Staszak}, {Telezhinsky},
  {Tucci}, {Tyler}, {Varlotta}, {Vassiliev}, {Wakely}, {Weinstein}, {Welsing},
  {Wilhelm}, {Williams}, \& {Zitzer}}]{2015ApJ...799....7A}
{Aliu}, E., {Archer}, A., {Aune}, T., {et~al.} 2015, \apj, 799, 7

\bibitem[{{Angel} \& {Stockman}(1980)}]{1980ARA&A..18..321A}
{Angel}, J.~R.~P., \& {Stockman}, H.~S. 1980, \araa, 18, 321

\bibitem[{{Archambault} {et~al.}(2015){Archambault}, {Archer}, {Beilicke},
  {Benbow}, {Bird}, {Biteau}, {Bouvier}, {Bugaev}, {Cardenzana}, {Cerruti},
  {Chen}, {Ciupik}, {Connolly}, {Cui}, {Dickinson}, {Dumm}, {Eisch}, {Errando},
  {Falcone}, {Feng}, {Finley}, {Fleischhack}, {Fortin}, {Fortson}, {Furniss},
  {Gillanders}, {Griffin}, {Griffiths}, {Grube}, {Gyuk}, {H{\aa}kansson},
  {Hanna}, {Holder}, {Humensky}, {Johnson}, {Kaaret}, {Kar}, {Kertzman},
  {Khassen}, {Kieda}, {Krause}, {Krennrich}, {Kumar}, {Lang}, {Maier},
  {McArthur}, {McCann}, {Meagher}, {Millis}, {Moriarty}, {Mukherjee}, {Nieto},
  {O'Faol{\'a}in de Bhr{\'o}ithe}, {Ong}, {Otte}, {Park}, {Pohl}, {Popkow},
  {Prokoph}, {Pueschel}, {Quinn}, {Ragan}, {Reyes}, {Reynolds}, {Richards},
  {Roache}, {Santander}, {Sembroski}, {Shahinyan}, {Smith}, {Staszak},
  {Telezhinsky}, {Tucci}, {Tyler}, {Varlotta}, {Vincent}, {Wakely},
  {Weinstein}, {Welsing}, {Wilhelm}, {Williams}, {Zitzer}, {Veritas
  Collaboration}, \& {Hughes}}]{2015ApJ...808..110A}
{Archambault}, S., {Archer}, A., {Beilicke}, M., {et~al.} 2015, \apj, 808, 110

\bibitem[{{Baumert}(1980)}]{1980PASP...92..156B}
{Baumert}, J.~H. 1980, \pasp, 92, 156

\bibitem[{{Beckmann} {et~al.}(2002){Beckmann}, {Wolter}, {Celotti},
  {Costamante}, {Ghisellini}, {Maccacaro}, \&
  {Tagliaferri}}]{2002A&A...383..410B}
{Beckmann}, V., {Wolter}, A., {Celotti}, A., {et~al.} 2002, \aap, 383, 410

\bibitem[{{Bonning} {et~al.}(2012){Bonning}, {Urry}, {Bailyn}, {Buxton},
  {Chatterjee}, {Coppi}, {Fossati}, {Isler}, \&
  {Maraschi}}]{2012ApJ...756...13B}
{Bonning}, E., {Urry}, C.~M., {Bailyn}, C., {et~al.} 2012, \apj, 756, 13

\bibitem[{{Bonnoli} {et~al.}(2011){Bonnoli}, {Ghisellini}, {Foschini},
  {Tavecchio}, \& {Ghirlanda}}]{2011MNRAS.410..368B}
{Bonnoli}, G., {Ghisellini}, G., {Foschini}, L., {Tavecchio}, F., \&
  {Ghirlanda}, G. 2011, \mnras, 410, 368

\bibitem[{{Bottacini} {et~al.}(2010){Bottacini}, {B{\"o}ttcher}, {Schady},
  {Rau}, {Zhang}, {Ajello}, {Fendt}, \& {Greiner}}]{2010ApJ...719L.162B}
{Bottacini}, E., {B{\"o}ttcher}, M., {Schady}, P., {et~al.} 2010, \apjl, 719,
  L162

\bibitem[{{B{\"o}ttcher}(2007)}]{2007Ap&SS.309...95B}
{B{\"o}ttcher}, M. 2007, \apss, 309, 95

\bibitem[{{B{\"o}ttcher} {et~al.}(2005){B{\"o}ttcher}, {Harvey}, {Joshi},
  {Villata}, {Raiteri}, {Bramel}, {Mukherjee}, {Savolainen}, {Cui}, {Fossati},
  {Smith}, {Able}, {Aller}, {Aller}, {Arkharov}, {Augusteijn}, {Baliyan},
  {Barnaby}, {Berdyugin}, {Ben{\'{\i}}tez}, {Boltwood}, {Carini}, {Carosati},
  {Ciprini}, {Coloma}, {Crapanzano}, {de Diego}, {Di Paola}, {Dolci}, {Fan},
  {Frasca}, {Hagen-Thorn}, {Horan}, {Ibrahimov}, {Kimeridze}, {Kovalev},
  {Kovalev}, {Kurtanidze}, {L{\"a}hteenm{\"a}ki}, {Lanteri}, {Larionov},
  {Larionova}, {Lindfors}, {Marilli}, {Mirabal}, {Nikolashvili}, {Nilsson},
  {Ohlert}, {Ohnishi}, {Oksanen}, {Ostorero}, {Oyer}, {Papadakis}, {Pasanen},
  {Poteet}, {Pursimo}, {Sadakane}, {Sigua}, {Takalo}, {Tartar},
  {Ter{\"a}sranta}, {Tosti}, {Walters}, {Wiik}, {Wilking}, {Wills}, {Xilouris},
  {Fletcher}, {Gu}, {Lee}, {Pak}, \& {Yim}}]{2005ApJ...631..169B}
{B{\"o}ttcher}, M., {Harvey}, J., {Joshi}, M., {et~al.} 2005, \apj, 631, 169

\bibitem[{{B{\"o}ttcher} {et~al.}(2009){B{\"o}ttcher}, {Fultz}, {Aller},
  {Aller}, {Apodaca}, {Arkharov}, {Bach}, {Bachev}, {Berdyugin}, {Buemi},
  {Calcidese}, {Carosati}, {Charlot}, {Ciprini}, {Paola}, {Dolci}, {Efimova},
  {Scurrats}, {Frasca}, {Gupta}, {Hagen-Thorn}, {Heidt}, {Hiriart},
  {Konstantinova}, {Kopatskaya}, {L{\"a}hteenm{\"a}ki}, {Lanteri}, {Larionov},
  {LeCampion}, {Leto}, {Lindfors}, {Mihov}, {Nieppola}, {Nilsson}, {Ovcharov},
  {P{\"a}{\"a}kk{\"o}nen}, {Pasanen}, {Ragozzine}, {Raiteri}, {Ros}, {Sadun},
  {Sanchez}, {Semkov}, {Sorcia}, {Strigachev}, {Takalo}, {Tornikoski},
  {Trigilio}, {Umana}, {Valcheva}, {Villata}, {Volvach}, {Wu}, \&
  {Zhou}}]{2009ApJ...694..174B}
{B{\"o}ttcher}, M., {Fultz}, K., {Aller}, H.~D., {et~al.} 2009, \apj, 694, 174

\bibitem[{{Browne}(1971)}]{1971Natur.231..515B}
{Browne}, I.~W.~A. 1971, \nat, 231, 515

\bibitem[{{Capetti} {et~al.}(2010){Capetti}, {Raiteri}, \&
  {Buttiglione}}]{2010A&A...516A..59C}
{Capetti}, A., {Raiteri}, C.~M., \& {Buttiglione}, S. 2010, \aap, 516, A59

\bibitem[{{Carangelo} {et~al.}(2003){Carangelo}, {Falomo}, {Kotilainen},
  {Treves}, \& {Ulrich}}]{2003A&A...412..651C}
{Carangelo}, N., {Falomo}, R., {Kotilainen}, J., {Treves}, A., \& {Ulrich},
  M.-H. 2003, \aap, 412, 651

\bibitem[{{Carini} \& {Miller}(1991)}]{1991BAAS...23.1420C}
{Carini}, M.~T., \& {Miller}, H.~R. 1991, in \baas, Vol.~23, Bulletin of the
  American Astronomical Society, 1420

\bibitem[{{Carini} {et~al.}(1992){Carini}, {Miller}, {Noble}, \&
  {Goodrich}}]{1992AJ....104...15C}
{Carini}, M.~T., {Miller}, H.~R., {Noble}, J.~C., \& {Goodrich}, B.~D. 1992,
  \aj, 104, 15

\bibitem[{{Carswell} {et~al.}(1974){Carswell}, {Strittmatter}, {Williams},
  {Kinman}, \& {Serkowski}}]{1974ApJ...190L.101C}
{Carswell}, R.~F., {Strittmatter}, P.~A., {Williams}, R.~E., {Kinman}, T.~D.,
  \& {Serkowski}, K. 1974, \apjl, 190, L101

\bibitem[{{Chen} {et~al.}(2014){Chen}, {Hu}, {Guo}, \&
  {Du}}]{2014Ap&SS.349..909C}
{Chen}, X., {Hu}, S.~M., {Guo}, D.~F., \& {Du}, J.~J. 2014, \apss, 349, 909

\bibitem[{{Cheng} {et~al.}(2013){Cheng}, {Zhang}, \&
  {Xu}}]{2013MNRAS.429.2773C}
{Cheng}, X.-L., {Zhang}, Y.-H., \& {Xu}, L. 2013, \mnras, 429, 2773

\bibitem[{{Chiang} \& {B{\"o}ttcher}(2002)}]{2002ApJ...564...92C}
{Chiang}, J., \& {B{\"o}ttcher}, M. 2002, \apj, 564, 92

\bibitem[{{Ciprini} {et~al.}(2007){Ciprini}, {Takalo}, {Tosti}, {Raiteri},
  {Fiorucci}, {Villata}, {Nucciarelli}, {Lanteri}, {Nilsson}, \&
  {Ros}}]{2007A&A...467..465C}
{Ciprini}, S., {Takalo}, L.~O., {Tosti}, G., {et~al.} 2007, \aap, 467, 465

\bibitem[{{Condon} {et~al.}(1977){Condon}, {Hicks}, \&
  {Jauncey}}]{1977AJ.....82..692C}
{Condon}, J.~J., {Hicks}, P.~D., \& {Jauncey}, D.~L. 1977, \aj, 82, 692

\bibitem[{{Dai} {et~al.}(2015){Dai}, {Zeng}, {Jiang}, {Fan}, {Hu}, {Zhang},
  {Yang}, {Yan}, {Wang}, \& {Zhang}}]{2015ApJS..218...18D}
{Dai}, B.-z., {Zeng}, W., {Jiang}, Z.-j., {et~al.} 2015, \apjs, 218, 18

\bibitem[{{Dai} {et~al.}(2013){Dai}, {Wu}, {Zhu}, {Zhou}, {Ma}, {Yuan}, \&
  {Wang}}]{2013ApJS..204...22D}
{Dai}, Y., {Wu}, J., {Zhu}, Z.-H., {et~al.} 2013, \apjs, 204, 22

\bibitem[{{D'Ammando} {et~al.}(2013){D'Ammando}, {Antolini}, {Tosti}, {Finke},
  {Ciprini}, {Larsson}, {Ajello}, {Covino}, {Gasparrini}, {Gurwell}, {Hauser},
  {Romano}, {Schinzel}, {Wagner}, {Impiombato}, {Perri}, {Persic}, {Pian},
  {Polenta}, {Sbarufatti}, {Treves}, {Vercellone}, {Wehrle}, \&
  {Zook}}]{2013MNRAS.431.2481D}
{D'Ammando}, F., {Antolini}, E., {Tosti}, G., {et~al.} 2013, \mnras, 431, 2481

\bibitem[{{Danforth} {et~al.}(2010){Danforth}, {Keeney}, {Stocke}, {Shull}, \&
  {Yao}}]{2010ApJ...720..976D}
{Danforth}, C.~W., {Keeney}, B.~A., {Stocke}, J.~T., {Shull}, J.~M., \& {Yao},
  Y. 2010, \apj, 720, 976

\bibitem[{{de Diego}(2010)}]{2010AJ....139.1269D}
{de Diego}, J.~A. 2010, \aj, 139, 1269

\bibitem[{{de Diego} {et~al.}(1998){de Diego}, {Dultzin-Hacyan},
  {Ram{\'{\i}}rez}, \& {Ben{\'{\i}}tez}}]{1998ApJ...501...69D}
{de Diego}, J.~A., {Dultzin-Hacyan}, D., {Ram{\'{\i}}rez}, A., \&
  {Ben{\'{\i}}tez}, E. 1998, \apj, 501, 69

\bibitem[{{Dominici} {et~al.}(2004){Dominici}, {Abraham}, {Teixeira}, \&
  {Benevides-Soares}}]{2004AJ....128...47D}
{Dominici}, T.~P., {Abraham}, Z., {Teixeira}, R., \& {Benevides-Soares}, P.
  2004, \aj, 128, 47

\bibitem[{{Falomo}(1996)}]{1996MNRAS.283..241F}
{Falomo}, R. 1996, \mnras, 283, 241

\bibitem[{{Falomo} {et~al.}(2002){Falomo}, {Kotilainen}, \&
  {Treves}}]{2002ApJ...569L..35F}
{Falomo}, R., {Kotilainen}, J.~K., \& {Treves}, A. 2002, \apjl, 569, L35

\bibitem[{{Falomo} {et~al.}(1994){Falomo}, {Scarpa}, \&
  {Bersanelli}}]{1994ApJS...93..125F}
{Falomo}, R., {Scarpa}, R., \& {Bersanelli}, M. 1994, \apjs, 93, 125

\bibitem[{{Falomo} \& {Treves}(1990)}]{1990PASP..102.1120F}
{Falomo}, R., \& {Treves}, A. 1990, \pasp, 102, 1120

\bibitem[{{Fan}(1995)}]{1995Ap&SS.229..157F}
{Fan}, J.~H. 1995, \apss, 229, 157

\bibitem[{{Fan} \& {Lin}(1999)}]{1999ApJS..121..131F}
{Fan}, J.~H., \& {Lin}, R.~G. 1999, \apjs, 121, 131

\bibitem[{{Fan} \& {Lin}(2000)}]{2000ApJ...537..101F}
---. 2000, \apj, 537, 101

\bibitem[{{Fan} {et~al.}(1999){Fan}, {Xie}, \& {Bacon}}]{1999A&AS..136...13F}
{Fan}, J.~H., {Xie}, G.~Z., \& {Bacon}, R. 1999, \aaps, 136, 13

\bibitem[{{Fan} {et~al.}(1997){Fan}, {Xie}, {Lin}, {Qin}, {Li}, \&
  {Zhang}}]{1997A&AS..125..525F}
{Fan}, J.~H., {Xie}, G.~Z., {Lin}, R.~G., {et~al.} 1997, \aaps, 125

\bibitem[{{Fan} {et~al.}(2016){Fan}, {Yang}, {Liu}, {Luo}, {Lin}, {Yuan},
  {Xiao}, {Zhou}, {Hua}, \& {Pei}}]{2016ApJS..226...20F}
{Fan}, J.~H., {Yang}, J.~H., {Liu}, Y., {et~al.} 2016, \apjs, 226, 20

\bibitem[{{Fan} {et~al.}(1996){Fan}, {Burstein}, {Chen}, {Zhu}, {Jiang}, {Wu},
  {Yan}, {Zheng}, {Zhou}, {Fang}, {Chen}, {Deng}, {Chu}, {Hester}, {Windhorst},
  {Li}, {Lu}, {Sun}, {Chen}, {Tsay}, {Chiueh}, {Chou}, {Ko}, {Lin}, {Guo}, \&
  {Byun}}]{1996AJ....112..628F}
{Fan}, X., {Burstein}, D., {Chen}, J.-S., {et~al.} 1996, \aj, 112, 628

\bibitem[{{Fiorucci} {et~al.}(1997){Fiorucci}, {Tosti}, \&
  {Hurst}}]{1997A&AT...14..119F}
{Fiorucci}, M., {Tosti}, G., \& {Hurst}, G.~M. 1997, Astronomical and
  Astrophysical Transactions, 14, 119

\bibitem[{{Foschini} {et~al.}(2007){Foschini}, {Ghisellini}, {Tavecchio},
  {Treves}, {Maraschi}, {Gliozzi}, {Raiteri}, {Villata}, {Pian}, {Tagliaferri},
  {Tosti}, {Sambruna}, {Malaguti}, {Di Cocco}, \&
  {Giommi}}]{2007ApJ...657L..81F}
{Foschini}, L., {Ghisellini}, G., {Tavecchio}, F., {et~al.} 2007, \apjl, 657,
  L81

\bibitem[{{Fossati} {et~al.}(1998){Fossati}, {Maraschi}, {Celotti}, {Comastri},
  \& {Ghisellini}}]{1998MNRAS.299..433F}
{Fossati}, G., {Maraschi}, L., {Celotti}, A., {Comastri}, A., \& {Ghisellini},
  G. 1998, \mnras, 299, 433

\bibitem[{{Gaur} {et~al.}(2012{\natexlab{a}}){Gaur}, {Gupta}, \&
  {Wiita}}]{2012AJ....143...23G}
{Gaur}, H., {Gupta}, A.~C., \& {Wiita}, P.~J. 2012{\natexlab{a}}, \aj, 143, 23

\bibitem[{{Gaur} {et~al.}(2012{\natexlab{b}}){Gaur}, {Gupta}, {Strigachev},
  {Bachev}, {Semkov}, {Wiita}, {Peneva}, {Boeva}, {Slavcheva-Mihova}, {Mihov},
  {Latev}, \& {Pandey}}]{2012MNRAS.425.3002G}
{Gaur}, H., {Gupta}, A.~C., {Strigachev}, A., {et~al.} 2012{\natexlab{b}},
  \mnras, 425, 3002

\bibitem[{{Gaur} {et~al.}(2012{\natexlab{c}}){Gaur}, {Gupta}, {Strigachev},
  {Bachev}, {Semkov}, {Wiita}, {Peneva}, {Boeva}, {Kacharov}, {Mihov}, \&
  {Ovcharov}}]{2012MNRAS.420.3147G}
---. 2012{\natexlab{c}}, \mnras, 420, 3147

\bibitem[{{Gaur} {et~al.}(2015){Gaur}, {Gupta}, {Bachev}, {Strigachev},
  {Semkov}, {Wiita}, {Volvach}, {Gu}, {Agarwal}, {Agudo}, {Aller}, {Aller},
  {Kurtanidze}, {Kurtanidze}, {L{\"a}hteenm{\"a}ki}, {Peneva}, {Nikolashvili},
  {Sigua}, {Tornikoski}, \& {Volvach}}]{2015A&A...582A.103G}
{Gaur}, H., {Gupta}, A.~C., {Bachev}, R., {et~al.} 2015, \aap, 582, A103

\bibitem[{{Ghisellini} {et~al.}(2010){Ghisellini}, {Tavecchio}, {Foschini},
  {Ghirlanda}, {Maraschi}, \& {Celotti}}]{2010MNRAS.402..497G}
{Ghisellini}, G., {Tavecchio}, F., {Foschini}, L., {et~al.} 2010, \mnras, 402,
  497

\bibitem[{{Ghosh} {et~al.}(2000){Ghosh}, {Ramsey}, {Sadun}, \&
  {Soundararajaperumal}}]{2000ApJS..127...11G}
{Ghosh}, K.~K., {Ramsey}, B.~D., {Sadun}, A.~C., \& {Soundararajaperumal}, S.
  2000, \apjs, 127, 11

\bibitem[{{Ghosh} \& {Soundararajaperumal}(1995)}]{1995ApJS..100...37G}
{Ghosh}, K.~K., \& {Soundararajaperumal}, S. 1995, \apjs, 100, 37

\bibitem[{{Giommi} {et~al.}(1995){Giommi}, {Ansari}, \&
  {Micol}}]{1995A&AS..109..267G}
{Giommi}, P., {Ansari}, S.~G., \& {Micol}, A. 1995, \aaps, 109

\bibitem[{{Gopal-Krishna} {et~al.}(2011){Gopal-Krishna}, {Goyal}, {Joshi},
  {Karthick}, {Sagar}, {Wiita}, {Anupama}, \& {Sahu}}]{2011MNRAS.416..101G}
{Gopal-Krishna}, {Goyal}, A., {Joshi}, S., {et~al.} 2011, \mnras, 416, 101

\bibitem[{{Goyal} {et~al.}(2009){Goyal}, {Gopal-Krishna}, {Anupama}, {Sahu},
  {Sagar}, {Britzen}, {Karouzos}, {Aller}, \& {Aller}}]{2009MNRAS.399.1622G}
{Goyal}, A., {Gopal-Krishna}, {Anupama}, G.~C., {et~al.} 2009, \mnras, 399,
  1622

\bibitem[{{Gu} {et~al.}(2001){Gu}, {Cao}, \& {Jiang}}]{2001MNRAS.327.1111G}
{Gu}, M., {Cao}, X., \& {Jiang}, D.~R. 2001, \mnras, 327, 1111

\bibitem[{{Gu} {et~al.}(2006){Gu}, {Lee}, {Pak}, {Yim}, \&
  {Fletcher}}]{2006A&A...450...39G}
{Gu}, M.~F., {Lee}, C.-U., {Pak}, S., {Yim}, H.~S., \& {Fletcher}, A.~B. 2006,
  \aap, 450, 39

\bibitem[{{Gupta} {et~al.}(2004){Gupta}, {Banerjee}, {Ashok}, \&
  {Joshi}}]{2004A&A...422..505G}
{Gupta}, A.~C., {Banerjee}, D.~P.~K., {Ashok}, N.~M., \& {Joshi}, U.~C. 2004,
  \aap, 422, 505

\bibitem[{{Gupta} {et~al.}(2012){Gupta}, {Pandey}, {Singh}, {Rani}, {Pan},
  {Fan}, \& {Gupta}}]{2012NewA...17....8G}
{Gupta}, S.~P., {Pandey}, U.~S., {Singh}, K., {et~al.} 2012, New Astron., 17, 8

\bibitem[{{Hao} {et~al.}(2010){Hao}, {Wang}, {Jiang}, \&
  {Dai}}]{2010RAA....10..125H}
{Hao}, J.-M., {Wang}, B.-J., {Jiang}, Z.-J., \& {Dai}, B.-Z. 2010, Research in
  Astronomy and Astrophysics, 10, 125

\bibitem[{{Holder} {et~al.}(2003){Holder}, {Bond}, {Boyle}, {Bradbury},
  {Buckley}, {Carter-Lewis}, {Cui}, {Dowdall}, {Duke}, {de la Calle Perez},
  {Falcone}, {Fegan}, {Fegan}, {Finley}, {Fortson}, {Gaidos}, {Gibbs},
  {Gammell}, {Hall}, {Hall}, {Hillas}, {Horan}, {Jordan}, {Kertzman}, {Kieda},
  {Kildea}, {Knapp}, {Kosack}, {Krawczynski}, {Krennrich}, {LeBohec}, {Linton},
  {Lloyd-Evans}, {Moriarty}, {M{\"u}ller}, {Nagai}, {Ong}, {Page},
  {Pallassini}, {Petry}, {Power-Mooney}, {Quinn}, {Rebillot}, {Reynolds},
  {Rose}, {Schroedter}, {Sembroski}, {Swordy}, {Vassiliev}, {Wakely}, {Walker},
  \& {Weekes}}]{2003ApJ...583L...9H}
{Holder}, J., {Bond}, I.~H., {Boyle}, P.~J., {et~al.} 2003, \apjl, 583, L9

\bibitem[{{Hu} {et~al.}(2011){Hu}, {Wu}, {Guo}, {Zhou}, {Zhang}, \&
  {Zheng}}]{2011Ap&SS.333..213H}
{Hu}, S.~M., {Wu}, J., {Guo}, H.~Y., {et~al.} 2011, \apss, 333, 213

\bibitem[{{Hu} {et~al.}(2006{\natexlab{a}}){Hu}, {Wu}, {Zhao}, \&
  {Zhou}}]{2006MNRAS.373..209H}
{Hu}, S.~M., {Wu}, J.~H., {Zhao}, G., \& {Zhou}, X. 2006{\natexlab{a}}, \mnras,
  373, 209

\bibitem[{{Hu} {et~al.}(2006{\natexlab{b}}){Hu}, {Zhao}, {Guo}, {Zhang}, \&
  {Zheng}}]{2006MNRAS.371.1243H}
{Hu}, S.~M., {Zhao}, G., {Guo}, H.~Y., {Zhang}, X., \& {Zheng}, Y.~G.
  2006{\natexlab{b}}, \mnras, 371, 1243

\bibitem[{{Ikejiri} {et~al.}(2011){Ikejiri}, {Uemura}, {Sasada}, {Ito},
  {Yamanaka}, {Sakimoto}, {Arai}, {Fukazawa}, {Ohsugi}, {Kawabata}, {Yoshida},
  {Sato}, \& {Kino}}]{2011PASJ...63..639I}
{Ikejiri}, Y., {Uemura}, M., {Sasada}, M., {et~al.} 2011, \pasj, 63, 639

\bibitem[{{Impiombato} {et~al.}(2011){Impiombato}, {Covino}, {Treves},
  {Foschini}, {Pian}, {Tosti}, {Fugazza}, {Nicastro}, \&
  {Ciprini}}]{2011ApJS..192...12I}
{Impiombato}, D., {Covino}, S., {Treves}, A., {et~al.} 2011, \apjs, 192, 12

\bibitem[{{Isler} {et~al.}(2017){Isler}, {Urry}, {Coppi}, {Bailyn}, {Brady},
  {MacPherson}, {Buxton}, \& {Hasan}}]{2017ApJ...844..107I}
{Isler}, J.~C., {Urry}, C.~M., {Coppi}, P., {et~al.} 2017, \apj, 844, 107

\bibitem[{{Jang} \& {Miller}(1997)}]{1997AJ....114..565J}
{Jang}, M., \& {Miller}, H.~R. 1997, \aj, 114, 565

\bibitem[{{Jorstad} {et~al.}(2001){Jorstad}, {Marscher}, {Mattox}, {Wehrle},
  {Bloom}, \& {Yurchenko}}]{2001ApJS..134..181J}
{Jorstad}, S.~G., {Marscher}, A.~P., {Mattox}, J.~R., {et~al.} 2001, \apjs,
  134, 181

\bibitem[{{Katajainen} {et~al.}(2000){Katajainen}, {Takalo},
  {Sillanp{\"a}{\"a}}, {Nilsson}, {Pursimo}, {Hanski}, {Hein{\"a}m{\"a}ki},
  {Kotoneva}, {Lainela}, {Nurmi}, {Pietil{\"a}}, {Rekola}, {Riehokainen},
  {Teerikorpi}, {Valtaoja}, \& {L{\"a}hteenm{\"a}ki}}]{2000A&AS..143..357K}
{Katajainen}, S., {Takalo}, L.~O., {Sillanp{\"a}{\"a}}, A., {et~al.} 2000,
  \aaps, 143, 357

\bibitem[{{Kirk} {et~al.}(1998){Kirk}, {Rieger}, \&
  {Mastichiadis}}]{1998A&A...333..452K}
{Kirk}, J.~G., {Rieger}, F.~M., \& {Mastichiadis}, A. 1998, \aap, 333, 452

\bibitem[{{Krawczynski} {et~al.}(2004){Krawczynski}, {Hughes}, {Horan},
  {Aharonian}, {Aller}, {Aller}, {Boltwood}, {Buckley}, {Coppi}, {Fossati},
  {G{\"o}tting}, {Holder}, {Horns}, {Kurtanidze}, {Marscher}, {Nikolashvili},
  {Remillard}, {Sadun}, \& {Schr{\"o}der}}]{2004ApJ...601..151K}
{Krawczynski}, H., {Hughes}, S.~B., {Horan}, D., {et~al.} 2004, \apj, 601, 151

\bibitem[{{Kurtanidze} {et~al.}(2009){Kurtanidze}, {Tetradze}, {Richter},
  {Nikolashvili}, {Kimeridze}, \& {Sigua}}]{2009ASPC..408..266K}
{Kurtanidze}, O.~M., {Tetradze}, S.~D., {Richter}, G.~M., {et~al.} 2009, in
  Astronomical Society of the Pacific Conference Series, Vol. 408, The
  Starburst-AGN Connection, ed. W.~{Wang}, Z.~{Yang}, Z.~{Luo}, \& Z.~{Chen},
  266

\bibitem[{{Lanzetta} {et~al.}(1993){Lanzetta}, {Turnshek}, \&
  {Sandoval}}]{1993ApJS...84..109L}
{Lanzetta}, K.~M., {Turnshek}, D.~A., \& {Sandoval}, J. 1993, \apjs, 84, 109

\bibitem[{{Marscher} {et~al.}(2002){Marscher}, {Jorstad}, {Mattox}, \&
  {Wehrle}}]{2002ApJ...577...85M}
{Marscher}, A.~P., {Jorstad}, S.~G., {Mattox}, J.~R., \& {Wehrle}, A.~E. 2002,
  \apj, 577, 85

\bibitem[{{Massaro} {et~al.}(1999){Massaro}, {Maesano}, {Montagni}, {Nesci},
  {Tosti}, {Fiorucci}, {Luciani}, {Takalo}, {Sillanp{\"a}{\"a}}, {Katajainen},
  {Kein{\"a}nen}, {Pursimo}, {Villata}, {Raiteri}, {de Francesco}, {Sobrito},
  {Efimov}, \& {Shakhovskoy}}]{1999A&A...342L..49M}
{Massaro}, E., {Maesano}, M., {Montagni}, F., {et~al.} 1999, \aap, 342, L49

\bibitem[{{Mastichiadis} \& {Kirk}(2002)}]{2002PASA...19..138M}
{Mastichiadis}, A., \& {Kirk}, J.~G. 2002, PASA, 19, 138

\bibitem[{{Meng} {et~al.}(2017){Meng}, {Wu}, {Webb}, {Zhang}, \&
  {Dai}}]{2017MNRAS.469.3588M}
{Meng}, N., {Wu}, J., {Webb}, J.~R., {Zhang}, X., \& {Dai}, Y. 2017, \mnras,
  469, 3588

\bibitem[{{Miller}(1975)}]{1975ApJ...201L.109M}
{Miller}, H.~R. 1975, \apjl, 201, L109

\bibitem[{{Miller} \& {McGimsey}(1978)}]{1978ApJ...220...19M}
{Miller}, H.~R., \& {McGimsey}, B.~Q. 1978, \apj, 220, 19

\bibitem[{{Miller} {et~al.}(1978){Miller}, {French}, \&
  {Hawley}}]{1978bllo.conf..176M}
{Miller}, J.~S., {French}, H.~B., \& {Hawley}, S.~A. 1978, in BL Lac Objects,
  ed. A.~M. {Wolfe}, 176--187

\bibitem[{{Miller} \& {Hawley}(1977)}]{1977ApJ...212L..47M}
{Miller}, J.~S., \& {Hawley}, S.~A. 1977, \apjl, 212, L47

\bibitem[{{Nesci} {et~al.}(1998){Nesci}, {Maesano}, {Massaro}, {Montagni},
  {Tosti}, \& {Fiorucci}}]{1998A&A...332L...1N}
{Nesci}, R., {Maesano}, M., {Massaro}, E., {et~al.} 1998, \aap, 332, L1

\bibitem[{{Nieppola} {et~al.}(2006){Nieppola}, {Tornikoski}, \&
  {Valtaoja}}]{2006A&A...445..441N}
{Nieppola}, E., {Tornikoski}, M., \& {Valtaoja}, E. 2006, \aap, 445, 441

\bibitem[{{Nilsson} {et~al.}(2012){Nilsson}, {Pursimo}, {Villforth},
  {Lindfors}, {Takalo}, \& {Sillanp{\"a}{\"a}}}]{2012A&A...547A...1N}
{Nilsson}, K., {Pursimo}, T., {Villforth}, C., {et~al.} 2012, \aap, 547, A1

\bibitem[{{Noble} {et~al.}(1997){Noble}, {Carini}, {Miller}, \&
  {Goodrich}}]{1997AJ....113.1995N}
{Noble}, J.~C., {Carini}, M.~T., {Miller}, H.~R., \& {Goodrich}, B. 1997, \aj,
  113, 1995

\bibitem[{{Oke}(1978)}]{1978ApJ...219L..97O}
{Oke}, J.~B. 1978, \apjl, 219, L97

\bibitem[{{Padovani} \& {Giommi}(1995)}]{1995ApJ...444..567P}
{Padovani}, P., \& {Giommi}, P. 1995, \apj, 444, 567

\bibitem[{{Papadakis} {et~al.}(2003){Papadakis}, {Boumis}, {Samaritakis}, \&
  {Papamastorakis}}]{2003A&A...397..565P}
{Papadakis}, I.~E., {Boumis}, P., {Samaritakis}, V., \& {Papamastorakis}, J.
  2003, \aap, 397, 565

\bibitem[{{Papadakis} {et~al.}(2007){Papadakis}, {Villata}, \&
  {Raiteri}}]{2007A&A...470..857P}
{Papadakis}, I.~E., {Villata}, M., \& {Raiteri}, C.~M. 2007, \aap, 470, 857

\bibitem[{{Perri} {et~al.}(2003){Perri}, {Massaro}, {Giommi}, {Capalbi},
  {Nesci}, {Tagliaferri}, {Ghisellini}, {Ravasio}, \&
  {Miller}}]{2003A&A...407..453P}
{Perri}, M., {Massaro}, E., {Giommi}, P., {et~al.} 2003, \aap, 407, 453

\bibitem[{{Persic} \& {Salucci}(1986)}]{1986ASSL..121..675P}
{Persic}, M., \& {Salucci}, P. 1986, in Astrophysics and Space Science Library,
  Vol. 121, Structure and Evolution of Active Galactic Nuclei, ed.
  G.~{Giuricin}, M.~{Mezzetti}, M.~{Ramella}, \& F.~{Mardirossian}, 675--677

\bibitem[{{Pian} {et~al.}(1994){Pian}, {Falomo}, {Scarpa}, \&
  {Treves}}]{1994ApJ...432..547P}
{Pian}, E., {Falomo}, R., {Scarpa}, R., \& {Treves}, A. 1994, \apj, 432, 547

\bibitem[{{Pica} \& {Smith}(1983)}]{1983ApJ...272...11P}
{Pica}, A.~J., \& {Smith}, A.~G. 1983, \apj, 272, 11

\bibitem[{{Pica} {et~al.}(1988){Pica}, {Smith}, {Webb}, {Leacock}, {Clements},
  \& {Gombola}}]{1988AJ.....96.1215P}
{Pica}, A.~J., {Smith}, A.~G., {Webb}, J.~R., {et~al.} 1988, \aj, 96, 1215

\bibitem[{{Pollock} {et~al.}(1979){Pollock}, {Pica}, {Smith}, {Leacock},
  {Edwards}, \& {Scott}}]{1979AJ.....84.1658P}
{Pollock}, J.~T., {Pica}, A.~J., {Smith}, A.~G., {et~al.} 1979, \aj, 84, 1658

\bibitem[{{Poon} {et~al.}(2009){Poon}, {Fan}, \& {Fu}}]{2009ApJS..185..511P}
{Poon}, H., {Fan}, J.~H., \& {Fu}, J.~N. 2009, \apjs, 185, 511

\bibitem[{{Pushkarev} {et~al.}(2009){Pushkarev}, {Kovalev}, {Lister}, \&
  {Savolainen}}]{2009A&A...507L..33P}
{Pushkarev}, A.~B., {Kovalev}, Y.~Y., {Lister}, M.~L., \& {Savolainen}, T.
  2009, \aap, 507, L33

\bibitem[{{Qian} {et~al.}(2000){Qian}, {Tao}, \& {Fan}}]{2000PASJ...52.1075Q}
{Qian}, B., {Tao}, J., \& {Fan}, J. 2000, \pasj, 52, 1075

\bibitem[{{Raiteri} {et~al.}(2007){Raiteri}, {Villata}, {Larionov}, {Pursimo},
  {Ibrahimov}, {Nilsson}, {Aller}, {Kurtanidze}, {Foschini}, {Ohlert},
  {Papadakis}, {Sumitomo}, {Volvach}, {Aller}, {Arkharov}, {Bach}, {Berdyugin},
  {B{\"o}ttcher}, {Buemi}, {Calcidese}, {Charlot}, {Delgado S{\'a}nchez}, {di
  Paola}, {Djupvik}, {Dolci}, {Efimova}, {Fan}, {Forn{\'e}}, {Gomez}, {Gupta},
  {Hagen-Thorn}, {Hooks}, {Hovatta}, {Ishii}, {Kamada}, {Konstantinova},
  {Kopatskaya}, {Kovalev}, {Kovalev}, {L{\"a}hteenm{\"a}ki}, {Lanteri}, {Le
  Campion}, {Lee}, {Leto}, {Lin}, {Lindfors}, {Mingaliev}, {Mizoguchi},
  {Nicastro}, {Nikolashvili}, {Nishiyama}, {{\"O}stman}, {Ovcharov},
  {P{\"a}{\"a}kk{\"o}nen}, {Pasanen}, {Pian}, {Rector}, {Ros}, {Sadakane},
  {Selj}, {Semkov}, {Sharapov}, {Somero}, {Stanev}, {Strigachev}, {Takalo},
  {Tanaka}, {Tavani}, {Torniainen}, {Tornikoski}, {Trigilio}, {Umana},
  {Vercellone}, {Valcheva}, {Volvach}, \& {Yamanaka}}]{2007A&A...473..819R}
{Raiteri}, C.~M., {Villata}, M., {Larionov}, V.~M., {et~al.} 2007, \aap, 473,
  819

\bibitem[{{Raiteri} {et~al.}(2008){Raiteri}, {Villata}, {Larionov}, {Gurwell},
  {Chen}, {Kurtanidze}, {Aller}, {B{\"o}ttcher}, {Calcidese}, {Hroch},
  {L{\"a}hteenm{\"a}ki}, {Lee}, {Nilsson}, {Ohlert}, {Papadakis}, {Agudo},
  {Aller}, {Angelakis}, {Arkharov}, {Bach}, {Bachev}, {Berdyugin}, {Buemi},
  {Carosati}, {Charlot}, {Chatzopoulos}, {Forn{\'e}}, {Frasca}, {Fuhrmann},
  {G{\'o}mez}, {Gupta}, {Hagen-Thorn}, {Hsiao}, {Jordan}, {Jorstad},
  {Konstantinova}, {Kopatskaya}, {Krichbaum}, {Lanteri}, {Larionova}, {Latev},
  {Le Campion}, {Leto}, {Lin}, {Marchili}, {Marilli}, {Marscher}, {McBreen},
  {Mihov}, {Nesci}, {Nicastro}, {Nikolashvili}, {Novak}, {Ovcharov}, {Pian},
  {Principe}, {Pursimo}, {Ragozzine}, {Ros}, {Sadun}, {Sagar}, {Semkov},
  {Smart}, {Smith}, {Strigachev}, {Takalo}, {Tavani}, {Tornikoski}, {Trigilio},
  {Uckert}, {Umana}, {Valcheva}, {Vercellone}, {Volvach}, \&
  {Wiesemeyer}}]{2008A&A...491..755R}
---. 2008, \aap, 491, 755

\bibitem[{{Raiteri} {et~al.}(2011){Raiteri}, {Villata}, {Aller}, {Gurwell},
  {Kurtanidze}, {L{\"a}hteenm{\"a}ki}, {Larionov}, {Romano}, {Vercellone},
  {Agudo}, {Aller}, {Arkharov}, {Bach}, {Ben{\'{\i}}tez}, {Berdyugin},
  {Blinov}, {Borisova}, {B{\"o}ttcher}, {Bravo Calle}, {Buemi}, {Calcidese},
  {Carosati}, {Casas}, {Chen}, {Efimova}, {G{\'o}mez}, {Gusbar}, {Hawkins},
  {Heidt}, {Hiriart}, {Hsiao}, {Jordan}, {Jorstad}, {Joshi}, {Kimeridze},
  {Koptelova}, {Konstantinova}, {Kopatskaya}, {Kurtanidze}, {Larionova},
  {Larionova}, {Leto}, {Li}, {Ligustri}, {Lindfors}, {Lister}, {Marscher},
  {Molina}, {Morozova}, {Nieppola}, {Nikolashvili}, {Nilsson}, {Palma},
  {Pasanen}, {Reinthal}, {Roberts}, {Ros}, {Roustazadeh}, {Sadun}, {Sakamoto},
  {Schwartz}, {Sigua}, {Sillanp{\"a}{\"a}}, {Takalo}, {Tammi}, {Taylor},
  {Tornikoski}, {Trigilio}, {Troitsky}, {Umana}, {Volvach}, \&
  {Yuldasheva}}]{2011A&A...534A..87R}
{Raiteri}, C.~M., {Villata}, M., {Aller}, M.~F., {et~al.} 2011, \aap, 534, A87

\bibitem[{{Raiteri} {et~al.}(2013){Raiteri}, {Villata}, {D'Ammando},
  {Larionov}, {Gurwell}, {Mirzaqulov}, {Smith}, {Acosta-Pulido}, {Agudo},
  {Ar{\'e}valo}, {Bachev}, {Ben{\'{\i}}tez}, {Berdyugin}, {Blinov}, {Borman},
  {B{\"o}ttcher}, {Bozhilov}, {Carnerero}, {Carosati}, {Casadio}, {Chen},
  {Doroshenko}, {Efimov}, {Efimova}, {Ehgamberdiev}, {G{\'o}mez},
  {Gonz{\'a}lez-Morales}, {Hiriart}, {Ibryamov}, {Jadhav}, {Jorstad}, {Joshi},
  {Kadenius}, {Klimanov}, {Kohli}, {Konstantinova}, {Kopatskaya}, {Koptelova},
  {Kimeridze}, {Kurtanidze}, {Larionova}, {Larionova}, {Ligustri}, {Lindfors},
  {Marscher}, {McBreen}, {McHardy}, {Metodieva}, {Molina}, {Morozova},
  {Nazarov}, {Nikolashvili}, {Nilsson}, {Okhmat}, {Ovcharov}, {Panwar},
  {Pasanen}, {Peneva}, {Phipps}, {Pulatova}, {Reinthal}, {Ros}, {Sadun},
  {Schwartz}, {Semkov}, {Sergeev}, {Sigua}, {Sillanp{\"a}{\"a}}, {Smith},
  {Stoyanov}, {Strigachev}, {Takalo}, {Taylor}, {Thum}, {Troitsky}, {Valcheva},
  {Wehrle}, \& {Wiesemeyer}}]{2013MNRAS.436.1530R}
{Raiteri}, C.~M., {Villata}, M., {D'Ammando}, F., {et~al.} 2013, \mnras, 436,
  1530

\bibitem[{{Raiteri} {et~al.}(2015){Raiteri}, {Stamerra}, {Villata}, {Larionov},
  {Acosta-Pulido}, {Ar{\'e}valo}, {Arkharov}, {Bachev}, {Ben{\'{\i}}tez},
  {Bozhilov}, {Borman}, {Buemi}, {Calcidese}, {Carnerero}, {Carosati},
  {Chigladze}, {Damljanovic}, {Di Paola}, {Doroshenko}, {Efimova},
  {Ehgamberdiev}, {Giroletti}, {Gonz{\'a}lez-Morales}, {Grinon-Marin},
  {Grishina}, {Hiriart}, {Ibryamov}, {Klimanov}, {Kopatskaya}, {Kurtanidze},
  {Kurtanidze}, {Kurtenkov}, {Larionova}, {Larionova}, {L{\'a}zaro},
  {L{\"a}hteenm{\"a}ki}, {Leto}, {Markovic}, {Mirzaqulov}, {Mokrushina},
  {Morozova}, {M{\'u}jica}, {Nazarov}, {Nikolashvili}, {Ohlert}, {Ovcharov},
  {Paiano}, {Pastor Yabar}, {Prandini}, {Ramakrishnan}, {Sadun}, {Semkov},
  {Sigua}, {Strigachev}, {Tammi}, {Tornikoski}, {Trigilio}, {Troitskaya},
  {Troitsky}, {Umana}, {Velasco}, \& {Vince}}]{2015MNRAS.454..353R}
{Raiteri}, C.~M., {Stamerra}, A., {Villata}, M., {et~al.} 2015, \mnras, 454,
  353

\bibitem[{{Rani} {et~al.}(2010){Rani}, {Gupta}, {Strigachev}, {Bachev},
  {Wiita}, {Semkov}, {Ovcharov}, {Mihov}, {Boeva}, {Peneva}, {Spassov},
  {Tsvetkova}, {Stoyanov}, \& {Valcheva}}]{2010MNRAS.404.1992R}
{Rani}, B., {Gupta}, A.~C., {Strigachev}, A., {et~al.} 2010, \mnras, 404, 1992

\bibitem[{{Romero} {et~al.}(1999){Romero}, {Cellone}, \&
  {Combi}}]{1999A&AS..135..477R}
{Romero}, G.~E., {Cellone}, S.~A., \& {Combi}, J.~A. 1999, \aaps, 135, 477

\bibitem[{{Romero} {et~al.}(2000){Romero}, {Cellone}, \&
  {Combi}}]{2000AJ....120.1192R}
---. 2000, \aj, 120, 1192

\bibitem[{{Sandage}(1967)}]{1967ApJ...150L.177S}
{Sandage}, A. 1967, \apjl, 150, L177

\bibitem[{{Sandrinelli} {et~al.}(2014){Sandrinelli}, {Covino}, \&
  {Treves}}]{2014A&A...562A..79S}
{Sandrinelli}, A., {Covino}, S., \& {Treves}, A. 2014, \aap, 562, A79

\bibitem[{{Sasada} {et~al.}(2010){Sasada}, {Uemura}, {Arai}, {Fukazawa},
  {Kawabata}, {Ohsugi}, {Yamashita}, {Isogai}, {Nagae}, {Uehara}, {Mizuno},
  {Katagiri}, {Takahashi}, {Sato}, \& {Kino}}]{2010PASJ...62..645S}
{Sasada}, M., {Uemura}, M., {Arai}, A., {et~al.} 2010, \pasj, 62, 645

\bibitem[{{Sbarufatti} {et~al.}(2005){Sbarufatti}, {Treves}, \&
  {Falomo}}]{2005ApJ...635..173S}
{Sbarufatti}, B., {Treves}, A., \& {Falomo}, R. 2005, \apj, 635, 173

\bibitem[{{Schachter} {et~al.}(1993){Schachter}, {Stocke}, {Perlman}, {Elvis},
  {Remillard}, {Granados}, {Luu}, {Huchra}, {Humphreys}, {Urry}, \&
  {Wallin}}]{1993ApJ...412..541S}
{Schachter}, J.~F., {Stocke}, J.~T., {Perlman}, E., {et~al.} 1993, \apj, 412,
  541

\bibitem[{{Schlafly} \& {Finkbeiner}(2011)}]{2011ApJ...737..103S}
{Schlafly}, E.~F., \& {Finkbeiner}, D.~P. 2011, \apj, 737, 103

\bibitem[{{Sillanp{\"a}{\"a}} {et~al.}(1988){Sillanp{\"a}{\"a}}, {Haarala}, \&
  {Korhonen}}]{1988A&AS...72..347S}
{Sillanp{\"a}{\"a}}, A., {Haarala}, S., \& {Korhonen}, T. 1988, \aaps, 72, 347

\bibitem[{{Sillanp{\"a}{\"a}} {et~al.}(1991){Sillanp{\"a}{\"a}}, {Mikkola}, \&
  {Valtaoja}}]{1991A&AS...88..225S}
{Sillanp{\"a}{\"a}}, A., {Mikkola}, S., \& {Valtaoja}, L. 1991, \aaps, 88, 225

\bibitem[{{Sokolov} {et~al.}(2004){Sokolov}, {Marscher}, \&
  {McHardy}}]{2004ApJ...613..725S}
{Sokolov}, A., {Marscher}, A.~P., \& {McHardy}, I.~M. 2004, \apj, 613, 725

\bibitem[{{Sorcia} {et~al.}(2013){Sorcia}, {Ben{\'{\i}}tez}, {Hiriart},
  {L{\'o}pez}, {Cabrera}, {M{\'u}jica}, {Heidt}, {Agudo}, {Nilsson}, \&
  {Mommert}}]{2013ApJS..206...11S}
{Sorcia}, M., {Ben{\'{\i}}tez}, E., {Hiriart}, D., {et~al.} 2013, \apjs, 206,
  11

\bibitem[{{Stalin} {et~al.}(2006){Stalin}, {Gopal-Krishna}, {Sagar}, {Wiita},
  {Mohan}, \& {Pandey}}]{2006MNRAS.366.1337S}
{Stalin}, C.~S., {Gopal-Krishna}, {Sagar}, R., {et~al.} 2006, \mnras, 366, 1337

\bibitem[{{Takalo} {et~al.}(1996){Takalo}, {Sillanpaeae}, {Pursimo}, {Lehto},
  {Nilsson}, {Teerikorpi}, {Heinaemaeki}, {Lainela}, {Kidger}, {de Diego},
  {Gonzalez-Perez}, {Rodriguez-Espinosa}, {Mahoney}, {Boltwood},
  {Dultzin-Hacyan}, {Benitez}, {Turner}, {Robertson}, {Honeycut}, {Efimov},
  {Shakhovskoy}, {Charles}, {Schramm}, {Borgeest}, {Linde}, {Weneit}, {Kuehl},
  {Schramm}, {Sadun}, {Grashuis}, {Heidt}, {Wagner}, {Bock}, {Kuemmel},
  {Pfeiffer}, {Heines}, {Fiorucci}, {Tosti}, {Ghisellini}, {Raiteri},
  {Villata}, {de Francesco}, {Bosio}, {Latini}, {Poyner}, {Aller}, {Aller},
  {Hughes}, {Valtaoja}, {Teraesranta}, \& {Tornikoski}}]{1996A&AS..120..313T}
{Takalo}, L.~O., {Sillanpaeae}, A., {Pursimo}, T., {et~al.} 1996, \aaps, 120,
  313

\bibitem[{{Tapia} {et~al.}(1977){Tapia}, {Craine}, {Gearhart}, {Pacht}, \&
  {Kraus}}]{1977ApJ...215L..71T}
{Tapia}, S., {Craine}, E.~R., {Gearhart}, M.~R., {Pacht}, E., \& {Kraus}, J.
  1977, \apjl, 215, L71

\bibitem[{{Tosti} {et~al.}(1998{\natexlab{a}}){Tosti}, {Fiorucci}, {Luciani},
  {Rizzi}, {Villata}, {Raiteri}, {de Francesco}, {Lanteri}, {Chiaberge},
  {Peila}, {Cavallone}, {Sobrito}, {Maesano}, {Massaro}, {Montagni}, {Nesci},
  {Ghisellini}, {Takalo}, {Sillanpaeae}, {Katajainen}, {Heinaemaeki},
  {Nilsson}, \& {Pursimo}}]{1998A&AS..130..109T}
{Tosti}, G., {Fiorucci}, M., {Luciani}, M., {et~al.} 1998{\natexlab{a}}, \aaps,
  130, 109

\bibitem[{{Tosti} {et~al.}(1998{\natexlab{b}}){Tosti}, {Fiorucci}, {Luciani},
  {Efimov}, {Shakhovskoy}, {Valtaoja}, {Teraesranta}, {Sillanpaeae}, {Takalo},
  {Villata}, {Raiteri}, {de Francesco}, \& {Sobrito}}]{1998A&A...339...41T}
---. 1998{\natexlab{b}}, \aap, 339, 41

\bibitem[{{Tosti} {et~al.}(2002){Tosti}, {Massaro}, {Nesci}, {Ciprini},
  {Nucciarelli}, {Maesano}, {Montagni}, {Raiteri}, {Villata}, {Lanteri}, \&
  {Ostorero}}]{2002A&A...395...11T}
{Tosti}, G., {Massaro}, E., {Nesci}, R., {et~al.} 2002, \aap, 395, 11

\bibitem[{{Ulrich} {et~al.}(1975){Ulrich}, {Kinman}, {Lynds}, {Rieke}, \&
  {Ekers}}]{1975ApJ...198..261U}
{Ulrich}, M.-H., {Kinman}, T.~D., {Lynds}, C.~R., {Rieke}, G.~H., \& {Ekers},
  R.~D. 1975, \apj, 198, 261

\bibitem[{{Ulrich} {et~al.}(1997){Ulrich}, {Maraschi}, \&
  {Urry}}]{1997ARA&A..35..445U}
{Ulrich}, M.-H., {Maraschi}, L., \& {Urry}, C.~M. 1997, \araa, 35, 445

\bibitem[{{Urry} \& {Padovani}(1995)}]{1995PASP..107..803U}
{Urry}, C.~M., \& {Padovani}, P. 1995, \pasp, 107, 803

\bibitem[{{Urry} {et~al.}(1996){Urry}, {Sambruna}, {Worrall}, {Kollgaard},
  {Feigelson}, {Perlman}, \& {Stocke}}]{1996ApJ...463..424U}
{Urry}, C.~M., {Sambruna}, R.~M., {Worrall}, D.~M., {et~al.} 1996, \apj, 463,
  424

\bibitem[{{Vagnetti} {et~al.}(2003){Vagnetti}, {Trevese}, \&
  {Nesci}}]{2003ApJ...590..123V}
{Vagnetti}, F., {Trevese}, D., \& {Nesci}, R. 2003, \apj, 590, 123

\bibitem[{{Vercellone} {et~al.}(2008){Vercellone}, {Chen}, {Giuliani},
  {Bulgarelli}, {Donnarumma}, {Lapshov}, {Tavani}, {Argan}, {Barbiellini},
  {Caraveo}, {Cocco}, {Costa}, {D'Ammando}, {Del Monte}, {De Paris}, {Di
  Cocco}, {Evangelista}, {Feroci}, {Fiorini}, {Froysland}, {Fuschino}, {Galli},
  {Gianotti}, {Labanti}, {Lazzarotto}, {Lipari}, {Longo}, {Marisaldi}, {Mauri},
  {Mereghetti}, {Morselli}, {Pacciani}, {Pellizzoni}, {Perotti}, {Picozza},
  {Prest}, {Pucella}, {Rapisarda}, {Soffitta}, {Trifoglio}, {Trois},
  {Vallazza}, {Vittorini}, {Zambra}, {Zanello}, {Pittori}, {Verrecchia},
  {Gasparrini}, {Cutini}, {Giommi}, {Antonelli}, {Colafrancesco}, \&
  {Salotti}}]{2008ApJ...676L..13V}
{Vercellone}, S., {Chen}, A.~W., {Giuliani}, A., {et~al.} 2008, \apjl, 676, L13

\bibitem[{{Vercellone} {et~al.}(2010){Vercellone}, {D'Ammando}, {Vittorini},
  {Donnarumma}, {Pucella}, {Tavani}, {Ferrari}, {Raiteri}, {Villata}, {Romano},
  {Krimm}, {Tiengo}, {Chen}, {Giovannini}, {Venturi}, {Giroletti}, {Kovalev},
  {Sokolovsky}, {Pushkarev}, {Lister}, {Argan}, {Barbiellini}, {Bulgarelli},
  {Caraveo}, {Cattaneo}, {Cocco}, {Costa}, {Del Monte}, {De Paris}, {Di Cocco},
  {Evangelista}, {Feroci}, {Fiorini}, {Fornari}, {Froysland}, {Fuschino},
  {Galli}, {Gianotti}, {Labanti}, {Lapshov}, {Lazzarotto}, {Lipari}, {Longo},
  {Giuliani}, {Marisaldi}, {Mereghetti}, {Morselli}, {Pellizzoni}, {Pacciani},
  {Perotti}, {Piano}, {Picozza}, {Pilia}, {Prest}, {Rapisarda}, {Rappoldi},
  {Sabatini}, {Soffitta}, {Striani}, {Trifoglio}, {Trois}, {Vallazza},
  {Zambra}, {Zanello}, {Pittori}, {Verrecchia}, {Santolamazza}, {Giommi},
  {Colafrancesco}, {Salotti}, {Agudo}, {Aller}, {Aller}, {Arkharov}, {Bach},
  {Bachev}, {Beltrame}, {Ben{\'{\i}}tez}, {B{\"o}ttcher}, {Buemi}, {Calcidese},
  {Capezzali}, {Carosati}, {Chen}, {Da Rio}, {Di Paola}, {Dolci}, {Dultzin},
  {Forn{\'e}}, {G{\'o}mez}, {Gurwell}, {Hagen-Thorn}, {Halkola}, {Heidt},
  {Hiriart}, {Hovatta}, {Hsiao}, {Jorstad}, {Kimeridze}, {Konstantinova},
  {Kopatskaya}, {Koptelova}, {Kurtanidze}, {L{\"a}hteenm{\"a}ki}, {Larionov},
  {Leto}, {Ligustri}, {Lindfors}, {Lopez}, {Marscher}, {Mujica},
  {Nikolashvili}, {Nilsson}, {Mommert}, {Palma}, {Pasanen}, {Roca-Sogorb},
  {Ros}, {Roustazadeh}, {Sadun}, {Saino}, {Sigua}, {Sorcia}, {Takalo},
  {Tornikoski}, {Trigilio}, {Turchetti}, \& {Umana}}]{2010ApJ...712..405V}
{Vercellone}, S., {D'Ammando}, F., {Vittorini}, V., {et~al.} 2010, \apj, 712,
  405

\bibitem[{{Villata} {et~al.}(2000{\natexlab{a}}){Villata}, {Raiteri},
  {Popescu}, {Sobrito}, {De Francesco}, {Lanteri}, \&
  {Ostorero}}]{2000A&AS..144..481V}
{Villata}, M., {Raiteri}, C.~M., {Popescu}, M.~D., {et~al.} 2000{\natexlab{a}},
  \aaps, 144, 481

\bibitem[{{Villata} {et~al.}(2000{\natexlab{b}}){Villata}, {Mattox}, {Massaro},
  {Nesci}, {Catalano}, {Frasca}, {Raiteri}, {Sobrito}, {Tosti}, {Nucciarelli},
  {Takalo}, {Sillanp{\"a}{\"a}}, {Karttunen}, {Maesano}, {Marilli}, {Ostorero},
  {Piironen}, \& {Sclavi}}]{2000A&A...363..108V}
{Villata}, M., {Mattox}, J.~R., {Massaro}, E., {et~al.} 2000{\natexlab{b}},
  \aap, 363, 108

\bibitem[{{Villata} {et~al.}(2002){Villata}, {Raiteri}, {Kurtanidze},
  {Nikolashvili}, {Ibrahimov}, {Papadakis}, {Tsinganos}, {Sadakane}, {Okada},
  {Takalo}, {Sillanp{\"a}{\"a}}, {Tosti}, {Ciprini}, {Frasca}, {Marilli},
  {Robb}, {Noble}, {Jorstad}, {Hagen-Thorn}, {Larionov}, {Nesci}, {Maesano},
  {Schwartz}, {Basler}, {Gorham}, {Iwamatsu}, {Kato}, {Pullen},
  {Ben{\'{\i}}tez}, {de Diego}, {Moilanen}, {Oksanen}, {Rodriguez}, {Sadun},
  {Kelly}, {Carini}, {Miller}, {Catalano}, {Dultzin-Hacyan}, {Fan}, {Ishioka},
  {Karttunen}, {Kein{\"a}nen}, {Kudryavtseva}, {Lainela}, {Lanteri},
  {Larionova}, {Matsumoto}, {Mattox}, {Montagni}, {Nucciarelli}, {Ostorero},
  {Papamastorakis}, {Pasanen}, {Sobrito}, \& {Uemura}}]{2002A&A...390..407V}
{Villata}, M., {Raiteri}, C.~M., {Kurtanidze}, O.~M., {et~al.} 2002, \aap, 390,
  407

\bibitem[{{Villata} {et~al.}(2004){Villata}, {Raiteri}, {Kurtanidze},
  {Nikolashvili}, {Ibrahimov}, {Papadakis}, {Tosti}, {Hroch}, {Takalo},
  {Sillanp{\"a}{\"a}}, {Hagen-Thorn}, {Larionov}, {Schwartz}, {Basler},
  {Brown}, {Balonek}, {Ben{\'{\i}}tez}, {Ram{\'{\i}}rez}, {Sadun}, {Boltwood},
  {Carini}, {Barnaby}, {Coloma}, {Ros}, {Dai}, {Xie}, {Mattox}, {Rodriguez},
  {Asfandiyarov}, {Atkerson}, {Beem}, {Bloom}, {Chanturiya}, {Ciprini},
  {Crapanzano}, {de Diego}, {Efimova}, {Gardiol}, {Guerra}, {Kahharov},
  {Kapanadze}, {Karttunen}, {Kato}, {Kimeridze}, {Kudryavtseva}, {Lainela},
  {Lanteri}, {Larionova}, {Maesano}, {Marchili}, {Massone}, {Monroe},
  {Montagni}, {Nesci}, {Nilsson}, {Noble}, {Nucciarelli}, {Ostorero},
  {Papamastorakis}, {Pasanen}, {Peters}, {Pursimo}, {Reig}, {Ryle}, {Sclavi},
  {Sigua}, {Uemura}, \& {Wills}}]{2004A&A...421..103V}
---. 2004, \aap, 421, 103

\bibitem[{{Villata} {et~al.}(2006){Villata}, {Raiteri}, {Balonek}, {Aller},
  {Jorstad}, {Kurtanidze}, {Nicastro}, {Nilsson}, {Aller}, {Arai}, {Arkharov},
  {Bach}, {Ben{\'{\i}}tez}, {Berdyugin}, {Buemi}, {B{\"o}ttcher}, {Carosati},
  {Casas}, {Caulet}, {Chen}, {Chiang}, {Chou}, {Ciprini}, {Coloma}, {di Rico},
  {D{\'{\i}}az}, {Efimova}, {Forsyth}, {Frasca}, {Fuhrmann}, {Gadway}, {Gupta},
  {Hagen-Thorn}, {Harvey}, {Heidt}, {Hernandez-Toledo}, {Hroch}, {Hu}, {Hudec},
  {Ibrahimov}, {Imada}, {Kamata}, {Kato}, {Katsuura}, {Konstantinova},
  {Kopatskaya}, {Kotaka}, {Kovalev}, {Kovalev}, {Krichbaum}, {Kubota},
  {Kurosaki}, {Lanteri}, {Larionov}, {Larionova}, {Laurikainen}, {Lee}, {Leto},
  {L{\"a}hteenm{\"a}ki}, {L{\'o}pez-Cruz}, {Marilli}, {Marscher}, {McHardy},
  {Mondal}, {Mullan}, {Napoleone}, {Nikolashvili}, {Ohlert}, {Postnikov},
  {Pursimo}, {Ragni}, {Ros}, {Sadakane}, {Sadun}, {Savolainen}, {Sergeeva},
  {Sigua}, {Sillanp{\"a}{\"a}}, {Sixtova}, {Sumitomo}, {Takalo},
  {Ter{\"a}sranta}, {Tornikoski}, {Trigilio}, {Umana}, {Volvach}, {Voss}, \&
  {Wortel}}]{2006A&A...453..817V}
{Villata}, M., {Raiteri}, C.~M., {Balonek}, T.~J., {et~al.} 2006, \aap, 453,
  817

\bibitem[{{Wagner} \& {Witzel}(1995)}]{1995ARA&A..33..163W}
{Wagner}, S.~J., \& {Witzel}, A. 1995, \araa, 33, 163

\bibitem[{{Webb} {et~al.}(1988){Webb}, {Smith}, {Leacock}, {Fitzgibbons},
  {Gombola}, \& {Shepherd}}]{1988AJ.....95..374W}
{Webb}, J.~R., {Smith}, A.~G., {Leacock}, R.~J., {et~al.} 1988, \aj, 95, 374

\bibitem[{{Webb} {et~al.}(1998){Webb}, {Freedman}, {Howard}, {Ma}, {Belfort},
  {Rave}, {Rumstay}, {Nicol}, {Krick}, {Oswalt}, {Marshall}, \&
  {Robishaw}}]{1998AJ....115.2244W}
{Webb}, J.~R., {Freedman}, I., {Howard}, E., {et~al.} 1998, \aj, 115, 2244

\bibitem[{{Weistrop} {et~al.}(1985){Weistrop}, {Shaffer}, {Hintzen}, \&
  {Romanishin}}]{1985ApJ...292..614W}
{Weistrop}, D., {Shaffer}, D.~B., {Hintzen}, P., \& {Romanishin}, W. 1985,
  \apj, 292, 614

\bibitem[{{Wierzcholska} {et~al.}(2015){Wierzcholska}, {Ostrowski}, {Stawarz},
  {Wagner}, \& {Hauser}}]{2015A&A...573A..69W}
{Wierzcholska}, A., {Ostrowski}, M., {Stawarz}, {\L}., {Wagner}, S., \&
  {Hauser}, M. 2015, \aap, 573, A69

\bibitem[{{Wills} \& {Wills}(1974)}]{1974ApJ...190L..97W}
{Wills}, B.~J., \& {Wills}, D. 1974, \apjl, 190, L97

\bibitem[{{Wolf}(1916)}]{1916AN....202..415W}
{Wolf}, M. 1916, Astronomische Nachrichten, 202, 415

\bibitem[{{Woo} \& {Urry}(2002)}]{2002ApJ...579..530W}
{Woo}, J.-H., \& {Urry}, C.~M. 2002, \apj, 579, 530

\bibitem[{{Wu} {et~al.}(2012){Wu}, {B{\"o}ttcher}, {Zhou}, {He}, {Ma}, \&
  {Jiang}}]{2012AJ....143..108W}
{Wu}, J., {B{\"o}ttcher}, M., {Zhou}, X., {et~al.} 2012, \aj, 143, 108

\bibitem[{{Wu} {et~al.}(2011){Wu}, {Zhou}, {Ma}, \&
  {Jiang}}]{2011MNRAS.418.1640W}
{Wu}, J., {Zhou}, X., {Ma}, J., \& {Jiang}, Z. 2011, \mnras, 418, 1640

\bibitem[{{Xie} {et~al.}(1987){Xie}, {Li}, {Bao}, {Hau}, {Zhou}, {Liu}, \&
  {Deng}}]{1987A&AS...67...17X}
{Xie}, G.-Z., {Li}, K.-H., {Bao}, M.-X., {et~al.} 1987, \aaps, 67, 17

\bibitem[{{Xie} {et~al.}(1991){Xie}, {Li}, {Cheng}, {Lu}, {Liu}, {Hao}, \&
  {Liu}}]{1991A&AS...87..461X}
{Xie}, G.~Z., {Li}, K.~H., {Cheng}, F.~Z., {et~al.} 1991, \aaps, 87, 461

\bibitem[{{Xie} {et~al.}(1992){Xie}, {Li}, {Liu}, {Lu}, {Wu}, {Fan}, {Zhu}, \&
  {Cheng}}]{1992ApJS...80..683X}
{Xie}, G.~Z., {Li}, K.~H., {Liu}, F.~K., {et~al.} 1992, \apjs, 80, 683

\bibitem[{{Xie} {et~al.}(1994){Xie}, {Li}, {Zhang}, {Liu}, {Fan}, \&
  {Wang}}]{1994A&AS..106..361X}
{Xie}, G.~Z., {Li}, K.~H., {Zhang}, Y.~H., {et~al.} 1994, \aaps, 106

\bibitem[{{Yan} {et~al.}(2000){Yan}, {Burstein}, {Fan}, {Zheng}, {Chen},
  {Byun}, {Chen}, {Chen}, {Deng}, {Deng}, {Fang}, {Hester}, {Jiang}, {Li},
  {Lin}, {Lu}, {Shang}, {Su}, {Sun}, {Tsay}, {Windhorst}, {Wu}, {Xia}, {Xu},
  {Xue}, {Zheng}, {Zhu}, \& {Zou}}]{2000PASP..112..691Y}
{Yan}, H., {Burstein}, D., {Fan}, X., {et~al.} 2000, \pasp, 112, 691

\bibitem[{{Yuan} {et~al.}(2015){Yuan}, {Fan}, \& {Pan}}]{2015AJ....150...67Y}
{Yuan}, Y.~H., {Fan}, J.~H., \& {Pan}, H.~J. 2015, \aj, 150, 67

\bibitem[{{Zekl} {et~al.}(1981){Zekl}, {Klare}, \&
  {Appenzeller}}]{1981A&A...103..342Z}
{Zekl}, H., {Klare}, G., \& {Appenzeller}, I. 1981, \aap, 103, 342

\bibitem[{{Zeng} {et~al.}(2018){Zeng}, {Zhao}, {Dai}, {Jiang}, {Geng}, {Yang},
  {Liu}, {Wang}, {Feng}, \& {Zhang}}]{2018PASP..130b4102Z}
{Zeng}, W., {Zhao}, Q.-J., {Dai}, B.-Z., {et~al.} 2018, \pasp, 130, 024102

\bibitem[{{Zhai} {et~al.}(2011){Zhai}, {Zheng}, \& {Wei}}]{2011A&A...531A..90Z}
{Zhai}, M., {Zheng}, W.~K., \& {Wei}, J.~Y. 2011, \aap, 531, A90

\bibitem[{{Zhang} \& {Li}(2017)}]{2017MNRAS.469.1682Z}
{Zhang}, Y.-H., \& {Li}, J.-C. 2017, \mnras, 469, 1682

\bibitem[{{Zhou} {et~al.}(2003){Zhou}, {Jiang}, {Ma}, {Xue}, {Wu}, {Chen},
  {Zhu}, {Sun}, \& {Windhorst}}]{2003A&A...397..361Z}
{Zhou}, X., {Jiang}, Z., {Ma}, J., {et~al.} 2003, \aap, 397, 361

\bibitem[{{Zwicky}(1966)}]{1966ApJ...143..192Z}
{Zwicky}, F. 1966, \apj, 143, 192

\end{thebibliography}
\bibliographystyle{apj}

\newpage

\begin{table*}
	\centering
	\caption{Central Wavelengths and Bandwidths of 5 Filters}
	\label{filters}
	\begin{tabular}{ccc}
	\hline
	\hline
	Filter & Central Wavelength & Bandwidth\\
	         &  (\AA )    & (\AA)\\
	\tableline
	$c$     &  4206  & 289 \\
	$e$       &  4885  & 372 \\
	$i $       &  6685  & 514 \\
	$m$      &  8013  & 287 \\
	$o$       &  9173  & 248 \\
	\hline
	 \hline
	\end{tabular}
\end{table*}

\begin{table*}
	\centering
	\caption{Observation details of ten objects}
	\label{sources}
	\begin{tabular}{lccclcllrr}
	\hline
	\hline
	Name & RA & Dec & z & Type & Band &Start date & End date & N$_{\mathrm{day}}$ & N$_{\mathrm{data}}$ \\
	\hline
	3C 66A & 02 22 39.61 & +43 02 07.8 & 0.444 &  IBL &    $   c  $& 2006 11 26 & 2011 02 27 & 237& 1623  \\
	             &                     &                     &           &             &   $    i  $ & 2005 01 29  & 2011 02 27  & 248 & 1790 \\
	             &                     &                     &           &           &$     o$   &  2006 11 26 &  2011 02 27  &  235 &   1649 \\ 
	   \hline
	PKS 0735+178 & 07 38 07.39 & +17 42 19.0 & 0.424 & LBL&$  c$ &  2009 12 18 & 2010 11 30 & 13 &29  \\
	             &                     &                     &           &        & $i    $      &   2009 12 18 & 2010 11 30  & 15 & 37\\
	             &                     &                     &           &        & $o  $      &   2009 12 18 & 2010 11 30  & 10 &23\\
	 \hline
	OI 090.4 & 07 57 06.64 &+09 56 34.8 & 0.266 & LBL &$ c $   &   2007 01 20 & 2011 03 30 & 19 &81\\
	             &                     &                     &           &     & $ i  $         &   2006 03 14 & 2011 03 30 & 21  &99\\
	             &                     &                     &           &      &$ o $        &   2007 01 20  &  2011 03 30  & 18  &  78\\
	 \hline
	Mrk 421 & 11 04 27.20 &+38 12 32.0 & 0.030 & HBL & $c   $  &   2007 04 05  & 2008 05 05 & 15 &108 \\
	             &                     &                     &           &      &$ m $        &   2005 01 30  & 2006 03 30  &  5  & 39 \\
	             &                     &                     &           &       &$ o $       &   2007 04 05  & 2008 05 05  & 16 & 128 \\     
	  \hline                  
	ON 231 & 12 21 31.69 & +28 13 58.5 & 0.102 & IBL & $c $ &  2009 05 23 & 2011 06 04 & 76 &338\\
	             &                     &                     &           &       & $i $       &  2009 05 23  & 2011 06 04 & 80 &380  \\
	             &                     &                     &           &       &  $ o $       &  2009 05 23  &  2011 05 15 & 78  & 360 \\
	 \hline            
	PG 1553+113 & 15 55 43.04 & +11 11 24.4 & 0.360 & HBL&$ c $ & 2009 05 26 & 2010 03 28 & 28&135 \\
	             &                     &                     &           &       &   $  i $       &    2009 05 26 & 2010 03 28 & 28&134 \\
	             &                     &                     &           &          &    $ o $     &   2009 05 26 & 2010 03 28 & 28& 131\\  
	 \hline                       
	OT 546 (phase 1) & 17 28 18.62 & +50 13 10.5 & 0.055 & HBL  &     $ e $       &   2005 03 14  & 2006 07 02  & 55 & 577 \\
	         &                     &                     &           &       &$  i $       &  2005 03 14   &2006 07 02    & 54 & 566  \\	
	             &                     &                     &           &        &   $m $      &  2005 03 14   &   2006 07 02  & 52 &  564 \\
	 \hline 
	 OT 546 (phase 2)    &  17 28 18.62 &  +50 13 10.5 &  0.055   & HBL &       $ c $       &   2007 02 03  & 2011 05 31  & 113 & 456 \\
	             &                     &                     &           &       &$  i $       &  2007 02 02  &  2011 05 31  & 121 & 549  \\   
	             &                     &                     &           &        &$  o$       &   2007 02 03   &  2011 03 23  &  110 & 466  \\
	  \hline                                            
	1ES 1959+650 & 19 59 59.85 & +65 08 54.7 & 0.047 & HBL &   $  c$& 2009 07 03 & 2011 06 13 & 81 & 336\\
	             &                     &                     &           &             &    $ i  $          &  2009 07 03  & 2011 06 13   & 81 & 349 \\
	             &                     &                     &           &             &   $o  $          &  2009 07 03  &  2011 05 14  & 71 & 256 \\ 
	 \hline                        
 BL Lacertae (phase 1)& 22 02 43.29 & +42 16 39.9 & 0.069 & LBL &  $  e$             &  2006 07 01    & 2006  11 16  &  10&  135   \\
	             &                     &                     &           &          &$  i$             &  2006 07 01   &  2006 11 23   & 11 & 158   \\ 	
	             &                     &                     &           &            & $m $         &  2006 07 01   &  2006 11 16    & 10   &  142  \\	
	\hline
BL Lacertae (phase 2)  &   22 02 43.29   & +42 16 39.9 &  0.069 &  LBL    & $c $ &   2006 12 04 & 2010 12 29 & 155 & 791 \\	             
	             &                     &                     &           &          &$  i$             &  2006 12 04   &  2010 12 29   & 162 & 877   \\ 
	             &                     &                     &           &            & $o  $          &  2006 12 04  & 2010 12 29   & 161 &   853 \\ 
	 \hline                                                
	3C 454.3 & 22 53 57.74 & +16 08 53.5 & 0.859 & FSRQ   &       $c  $      &   2007 08 17  & 2010 12 14 & 39 & 179 \\
	             &                     &                     &           &        &     $ i  $            &   2007 07 19    & 2010 12 14  & 42&  196  \\
	             &                     &                     &           &        &    $  o$            &   2007 08 17  &  2010 12 14  &  40 & 187  \\

	\hline
	 \hline
	\end{tabular}

\end{table*}

\begin{table*}
	\centering
	\caption{The Observation and Result of 3C 66A in the $c$ band} 
	\label{example}
	\begin{threeparttable}
	\begin{tabular}{ccccccccccccr}
	\hline
	\hline

Date & & Time &&Julian Date && Exp &&  $c$ && $c_{err}$ && dfmag \\
(UT) & &(UT) &&              &&   (s)   && (mag)  &&(mag)&&(mag)\\
\tableline
 2006 11 26 & &14:01:56.0 & & 2454066.08468 && 240 & & 15.292 && 0.022&&$-0.024$\\
 2006 11 26 &&14:18:10.0 & &2454066.09595  & &240  & &15.285  & &0.023&&  0.021\\
 2006 11 26 & &14:44:55.0 & & 2454066.11453 & & 240 && 15.322 && 0.021&&$-0.006$\\
 2006 11 26 & &14:57:37.0 & & 2454066.12334 & & 240 & & 15.310 && 0.021&&$-0.006$\\
 2006 11 26 & &15:10:47.0 & & 2454066.13249 & & 240 & & 15.284 && 0.022&& 0.007\\
 ...  && ...  & & ... && ... & & ... && ... &&... \\
\tableline
 \hline
\end{tabular} 
\vspace{3 pt}
\begin{tablenotes} 
	\item(The light curves for all 34 observations of the blazars are available in their entirety in a machine-readable form in the online journal. A portion is shown for guidance regarding its form and content.)
	\end{tablenotes}
\end{threeparttable}

\end{table*}

\begin{table*}
\centering
\caption{Galactic extinction values of each source in the passbands}
\label{extinction}
\begin{tabular}{cccccc}
\hline
Name&$A_{c}$&$A_{e}$&$A_{i}$&$A_{m}$&$A_{o}$\\
         & (mag)& (mag) & (mag) &(mag) & (mag)\\
\hline

3C 66A&0.314&0.261&0.172&0.127&0.101\\

PKS 0735+178&0.127&0.105&0.070&0.052&0.041\\

OI 090.4&0.082&0.069&0.045&0.033&0.027\\

ON 231&0.087&0.072&0.048&0.035&0.028\\

OT 546&0.109&0.091&0.060&0.045&0.035\\

1ES 1959+650&0.642&0.535&0.353&0.261&0.207\\

BL Lacertae&1.223&1.017&0.671&0.495&0.394\\

3C 454.3&0.392&0.326&0.215&0.159&0.126\\

\hline
\end{tabular}
\end{table*}

\begin{table*}
	\centering
	\caption{The statistical results of the variability behaviors}
	\label{observations}
	\begin{tabular}{llcccc}
	\hline
	\hline
	Name & Band & Maximum & Minimum & Average &Amplitude\\
	          &          &     (mag)   &   (mag)   &      (mag)   &   (mag) \\
	\tableline
	3C 66A  & $ c$  & 15.548 $\pm$ 0.014  & 14.047 $\pm$ 0.029 &14.798&1.501 \\
	              & $ i  $ & 14.882 $\pm$ 0.012  & 13.409 $\pm$ 0.012 &14.146& 1.473\\
	              & $ o $ & 14.465 $\pm$ 0.019  & 12.990 $\pm$ 0.016 &13.728&1.475\\
	               \hline
	PKS 0735+178  &$  c$ & 17.489 $\pm$ 0.136 & 16.554 $\pm$ 0.022 &17.022&0.935\\
	                          & $ i  $  & 16.675 $\pm$ 0.044 & 15.819 $\pm$ 0.030&16.247&0.856 \\
	                          & $ o $ & 16.191 $\pm$ 0.063 & 15.434 $\pm$ 0.037&15.813&0.757 \\
	                           \hline
	OI 090.4 &$ c$ & 17.712 $\pm$ 0.065    & 16.575 $\pm$ 0.020&17.144&1.137 \\
	            &    $ i $  & 16.933 $\pm$ 0.041  & 15.613 $\pm$ 0.086&16.273& 1.320 \\
	            &    $  o $& 16.362 $\pm$ 0.066  & 15.347 $\pm$ 0.026&15.855&1.015\\
	             \hline
	Mrk 421 &$ c  $ & 13.616 $\pm$ 0.057   & 13.159 $\pm$ 0.016 &13.388&0.457\\
	              & $ m $ & 13.250 $\pm$ 0.030  & 12.962 $\pm$ 0.027&13.106 &0.288\\
	              & $ o  $& 12.906 $\pm$ 0.102   & 12.569 $\pm$ 0.026&12.738 &0.337\\
	               \hline
	ON 231 &$ c $& 16.661 $\pm$ 0.033     & 15.105 $\pm$ 0.038&15.883&1.556 \\
	             & $ i  $& 15.976 $\pm$ 0.027   & 14.493 $\pm$ 0.021 &15.235&1.483  \\
	             & $ o $ & 15.510 $\pm$ 0.047  & 14.064 $\pm$ 0.018 &14.787& 1.446 \\
	              \hline
	OT 546 (phase 1) &  $e  $& 16.791 $\pm$ 0.170   & 16.344 $\pm$ 0.060&16.568&0.447 \\
	&    $i  $&  15.858$\pm$ 0.032    & 15.607  $\pm$ 0.021 & 15.751    & 0.241  \\
	              & $m $& 15.502 $\pm$ 0.039    & 15.287 $\pm$ 0.026 &15.395&0.215\\
	                \hline
	OT 546 (phase 2) &$ c $& 16.925 $\pm$ 0.072    & 16.162 $\pm$ 0.028&16.544&0.763 \\	            
	            &  $i  $& 16.016 $\pm$ 0.042   & 15.493 $\pm$0.025 & 15.806  & 0.523 \\	       
	            &  $o  $& 15.322 $\pm$ 0.042   & 14.898 $\pm$ 0.039 &15.110&0.424\\
	             \hline
	 1ES 1959+650 &$ c$ & 16.275 $\pm$ 0.026   &14.900 $\pm$ 0.017&15.588&1.375\\
	                          &   $i  $& 15.205 $\pm$ 0.016   & 14.159 $\pm$ 0.016&14.682&1.046\\
	                          &   $ o $& 14.663 $\pm$ 0.031& 13.737 $\pm$ 0.024 &14.200&0.926\\
	                           \hline
	  BL Lacertae (phase 1) & $ e $ & 16.067 $\pm$ 0.020     & 15.247 $\pm$ 0.070&15.657& 0.820\\
	    & $ i  $& 14.880  $\pm$ 0.028        & 14.119  $\pm$ 0.092 & 14.387& 0.761 \\
	                         & $m$ & 14.337 $\pm$ 0.034       & 13.572 $\pm$ 0.034&13.955&0.765\\
	                           \hline
	     BL Lacertae (phase 2)   &$ c $& 16.701 $\pm$ 0.055       & 15.039 $\pm$ 0.041&15.870&1.662 \\
	                      & $ i  $&15.108  $\pm$  0.012       & 13.704 $\pm$ 0.017 & 14.449 & 1.404 \\	                     
	                      &  $o$ & 14.181 $\pm$ 0.015       & 12.854 $\pm$ 0.014&13.518&1.327\\
	                       \hline
	3C 454.3 &$ c $&   16.840 $\pm$ 0.034             & 14.346 $\pm$ 0.013&15.593&2.494\\
	               &  $i  $& 16.096 $\pm$ 0.033             & 13.166 $\pm$ 0.021&14.631&2.930 \\
	               &  $o $ & 15.494 $\pm$ 0.044             & 12.758 $\pm$ 0.016&14.126&2.736\\
	\hline 
	 \hline
	\end{tabular}

\end{table*}

\begin{table*}
	\centering
	\caption{Time lags for some objects}
	\label{alldelay}
	\begin{tabular}{lcrr}
	\hline
	\hline
	  Source & Correlated Passbands & Delay & Error \\
	             &    &  (day)    & (day)\\
	\tableline
		3C 66A &$c-i$     &  8.125 & 7.954 \\
	             &$c-o$   &  15.002  & 8.283 \\
	             &$i-o$    & 8.632   & 7.354 \\
                   \hline
	 ON 231 &$c-i$     & 0.471 & 2.775 \\
	             &$c-o$   &  0.907  & 3.333 \\
	             &$i-o$    & 0.301   &2.783 \\ 
	               \hline
	 PG 1553+113 &$c-i$     &  0.898 & 2.068 \\
	             &$c-o$   &  1.325  & 2.512 \\
	             &$i-o$    & 0.456   & 2.810 \\  
	               \hline 
	  1ES 1959+650 &$c-i$     &  1.827 & 2.495 \\
	             &$c-o$   &  0.778  & 3.011 \\
	             &$i-o$    & $-0.816$   & 2.881 \\   
	               \hline
	 BL Lacertae (Overall) &$c-i$     &  $-0.200$ & 0.851 \\
	             &$c-o$   &  $-0.304$  & 0.907 \\
	             &$i-o$    & $-0.062$   &0.790 \\  
	               \hline
	 BL Lacertae (IDV)  &$e-i$     &  $-0.004$ & 0.006 \\
	            &$e-m$       &  $-0.003$  & 0.006 \\
	            &$i-m$       & 0.001  & 0.006 \\                                                
		
	\hline
	 \hline
	\end{tabular}
	
\end{table*}

\begin{table*}
	\centering
	\caption{Color-magnitude correlations fitting results.}
	\label{color}
	\begin{threeparttable}
	\begin{tabular}{llcrrrl}
	\hline
	\hline
	    Source  & Color &  Mag. & Slope & Intercept  & r  & Prob.  \\
	       (1)  &  (2)& (3)   & (4)&(5)&(6)&(7)\\
	\tableline
	3C 66A  & $ c-i$ & $c$         &   0.04 & 0.16 & 0.24 &  $7.56\times 10^{-4}$   \\
	              & $ c-o  $&    $c$    &  0.07 &  0.06& 0.35 &  $2.61\times 10^{-7}   $\\
	              & $ i-o $&   $i$        &   0.02&  0.09 &  0.25 &  $5.37\times 10^{-4}   $ \\
	               \hline
	PKS 0735+178 & $ c-i$ & $c$         &  0.15 & $-1.89$ & 0.76 &  $6.39\times 10^{-3}$ \\
	                         & $ c-o  $ &  $c$    & 0.28&  $-3.60$& 0.77 &  $1.43\times 10^{-2}$\\
	                         & $ i-o $ &  $i$        &  0.11&  $-1.34$ &  0.58 &$7.62\times 10^{-2} $  \\
	                          \hline
	OI 090.4 &$ c-i$&  $c$         &   0.02 & 0.32 & 0.10 & $7.65\times 10^{-1}$ \\
	            &   $ c-o  $&    $c$    &  0.11 &  $-0.59$& 0.40 & $ 1.16\times 10^{-1}$\\
	            &   $ i-o $ &  $i$        &   0.11&  $-1.27$ &  0.70 &$1.54\times 10^{-2}  $ \\
	             \hline
	Mrk 421 &  $ c-o  $ &   $c$    &  0.37 &  $-4.29$& 0.86 & $ 7.56\times 10^{-4}  $ \\  
	 \hline       
	ON 231 & $ c-i$ & $c$         &  0.16 & $-1.84$ & 0.78 & $ 9.95\times 10^{-16}$ \\
	             &$ c-o  $  & $c$    & 0.24&  $-2.55$& 0.78 & $ 4.08\times 10^{-15}$\\
	             &$ i-o $ &  $i$        &  0.09&  $-0.84$ &  0.59 &$4.42\times 10^{-8}  $ \\	
	              \hline             
	OT 546 (phase 1) &$ e-i$  &  $e$    &   0.51 &$-7.56$ & 0.67 &$ 8.19\times 10^{-8}$ \\
	            &  $e-m $ & $e$ &  0.56&  $-8.14$& 0.67 &  $1.38\times 10^{-7}$\\
	            &  $i-m $ &$i$  &   0.15& $-2.04$ &  0.34 & $1.51\times 10^{-2} $  \\
	          \hline
       OT 546 (phase 2) &  $ c-i$ & $c$         &   0.42 &$-6.18$ & 0.85 & $1.32\times 10^{-31} $\\
	              & $ c-o  $  &  $c$    &  0.49 &  $-6.76$& 0.86 &  $3.36\times 10^{-30}$\\
	              & $ i-o $ &  $i$        &   0.16& $-1.97$ &  0.36 & $2.37\times 10^{-4} $  \\
	             \hline	            
	 1ES 1959+650 &  $ c-i$ & $c$         &   0.27 &$-3.37$ & 0.95 &$ 3.85\times 10^{-38} $\\
	                         	 & $ c-o  $  &  $c$    &  0.37 &  $-4.46$& 0.98 & $ 2.12\times 10^{-47}$\\
	                          & $ i-o $ &  $i$        &   0.16& $-1.88$ &  0.89 &$ 1.62\times 10^{-23}  $ \\
	                          	 \hline	                           
	  BL Lacertae (phase 1)  &$e-i$  &  $e$    &   0.07 &$-0.03$ & 0.60 & $8.64\times 10^{-2}$ \\
	                      &  $e-m $ & $e$ &  0.11&  0.02& 0.68 & $ 4.18\times 10^{-2}$\\
	                      &  $i-m $ & $i$  &   0.03& 0.03 &  0.40 &$ 2.89\times 10^{-1}  $ \\	  
	                        \hline
	 BL Lacertae (phase 2)  &  $ c-i$ & $c$         &   0.15 &$-1.00$ & 0.43 & $4.16\times 10^{-4}$ \\
	                        & $ c-o  $  &  $c$    &  0.18&  $-0.54$& 0.71 & $ 2.99\times 10^{-19}$\\ 
	                        & $ i-o $ &  $i$        &   0.10& $-0.50$ &  0.46 &$ 3.87\times 10^{-6} $  \\	                    
	                      \hline	   	                           
	3C 454.3 &$c-i$ & $c$         &   $-0.13$ & 3.02 & 0.71 &$  9.80\times 10^{-6} $  \\
	               & $c-o  $  & $c$    &  $-0.10 $&  3.26& 0.49 &$  3.10\times 10^{-3}  $ \\
	               & $i-o $ &  $i$        &  $- 0.03$&  1.10 &  0.31 &  $8.02\times 10^{-2}$    \\	
	\hline
	 \hline
	\end{tabular}
	
	      \vspace{5 pt}
	\begin{tablenotes} 
	\item {\bfseries Note.} (1) Source name; (2) color; (3) magnitude; (4) slope; (5) intercept; (6) correlation coefficient; (7) null hypothesis probability.
	\end{tablenotes}
\end{threeparttable}
	
\end{table*}

\begin{figure*}
\centering
	\includegraphics[scale=0.7]{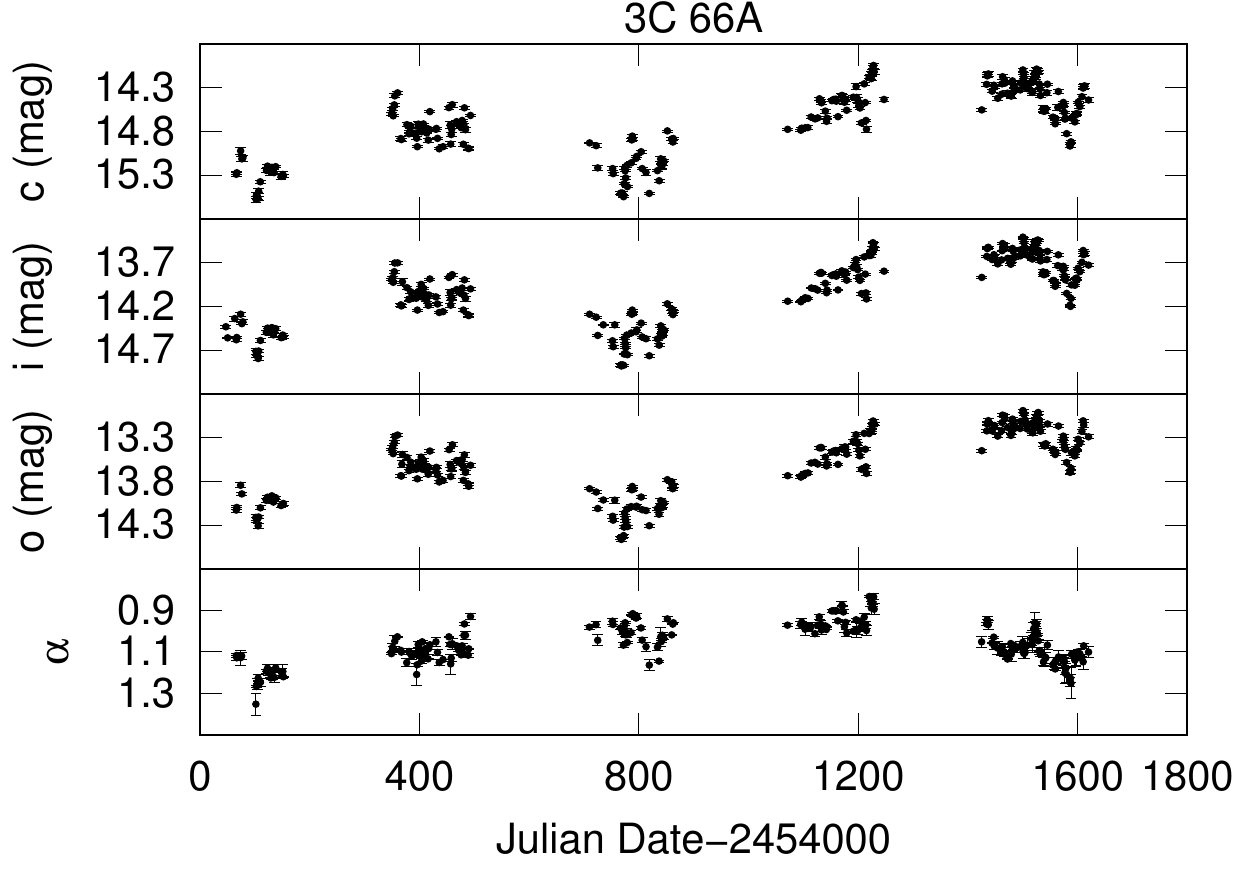}
	\includegraphics[scale=0.7]{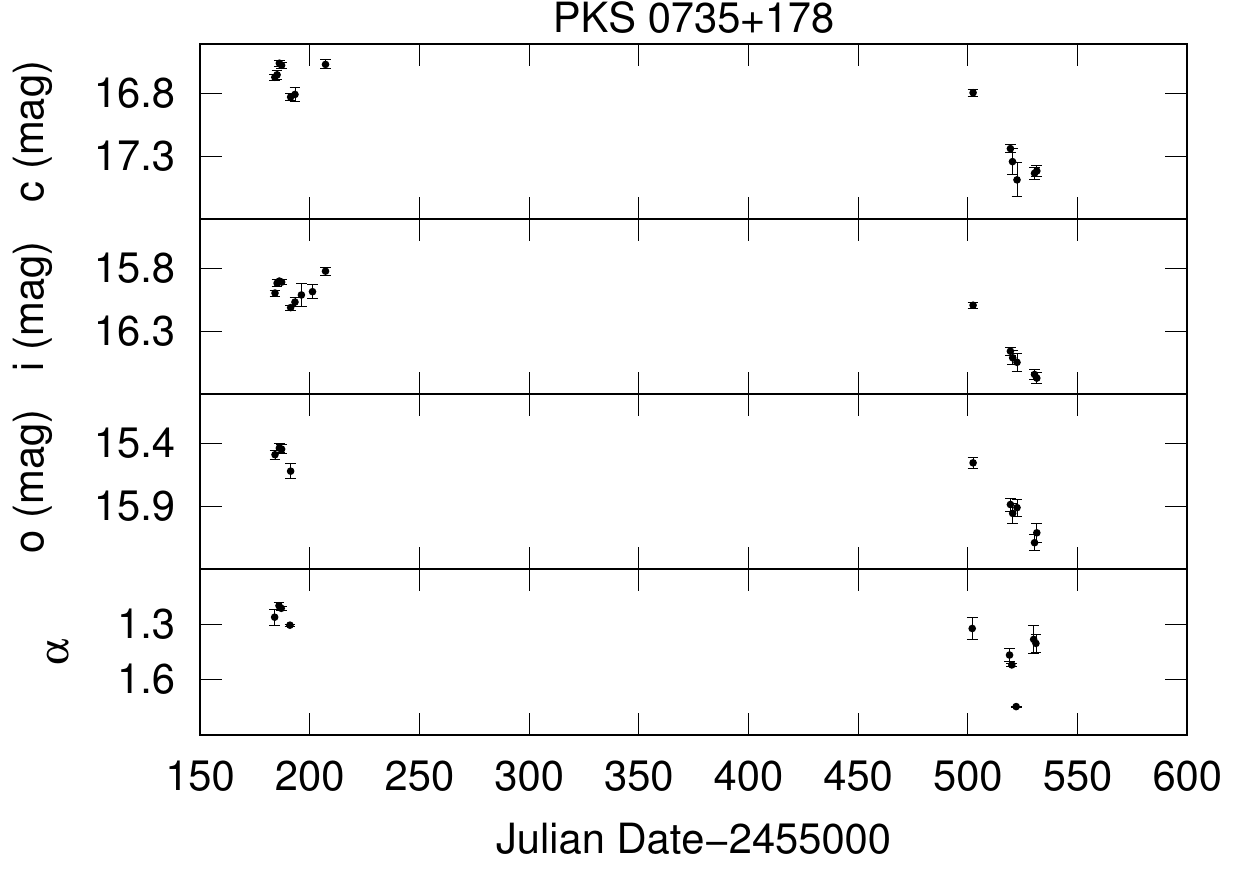}
	\includegraphics[scale=0.7]{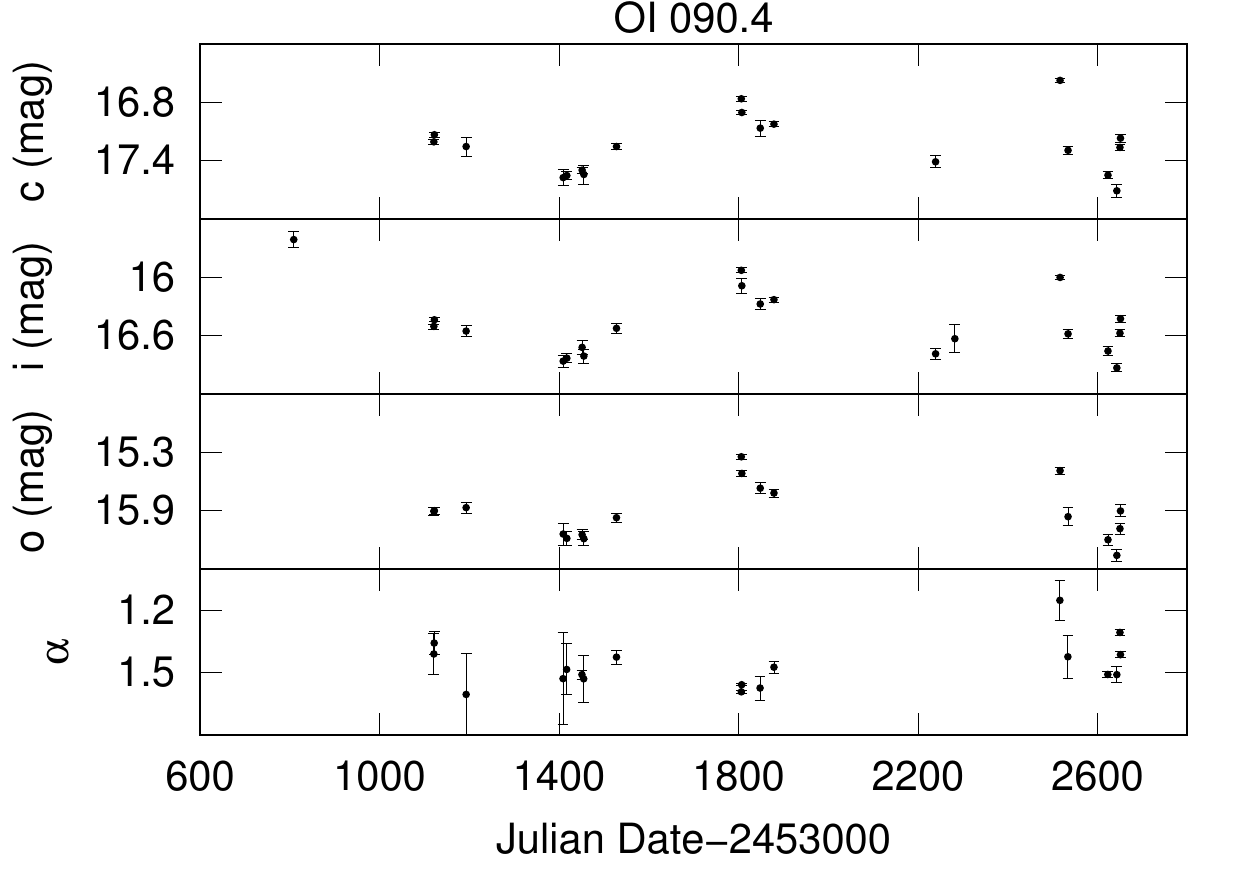}
	\includegraphics[scale=0.7]{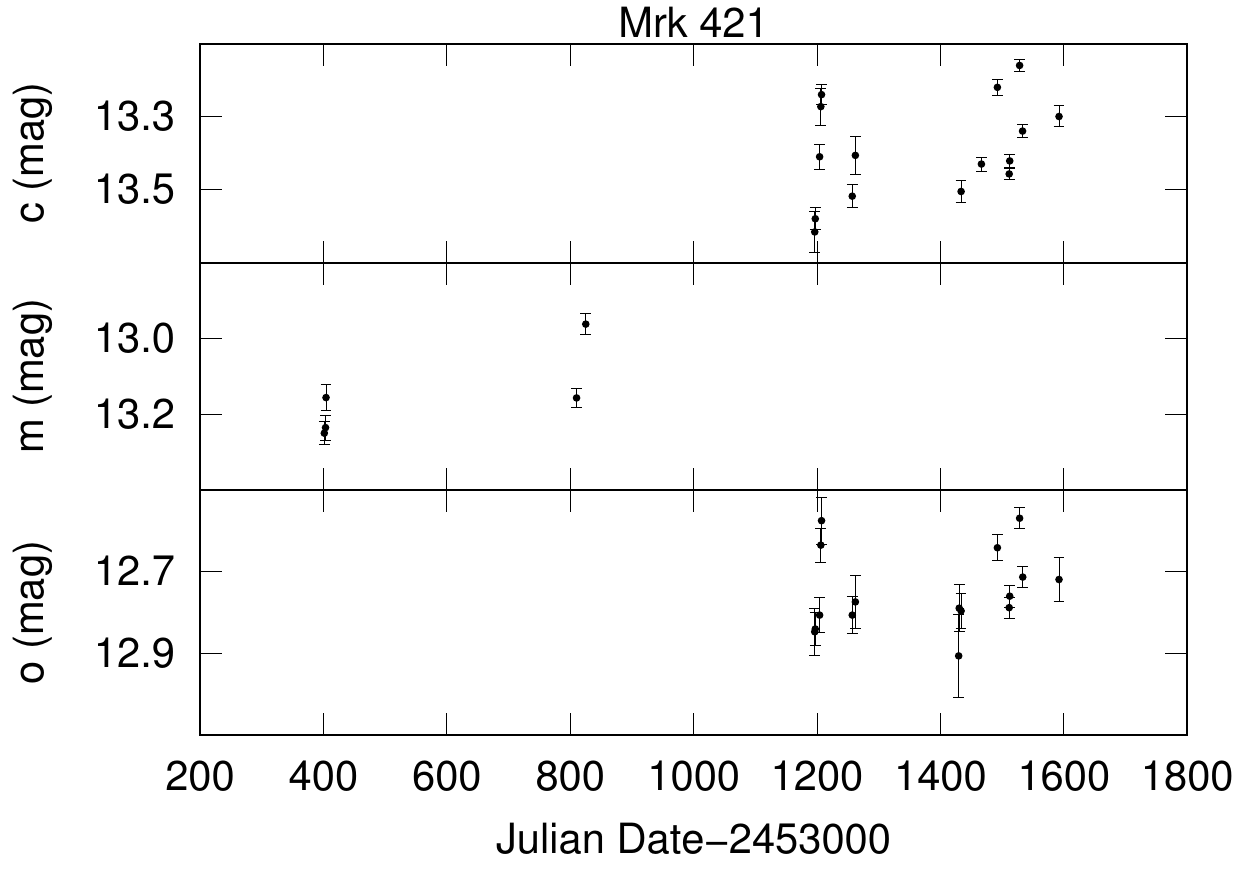}
	\includegraphics[scale=0.7]{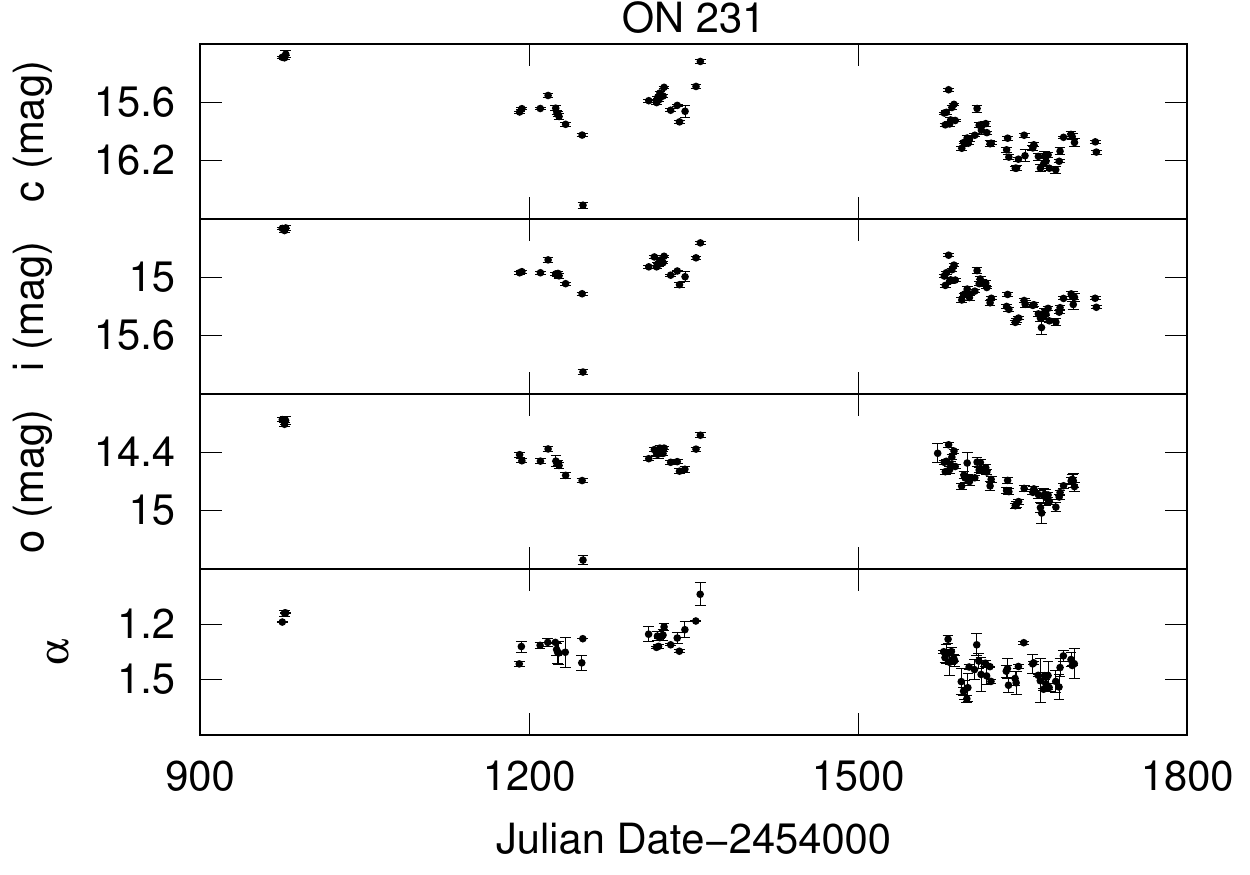}
	\includegraphics[scale=0.7]{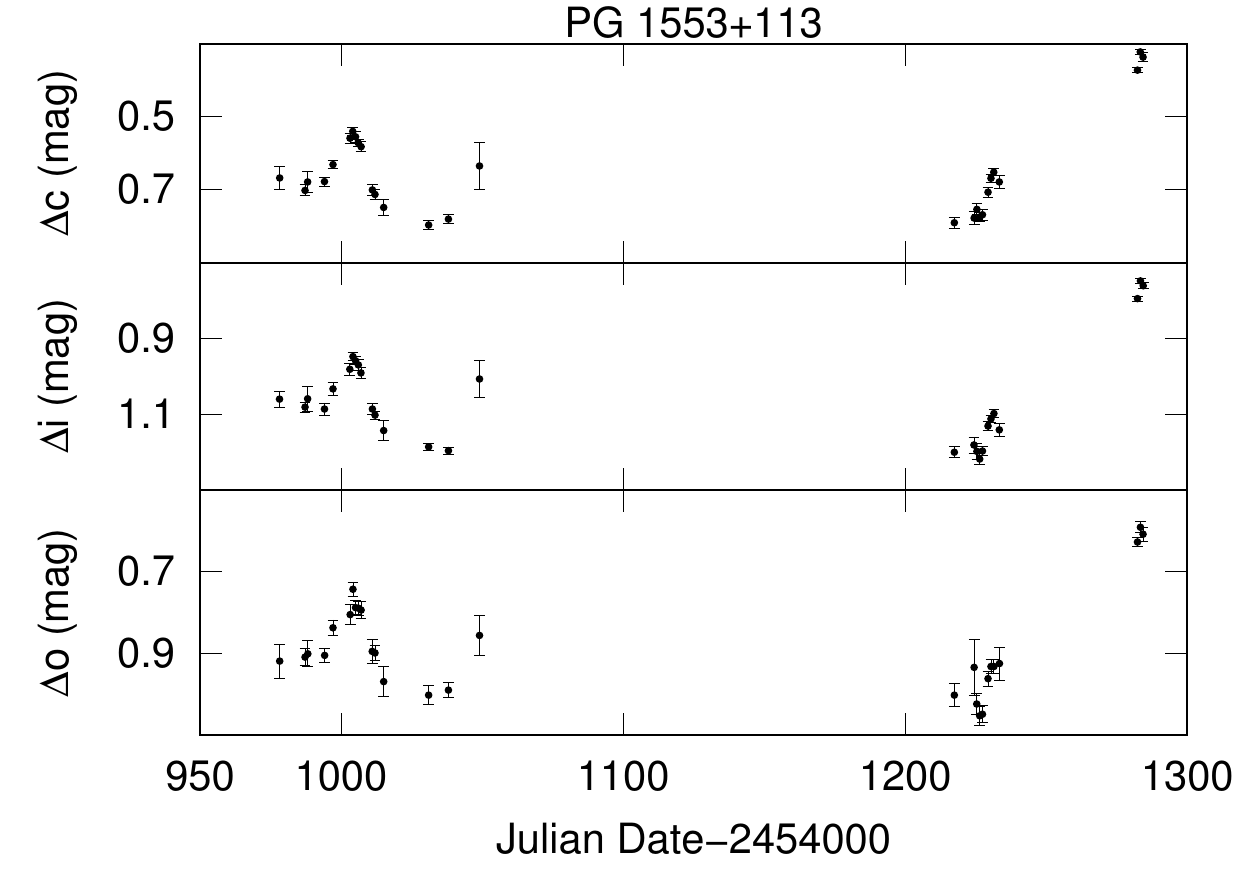}
    \caption{The lightcurves and spectral index diagrams of all targets. The dash lines separate each whole observation period into two phases for OT 546 and BL Lacertae. For Mrk 421 and PG 1553+113, we didn't calculate their spectral indices, see text for details.}
    \label{fig:all} 
\end{figure*}

\renewcommand{\thefigure}{\arabic{figure} }
\addtocounter{figure}{-1}
\begin{figure*}
\centering
	\includegraphics[scale=0.7]{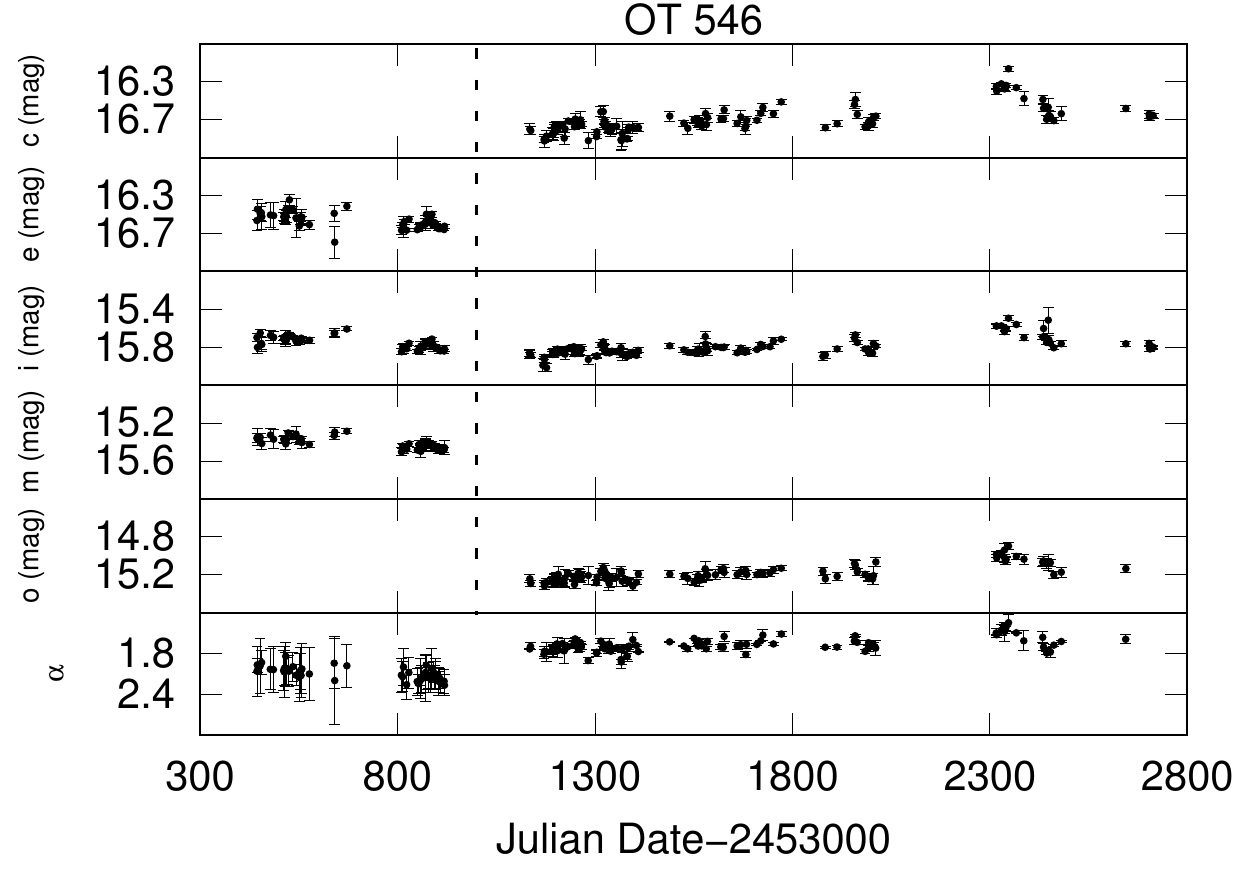}
	\includegraphics[scale=0.7]{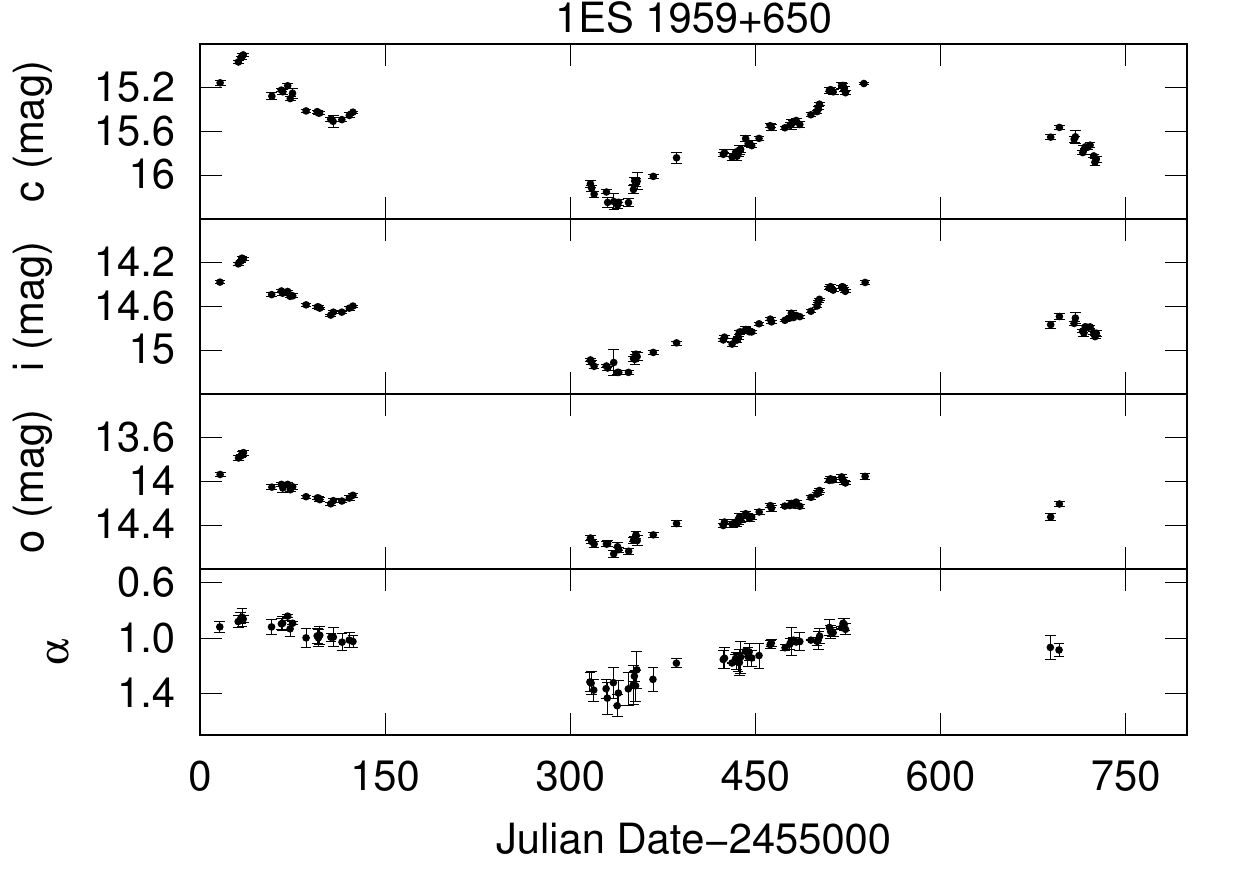}
	\includegraphics[scale=0.7]{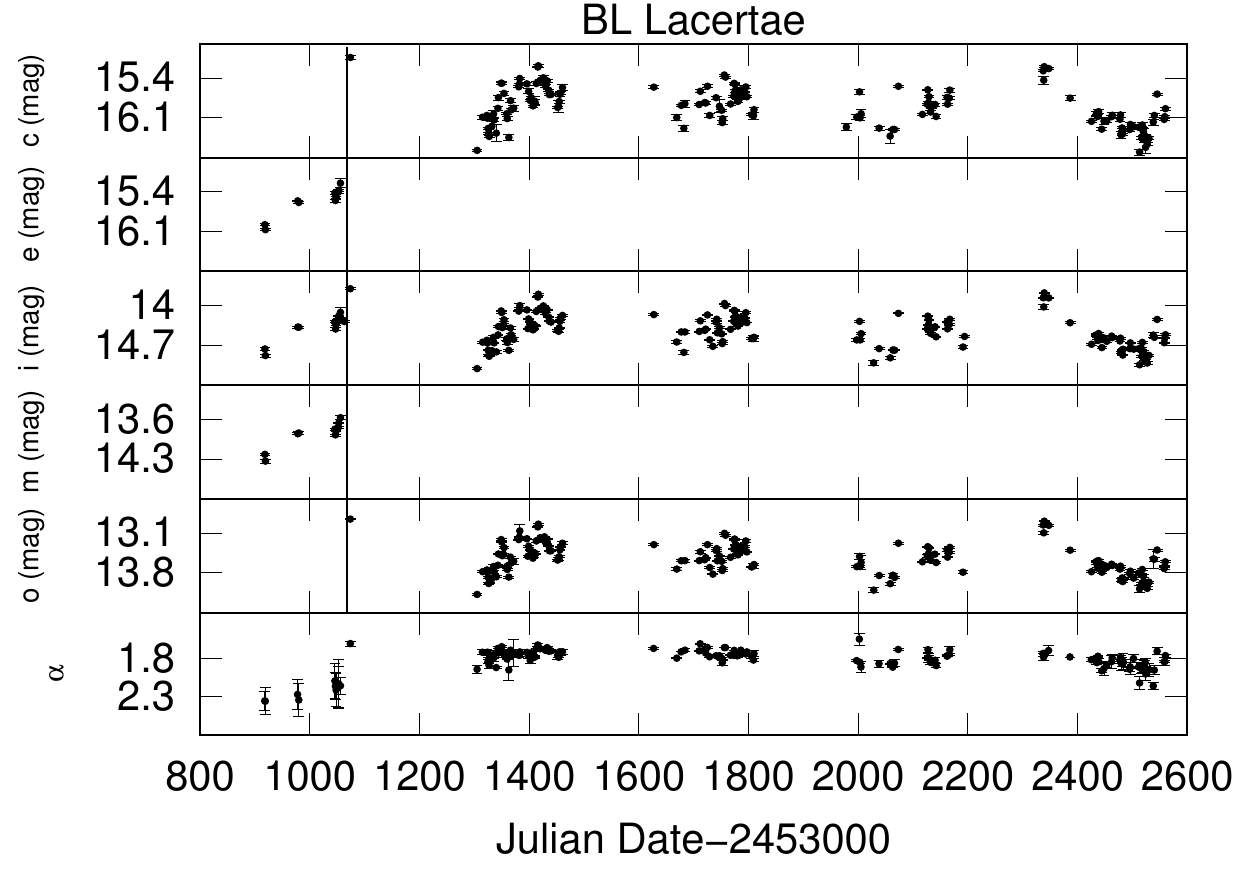}
	\includegraphics[scale=0.7]{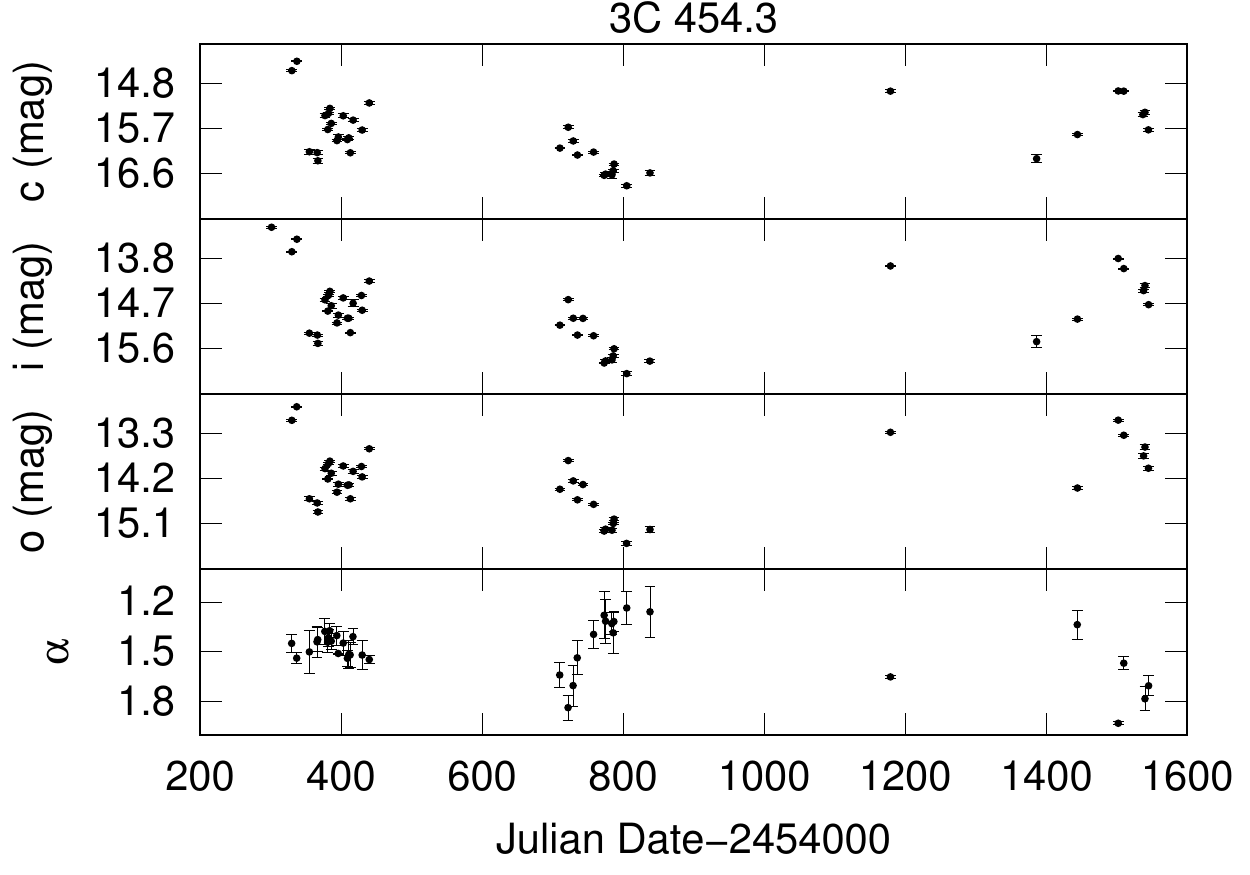}
    \caption{continued}
    \label{fig:all} 
\end{figure*}

\renewcommand{\thefigure}{\arabic{figure}}

\begin{figure*}[htbp]
\centering
\begin{minipage}[t]{0.24\textwidth}
	\includegraphics[scale=0.24]{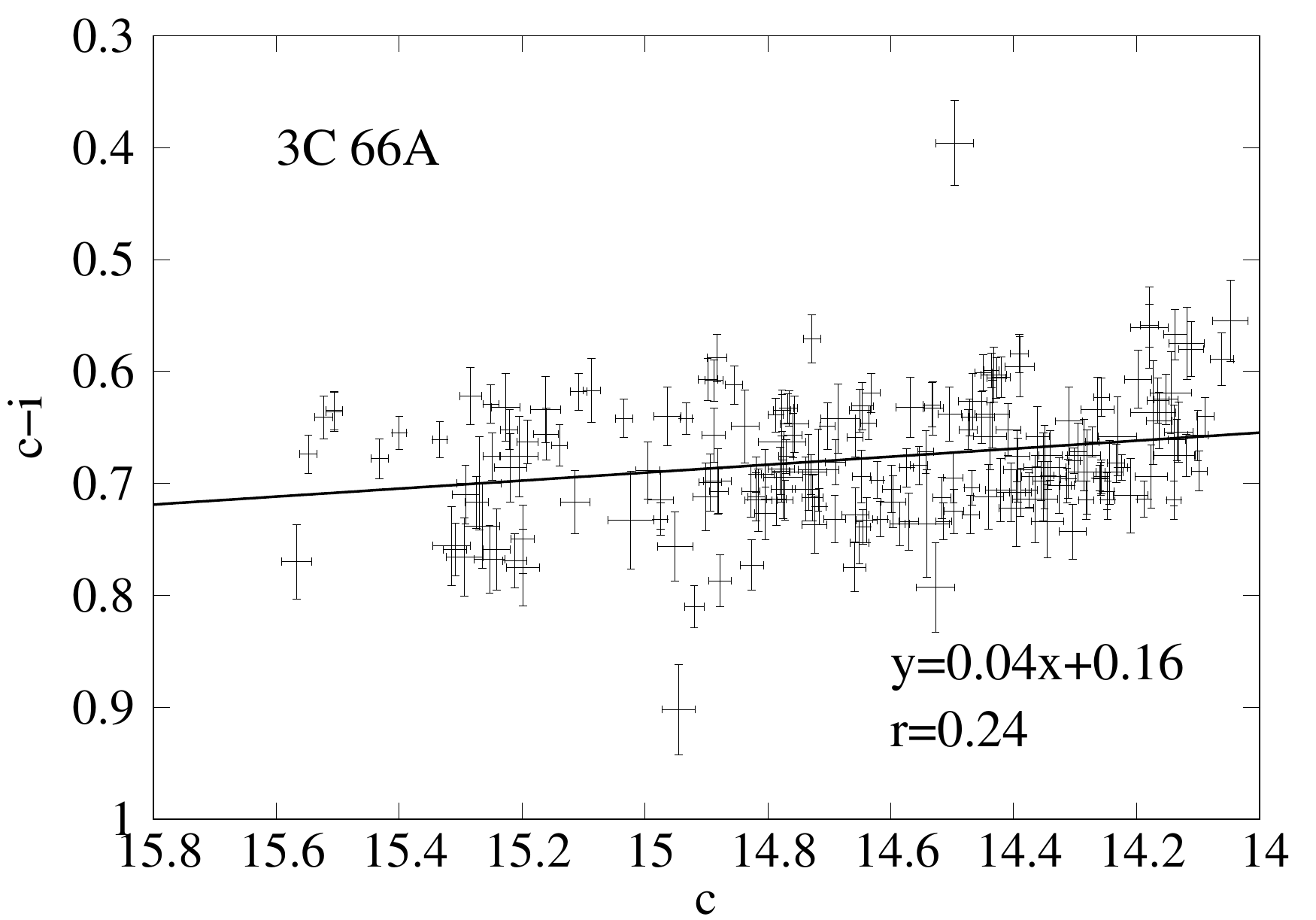}  
   \end{minipage}
\begin{minipage}[t]{0.24\textwidth}
	\includegraphics[scale=0.24]{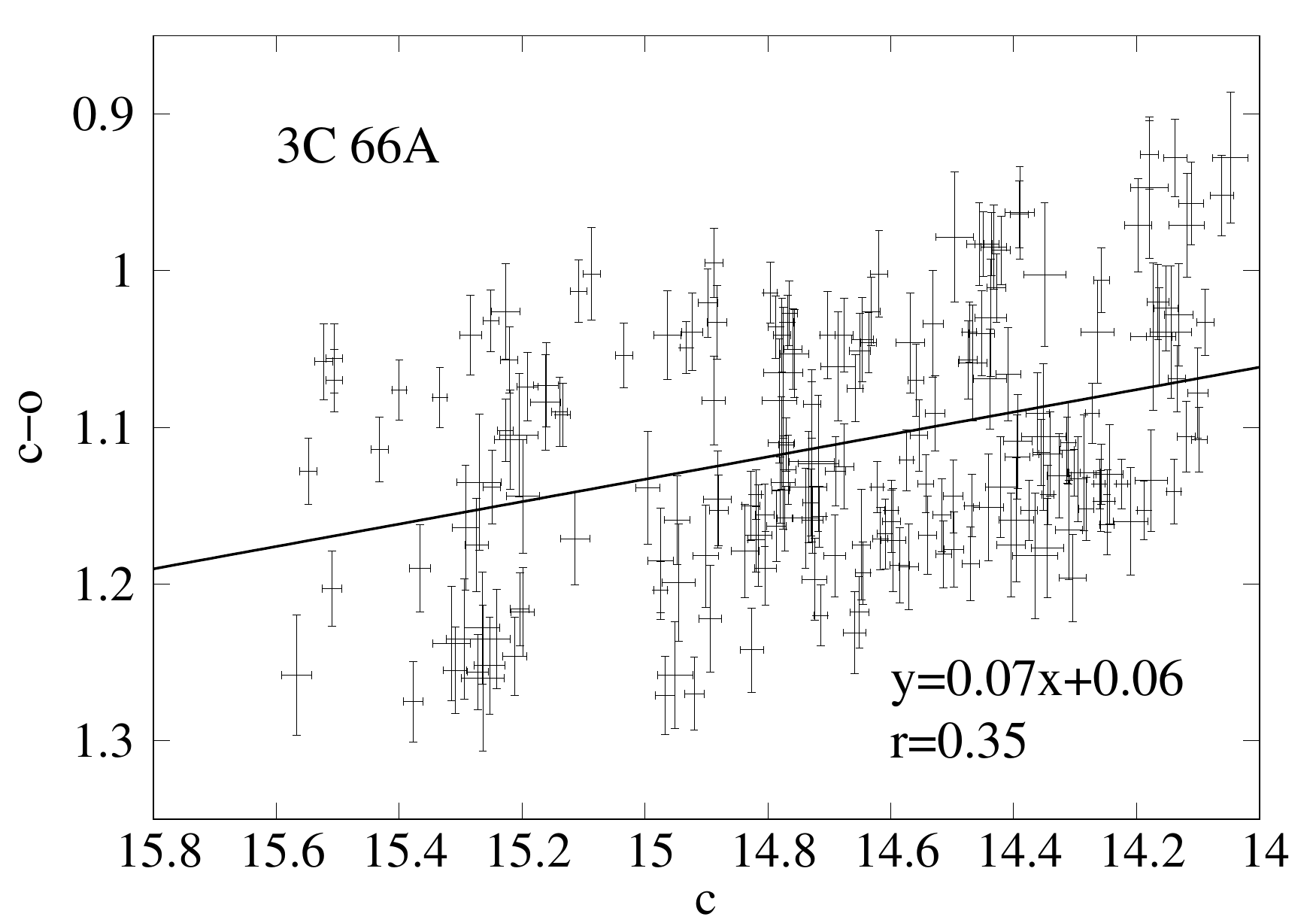}  
   \end{minipage}   
   \begin{minipage}[t]{0.24\textwidth}
	\includegraphics[scale=0.24]{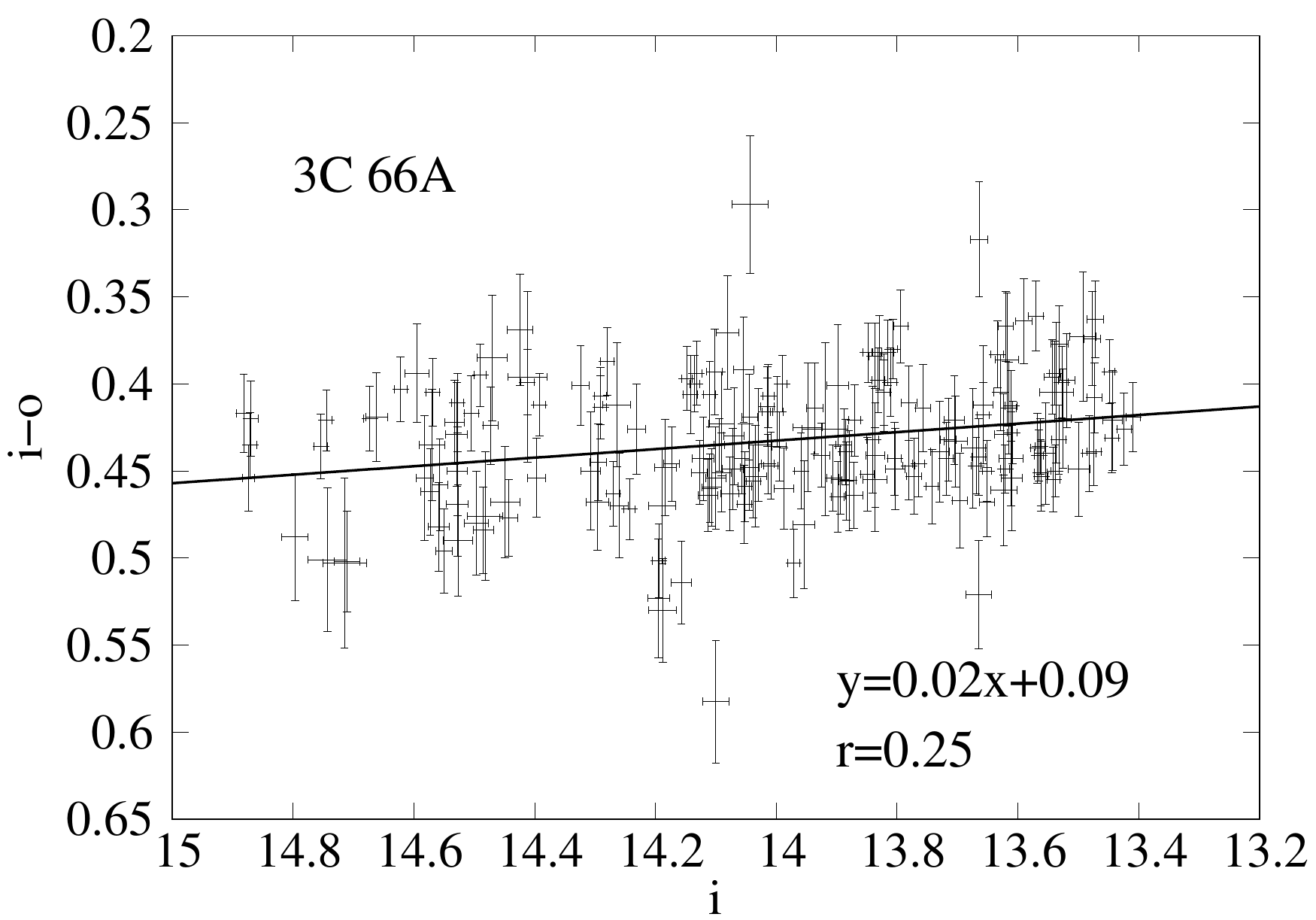}  
   \end{minipage}
\begin{minipage}[t]{0.24\textwidth}
	\includegraphics[scale=0.24]{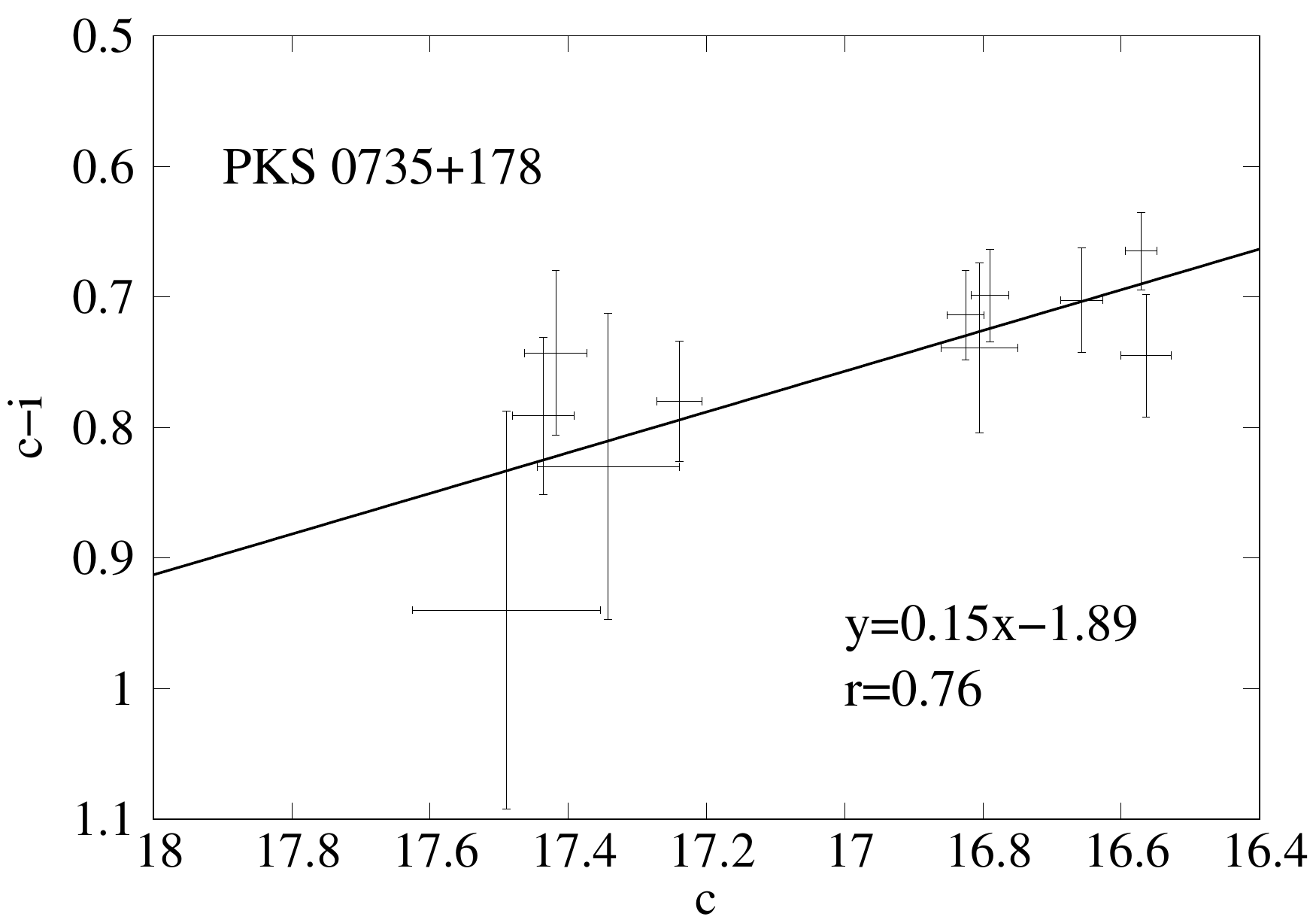}  
   \end{minipage}

   \begin{minipage}[t]{0.24\textwidth}
	\includegraphics[scale=0.24]{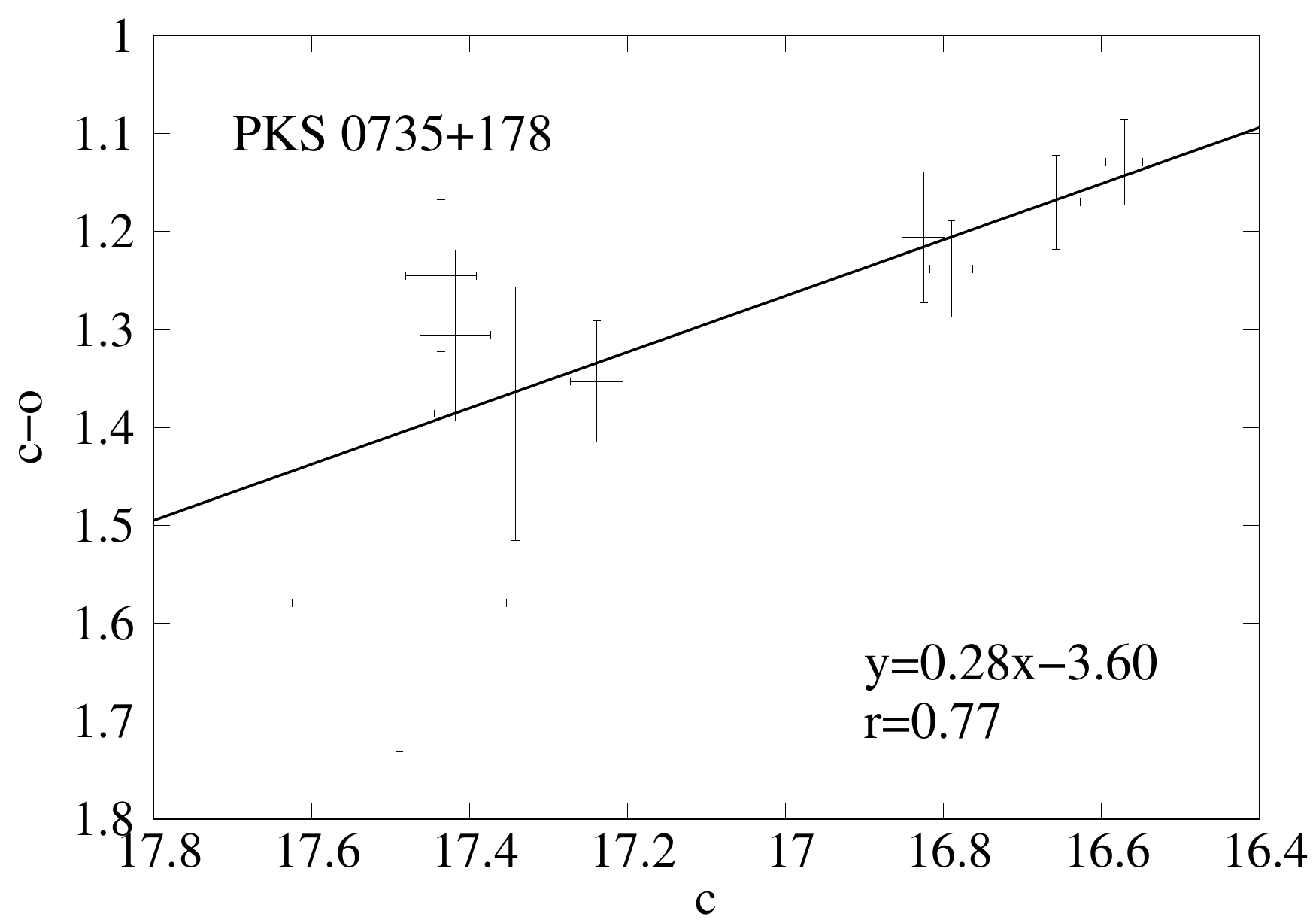}  
   \end{minipage}
\begin{minipage}[t]{0.24\textwidth}
	\includegraphics[scale=0.24]{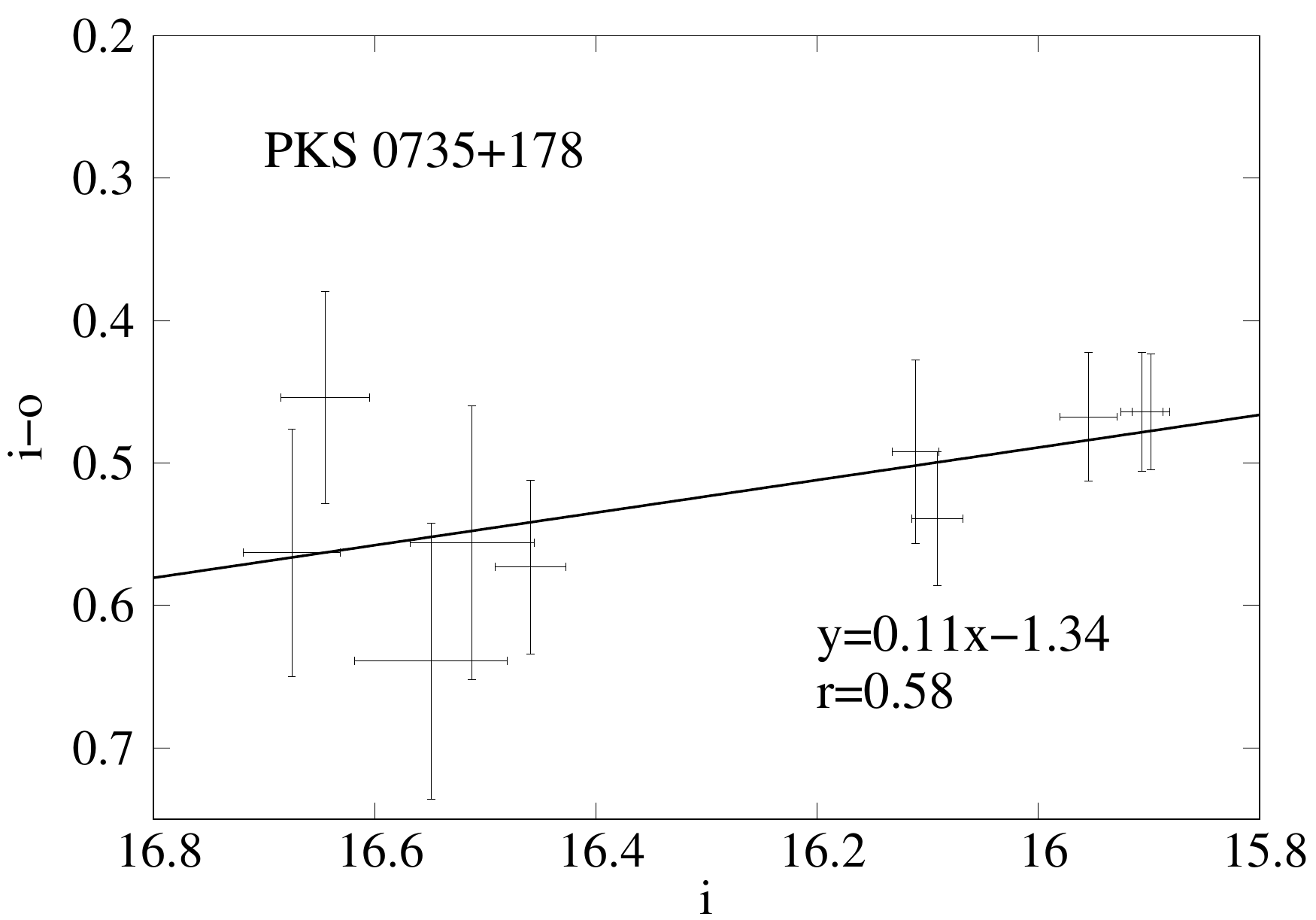}  
   \end{minipage}   
   \begin{minipage}[t]{0.24\textwidth}
	\includegraphics[scale=0.24]{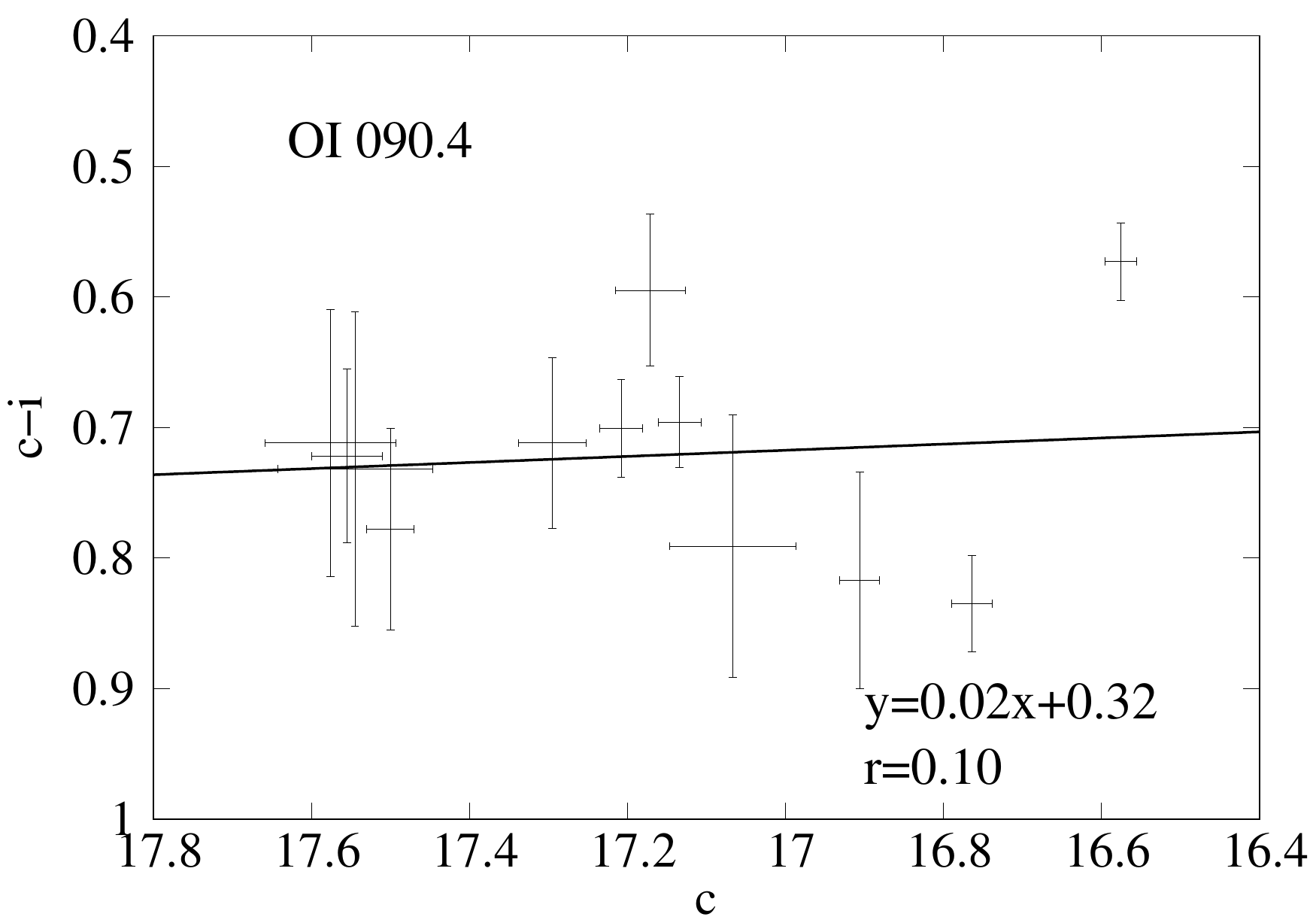}  
   \end{minipage}
\begin{minipage}[t]{0.24\textwidth}
	\includegraphics[scale=0.24]{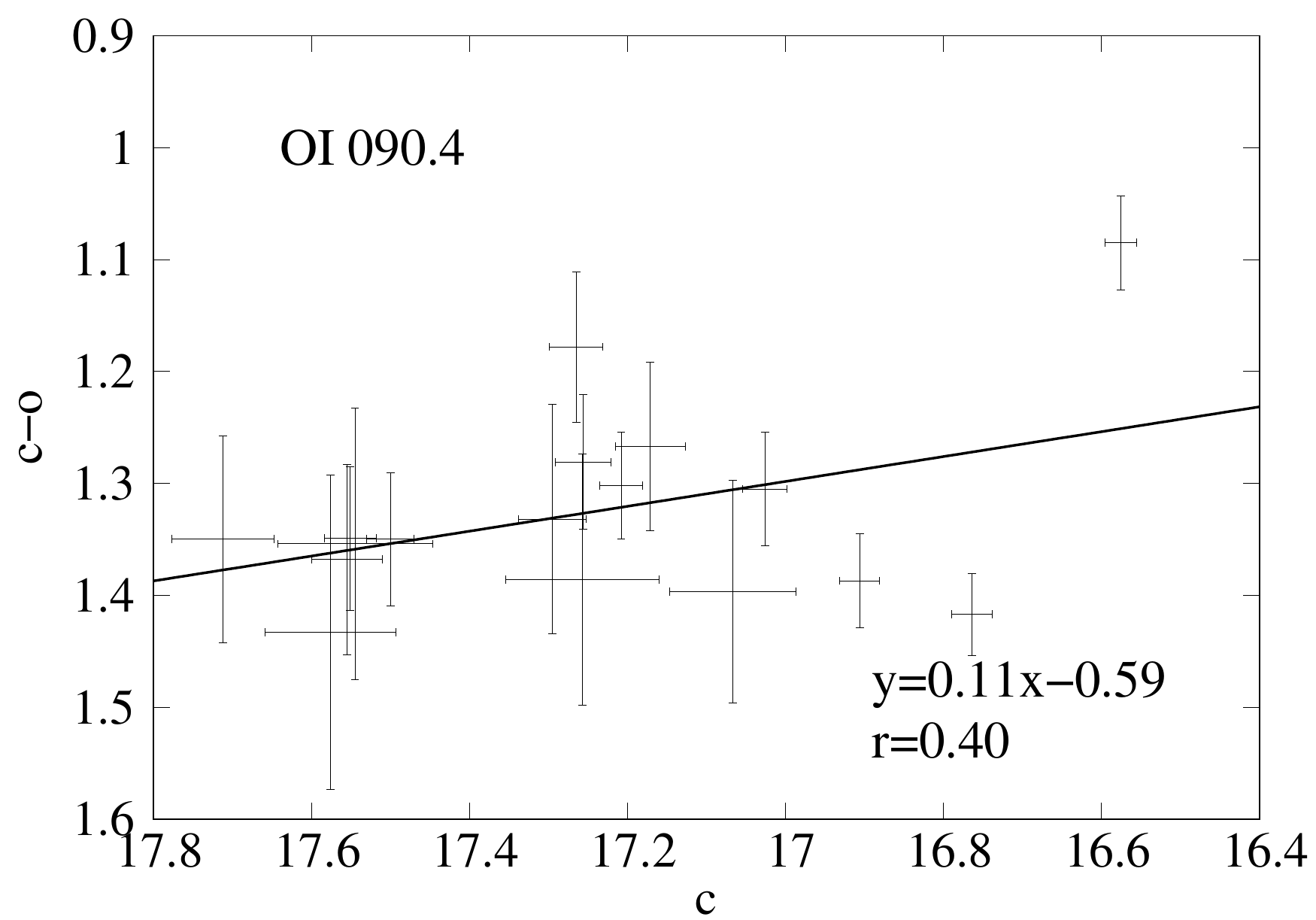}  
   \end{minipage}

      \begin{minipage}[t]{0.24\textwidth}
	\includegraphics[scale=0.24]{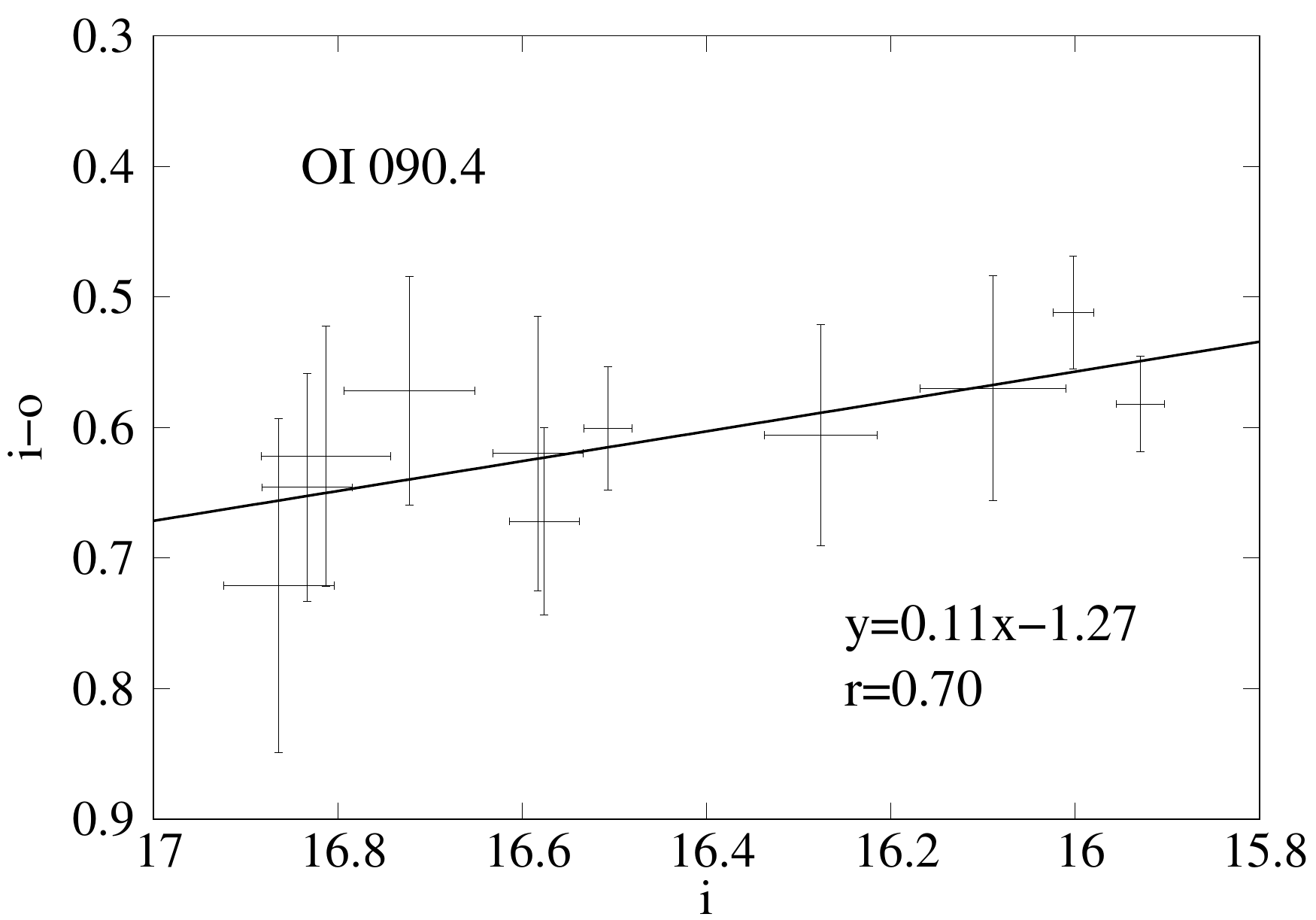}  
   \end{minipage}
\begin{minipage}[t]{0.24\textwidth}
	\includegraphics[scale=0.24]{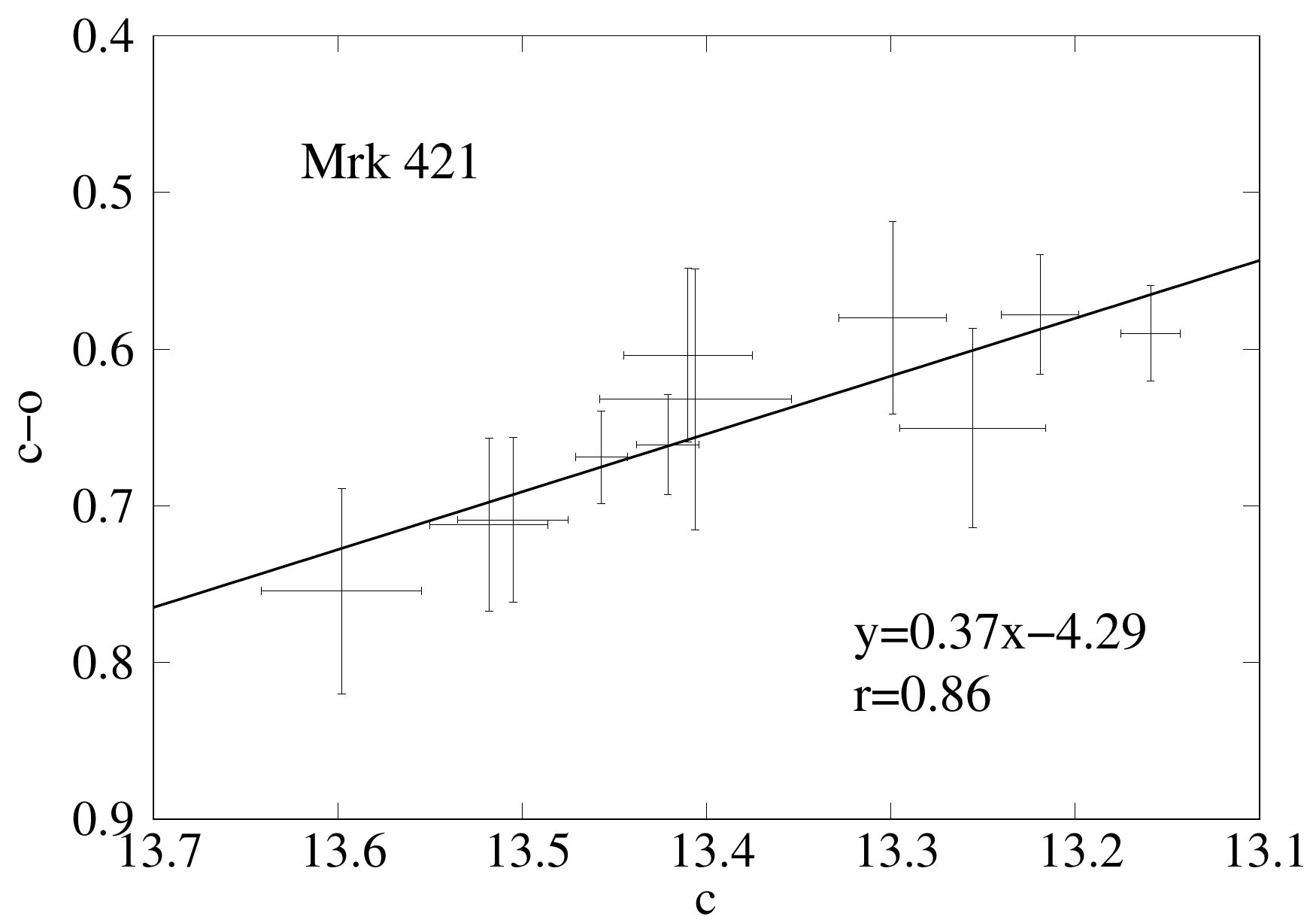}  
   \end{minipage}   
   \begin{minipage}[t]{0.24\textwidth}
	\includegraphics[scale=0.24]{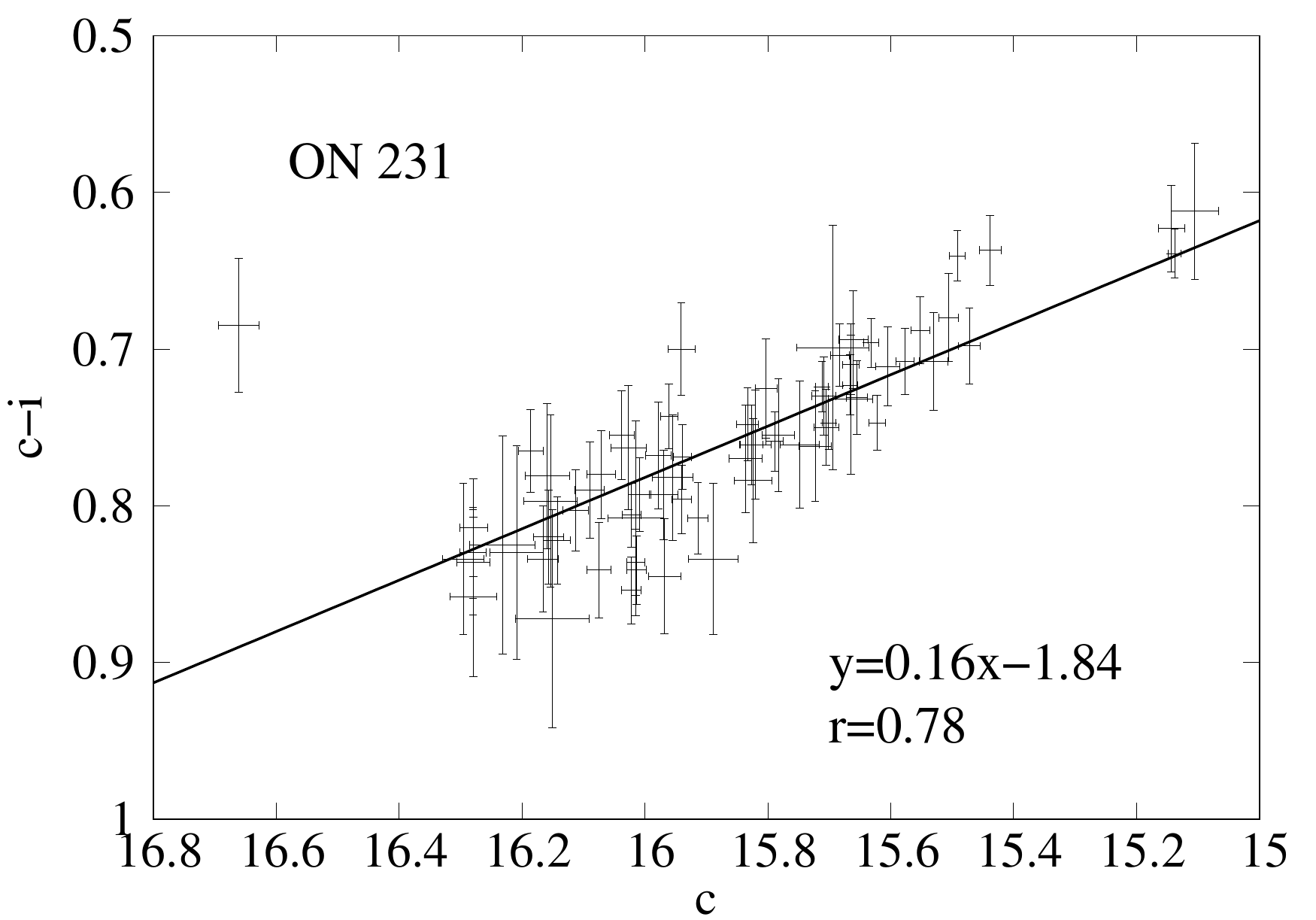}  
   \end{minipage}
\begin{minipage}[t]{0.24\textwidth}
	\includegraphics[scale=0.24]{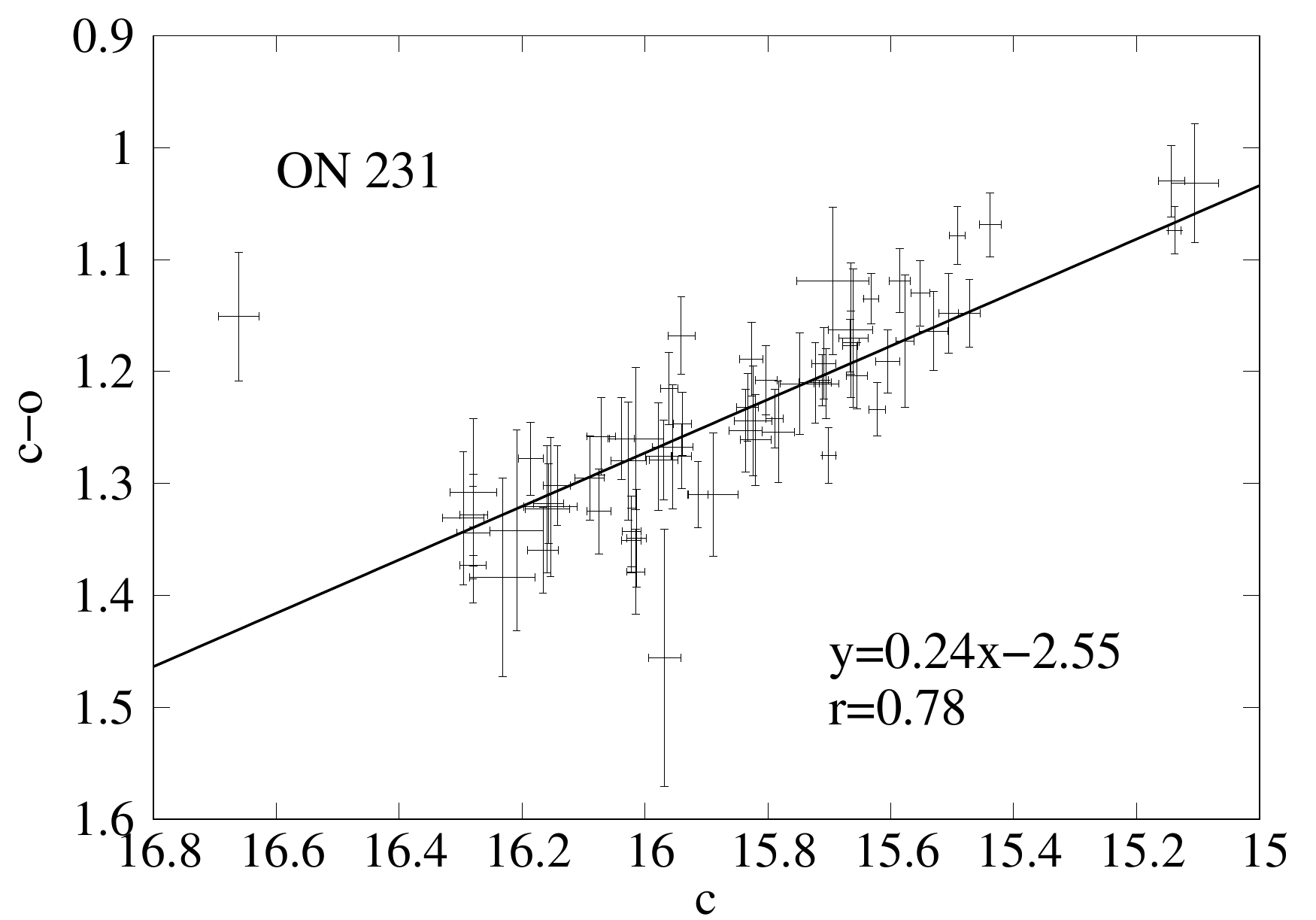}  
   \end{minipage}

      \begin{minipage}[t]{0.24\textwidth}
	\includegraphics[scale=0.24]{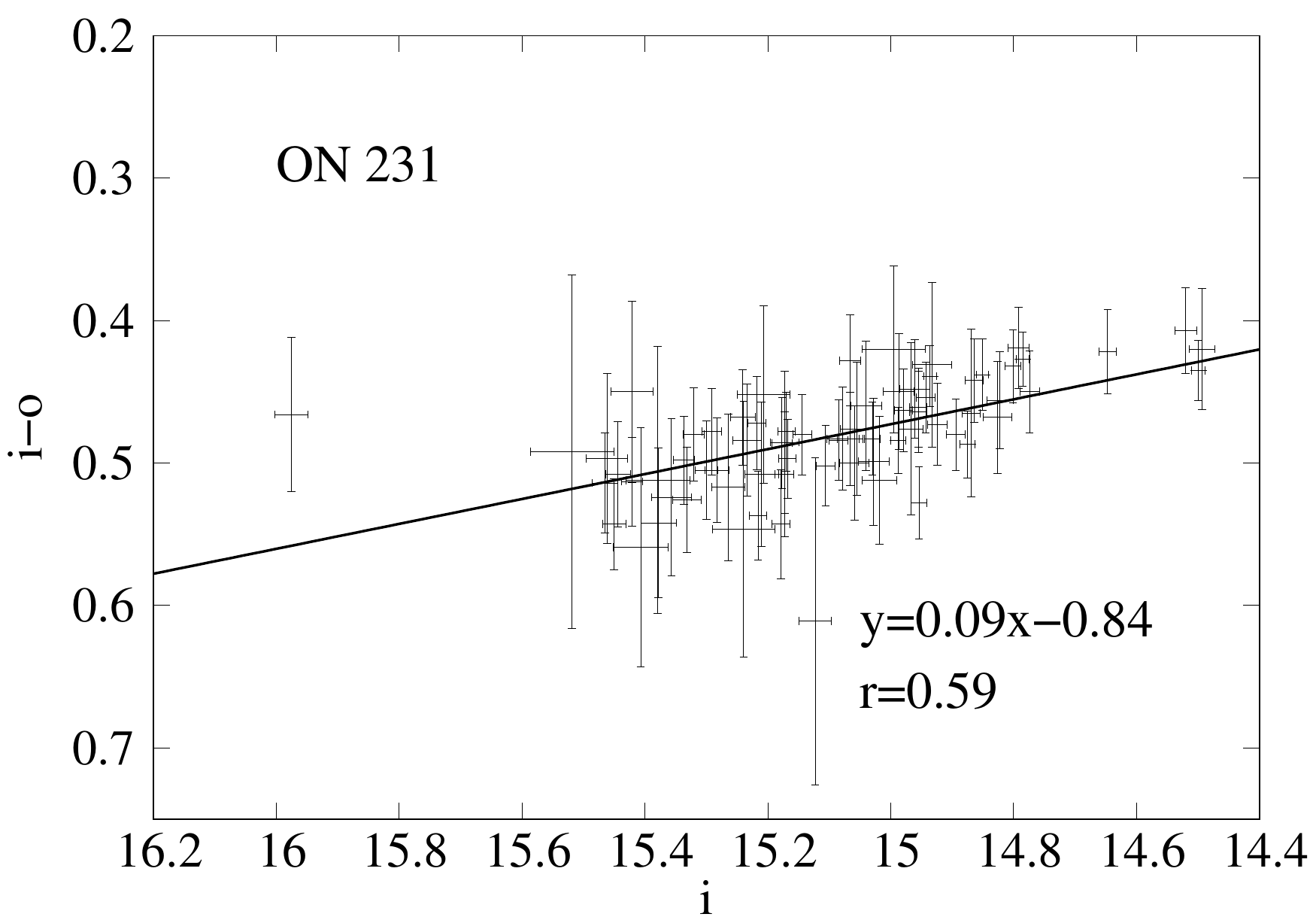}  
   \end{minipage}
          \begin{minipage}[t]{0.24\textwidth}
	\includegraphics[scale=0.24]{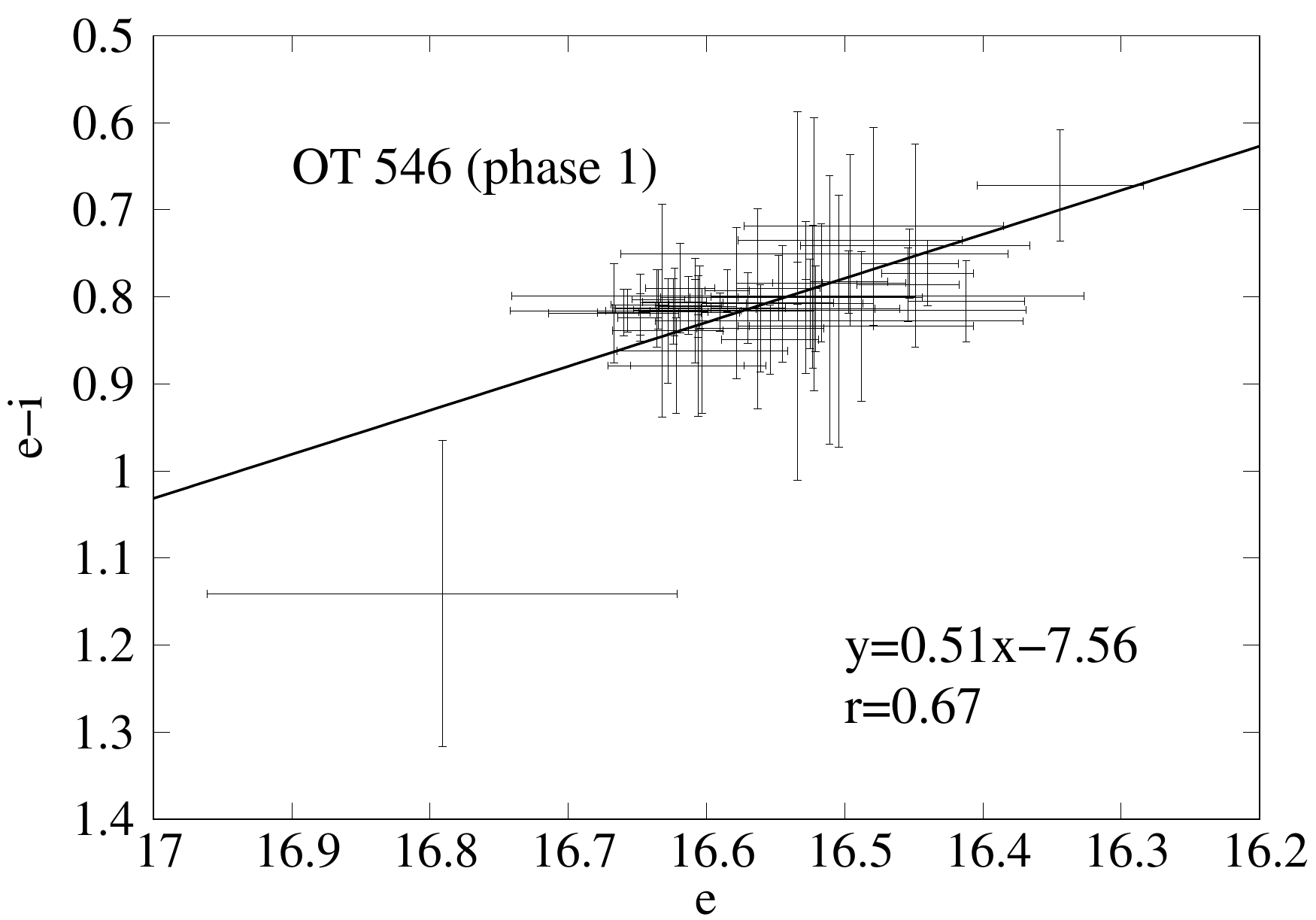}  
   \end{minipage}
\begin{minipage}[t]{0.24\textwidth}
	\includegraphics[scale=0.24]{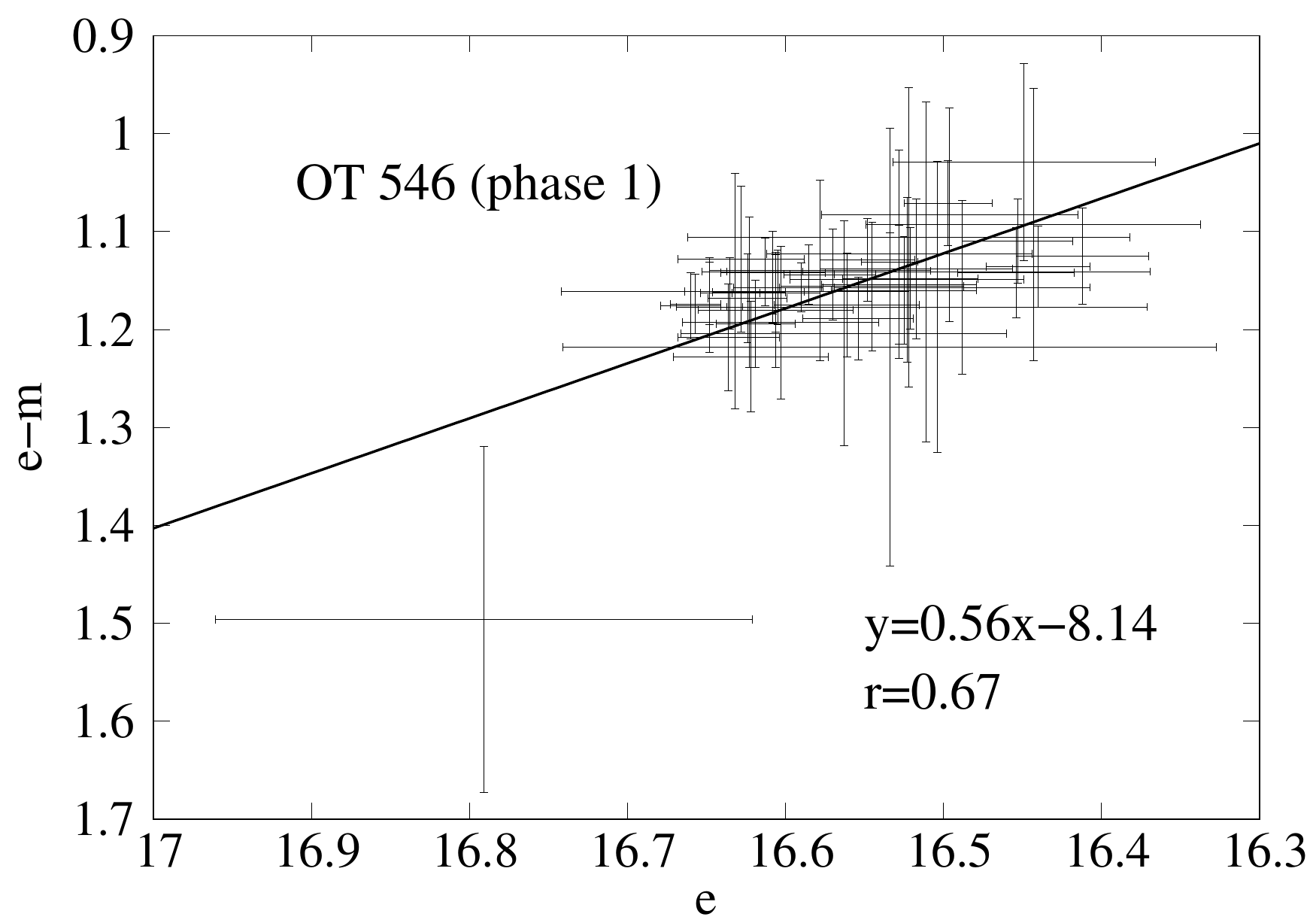}  
   \end{minipage}   
   \begin{minipage}[t]{0.24\textwidth}
	\includegraphics[scale=0.24]{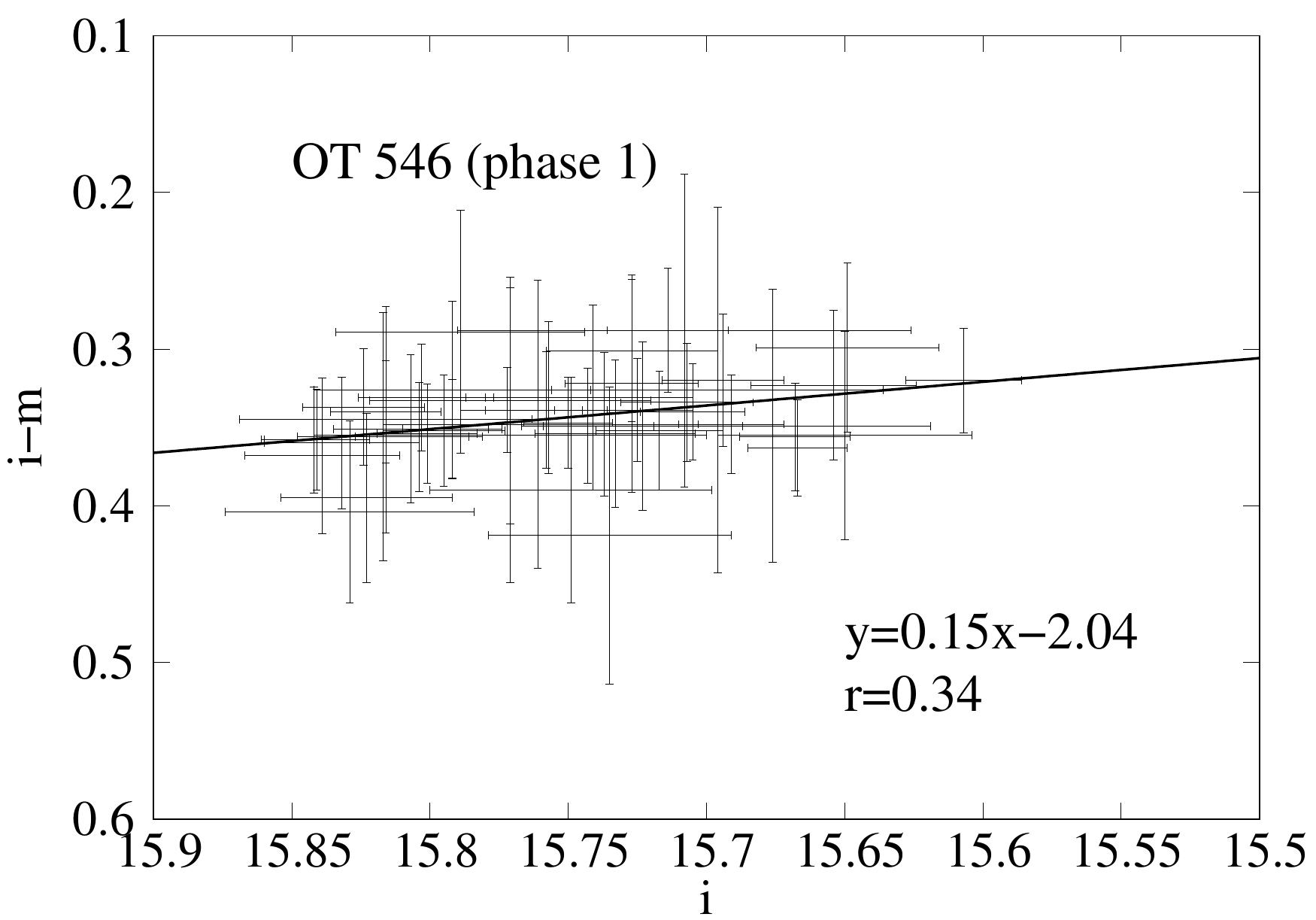}  
   \end{minipage}

\begin{minipage}[t]{0.24\textwidth}
	\includegraphics[scale=0.24]{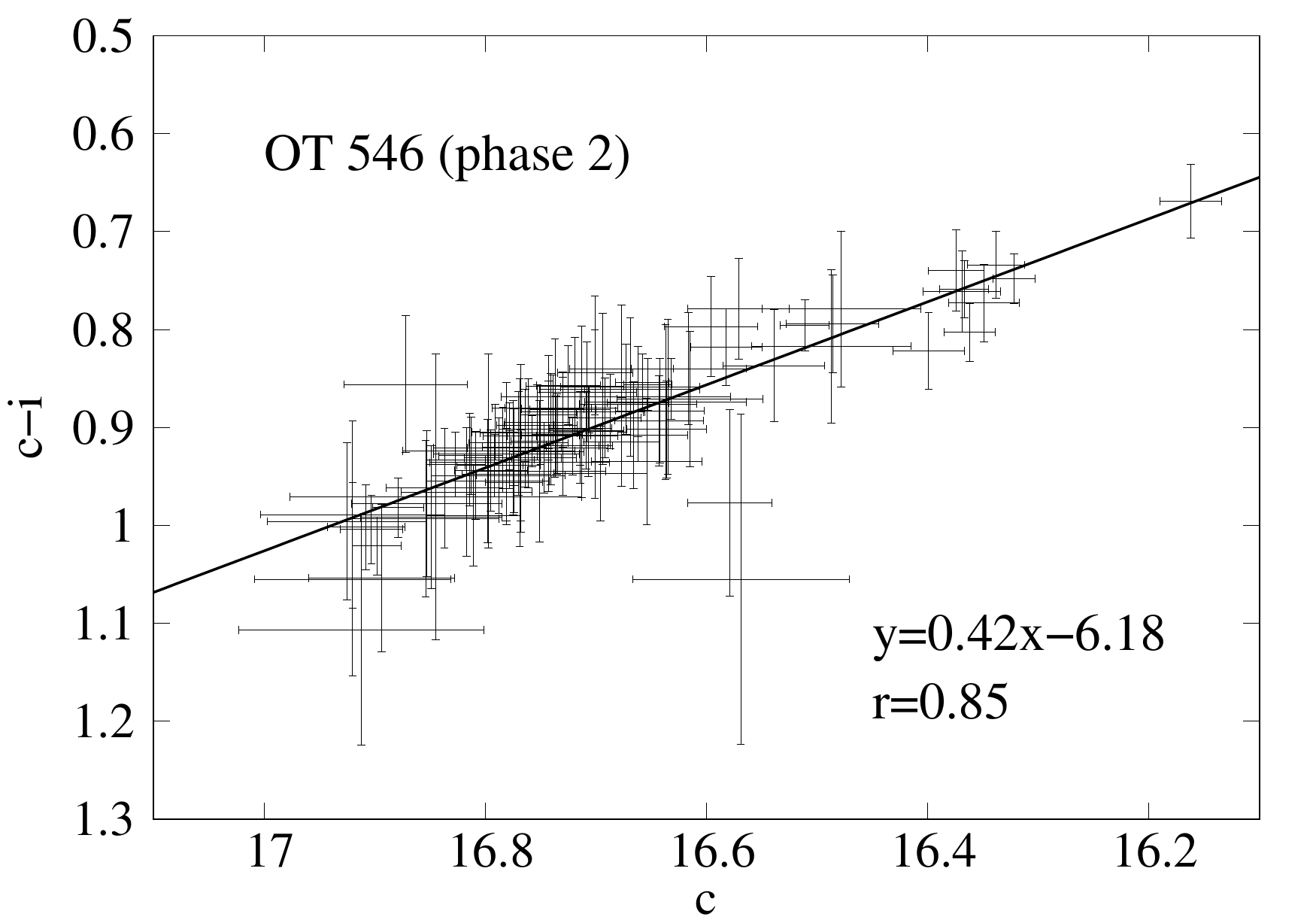}  
   \end{minipage}   
   \begin{minipage}[t]{0.24\textwidth}
	\includegraphics[scale=0.24]{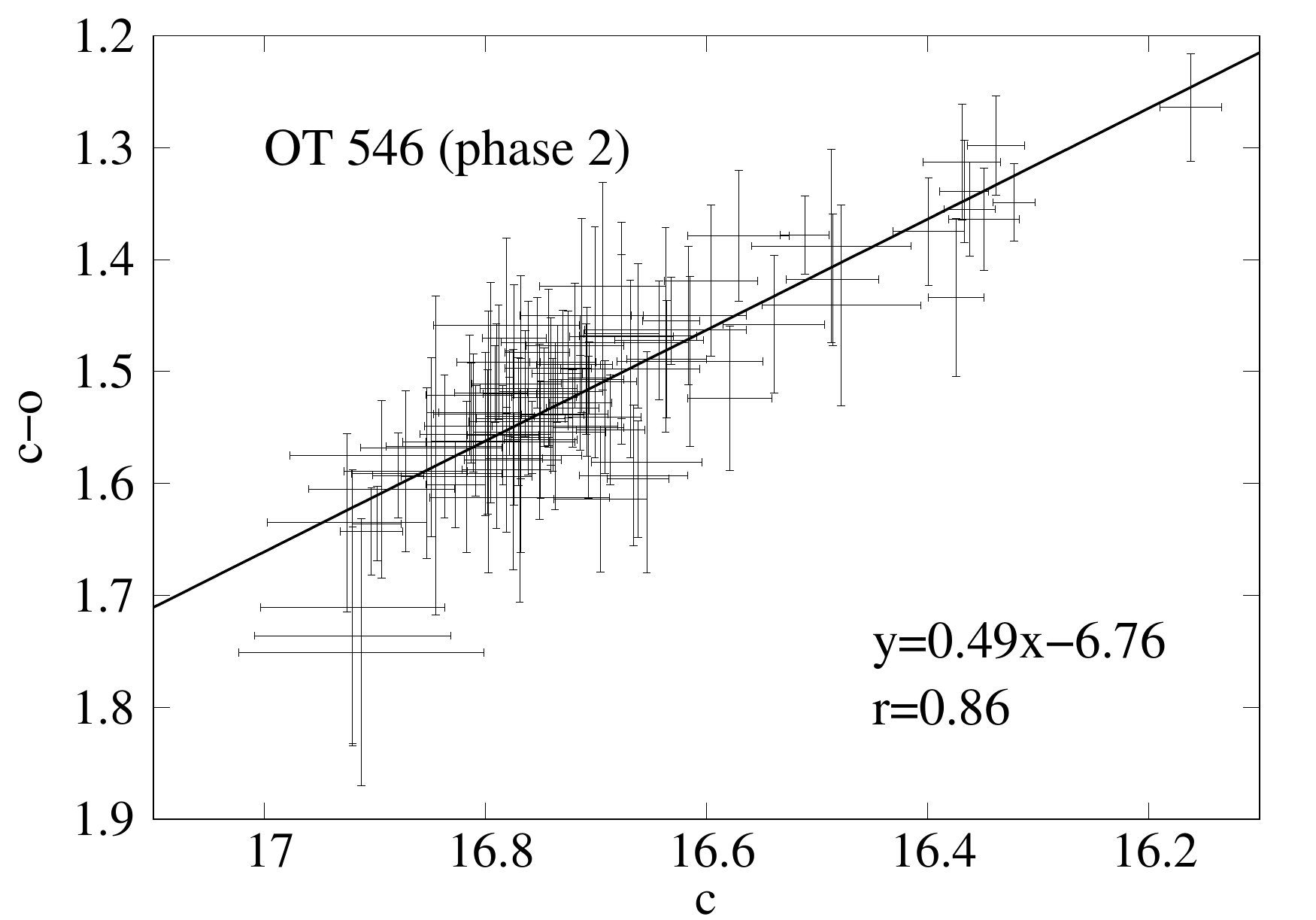}  
   \end{minipage}
\begin{minipage}[t]{0.24\textwidth}
	\includegraphics[scale=0.24]{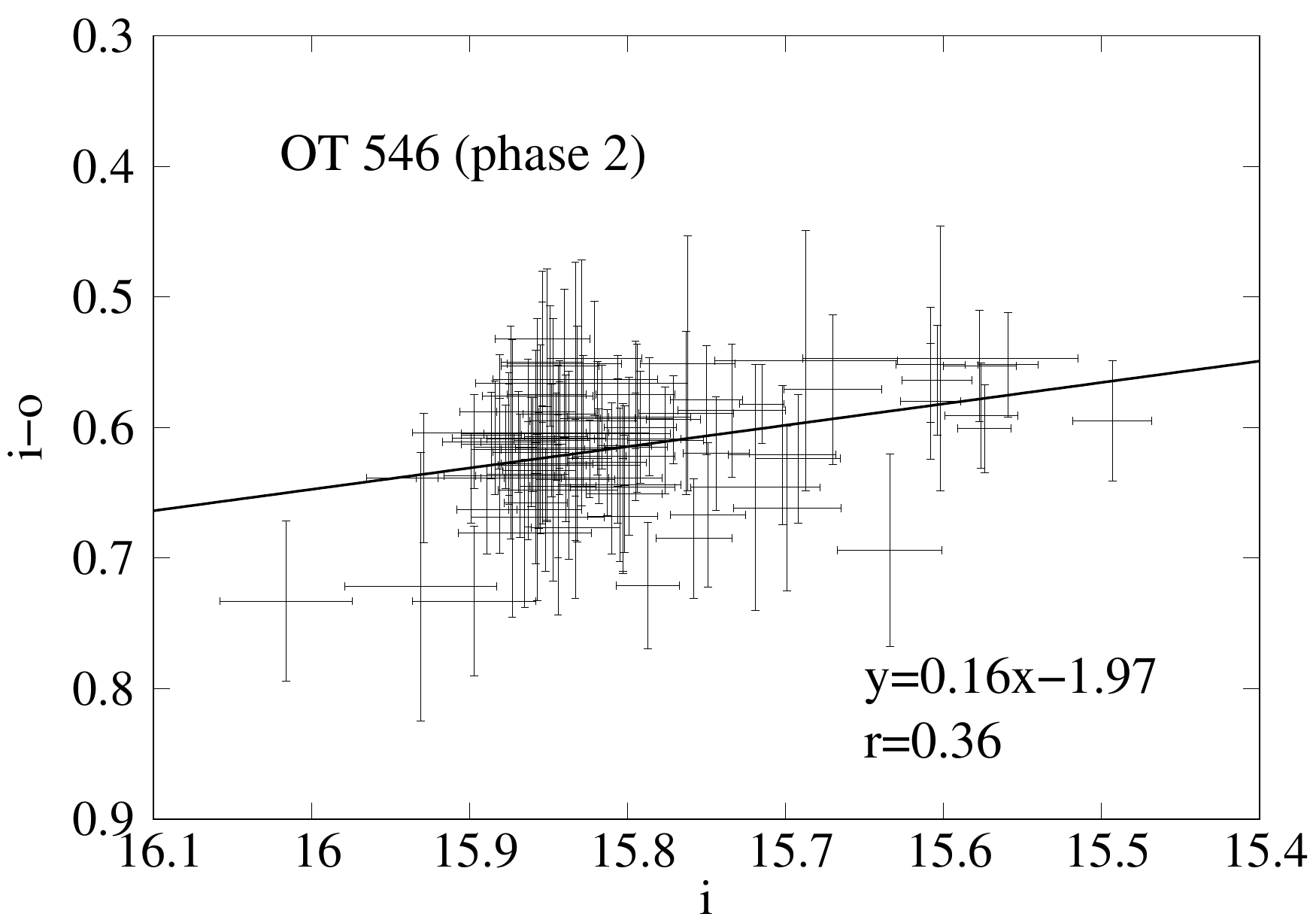}  
   \end{minipage}  
  \begin{minipage}[t]{0.24\textwidth}
	\includegraphics[scale=0.24]{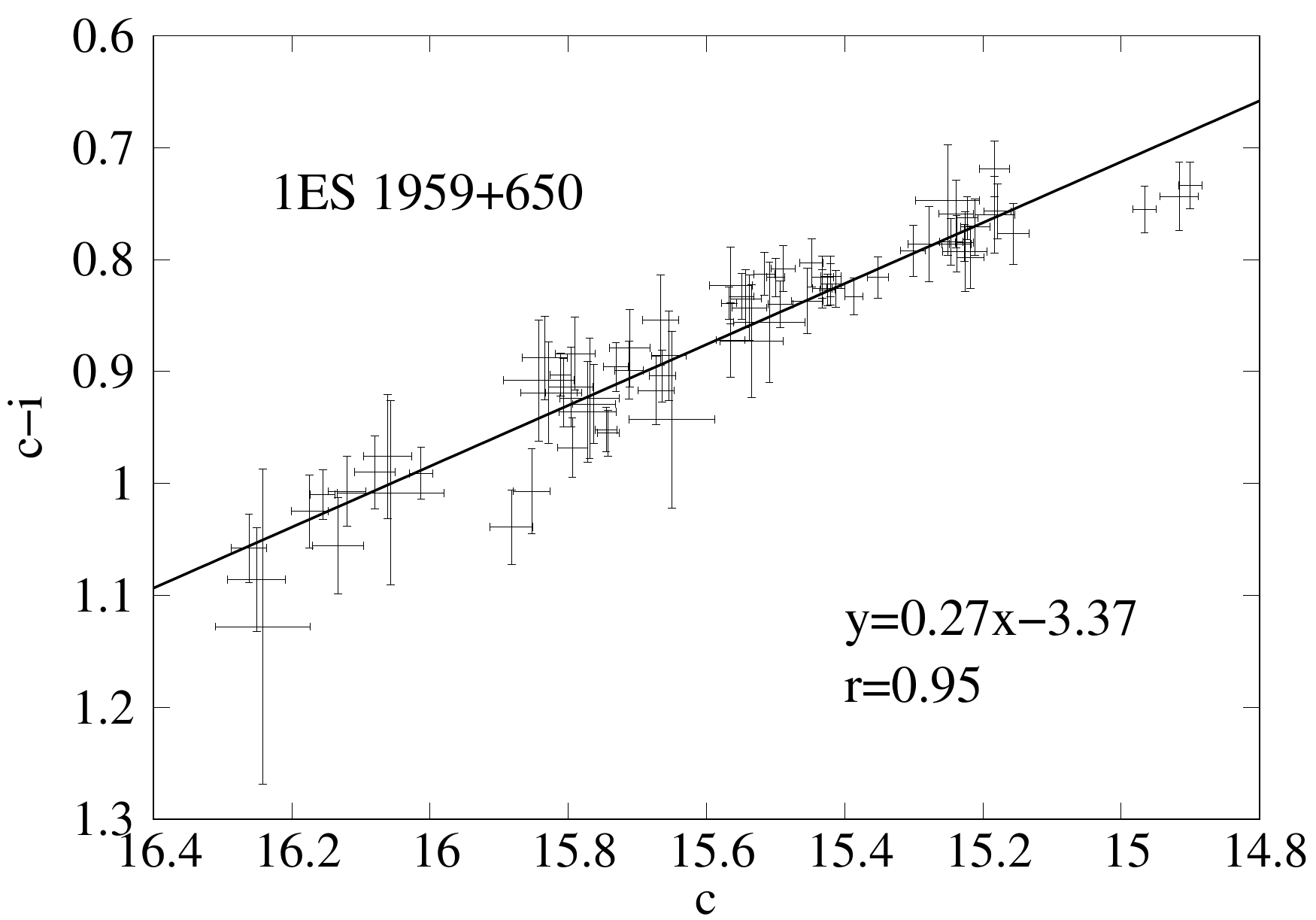}  
   \end{minipage}  
   
   \begin{minipage}[t]{0.24\textwidth}
	\includegraphics[scale=0.24]{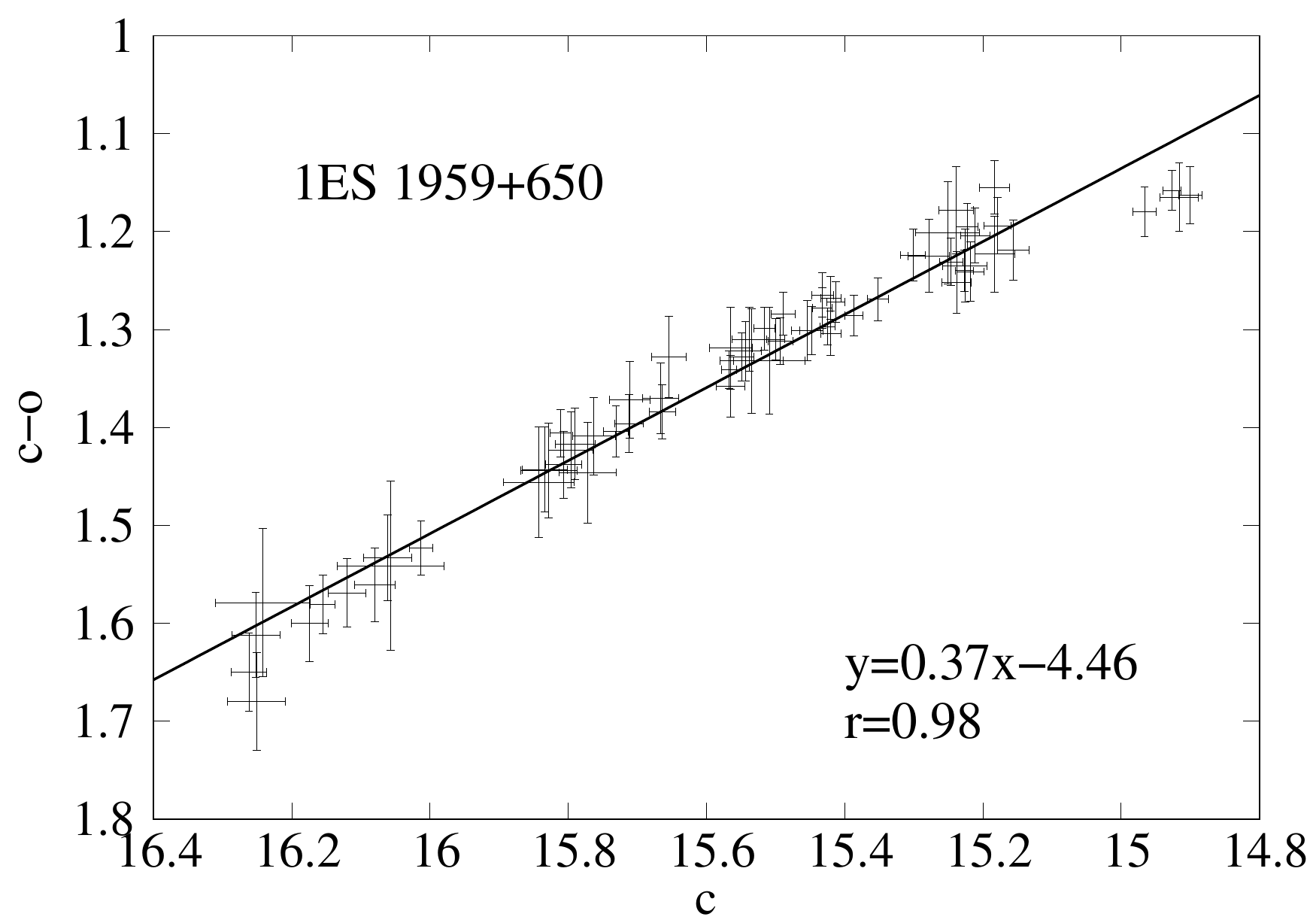}  
   \end{minipage}  
   \begin{minipage}[t]{0.24\textwidth}
	\includegraphics[scale=0.24]{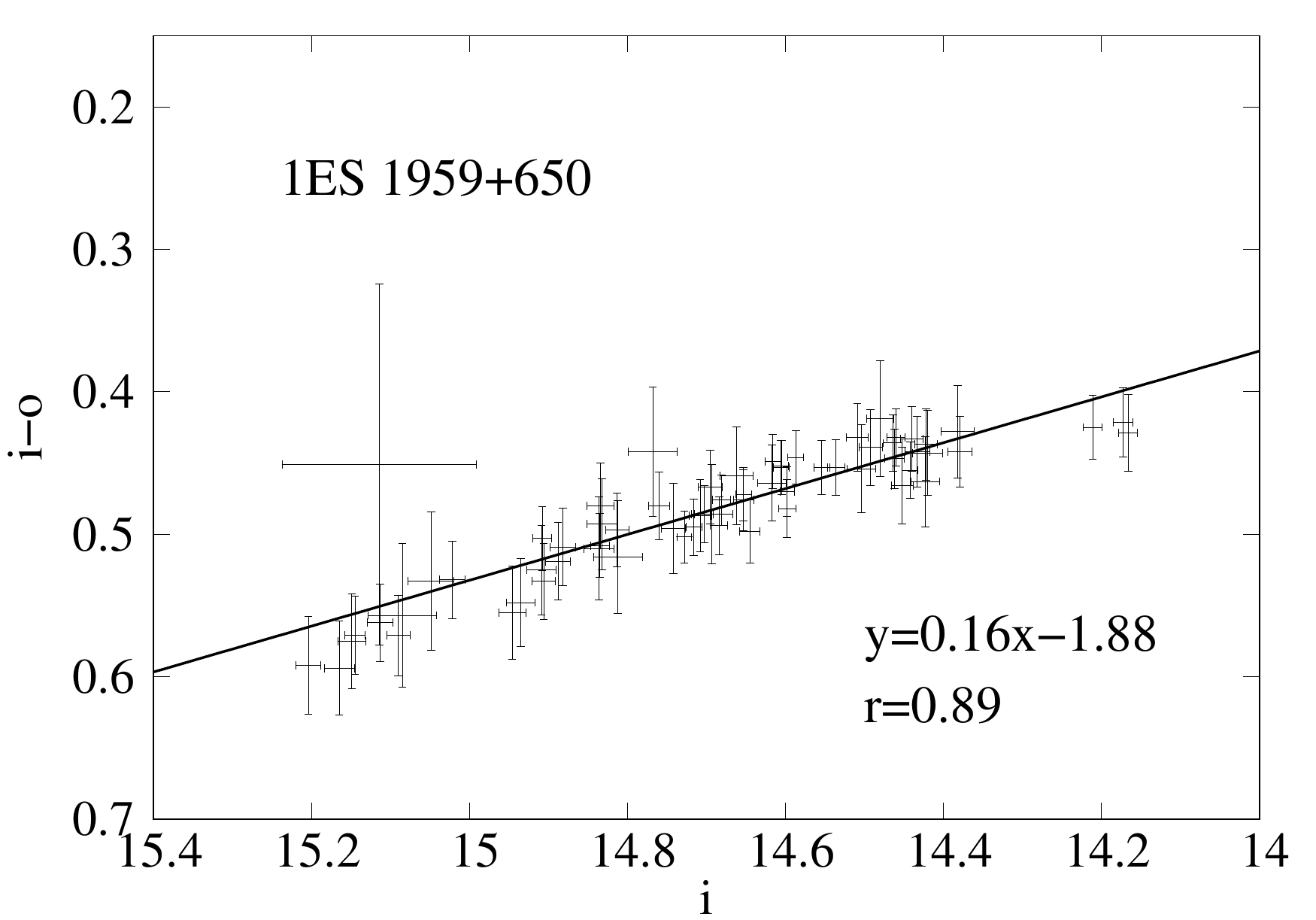}  
   \end{minipage}  
      \begin{minipage}[t]{0.24\textwidth}
	\includegraphics[scale=0.24]{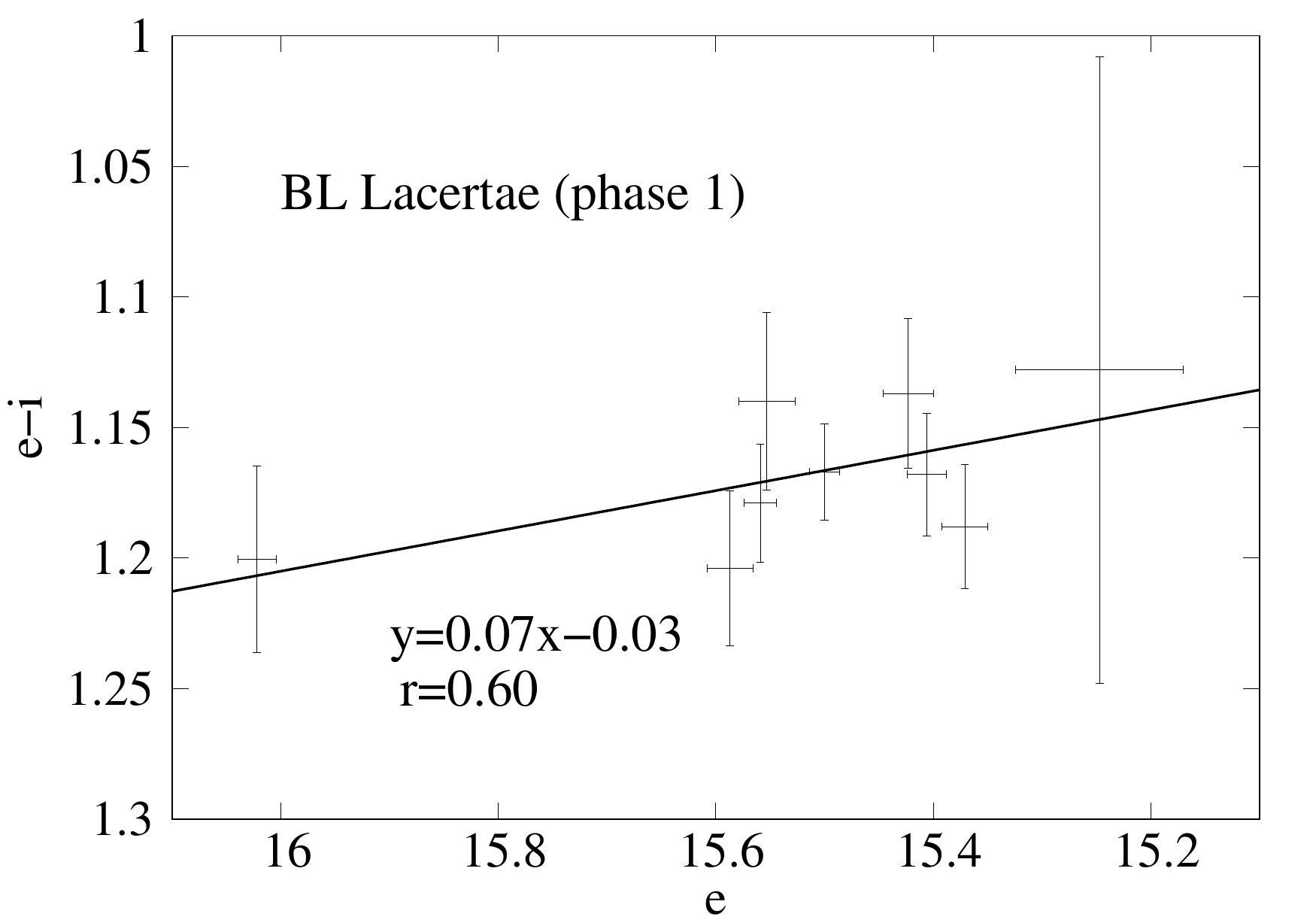}  
   \end{minipage}  
      \begin{minipage}[t]{0.24\textwidth}
	\includegraphics[scale=0.24]{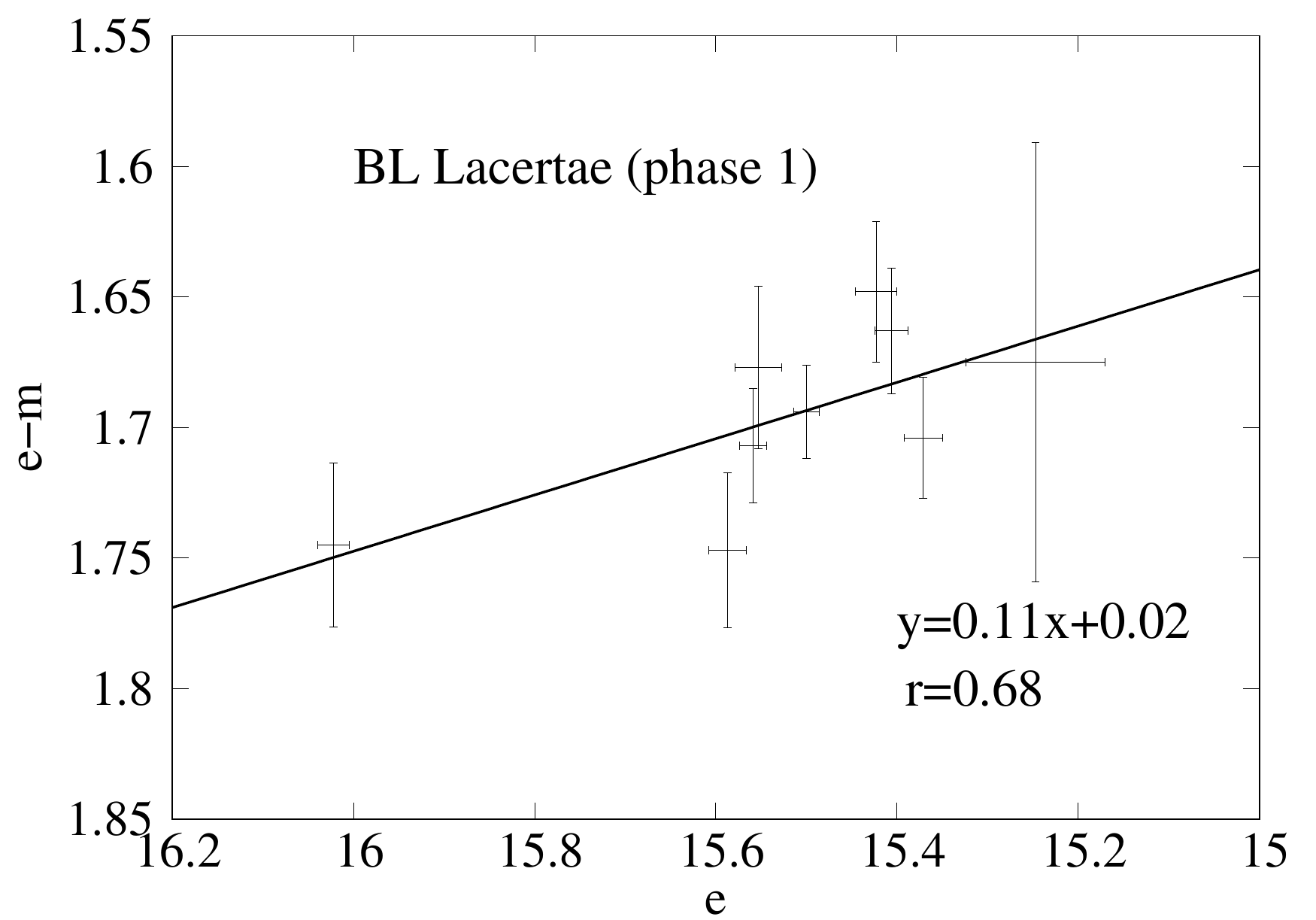}  
   \end{minipage}

      \begin{minipage}[t]{0.24\textwidth}
	\includegraphics[scale=0.24]{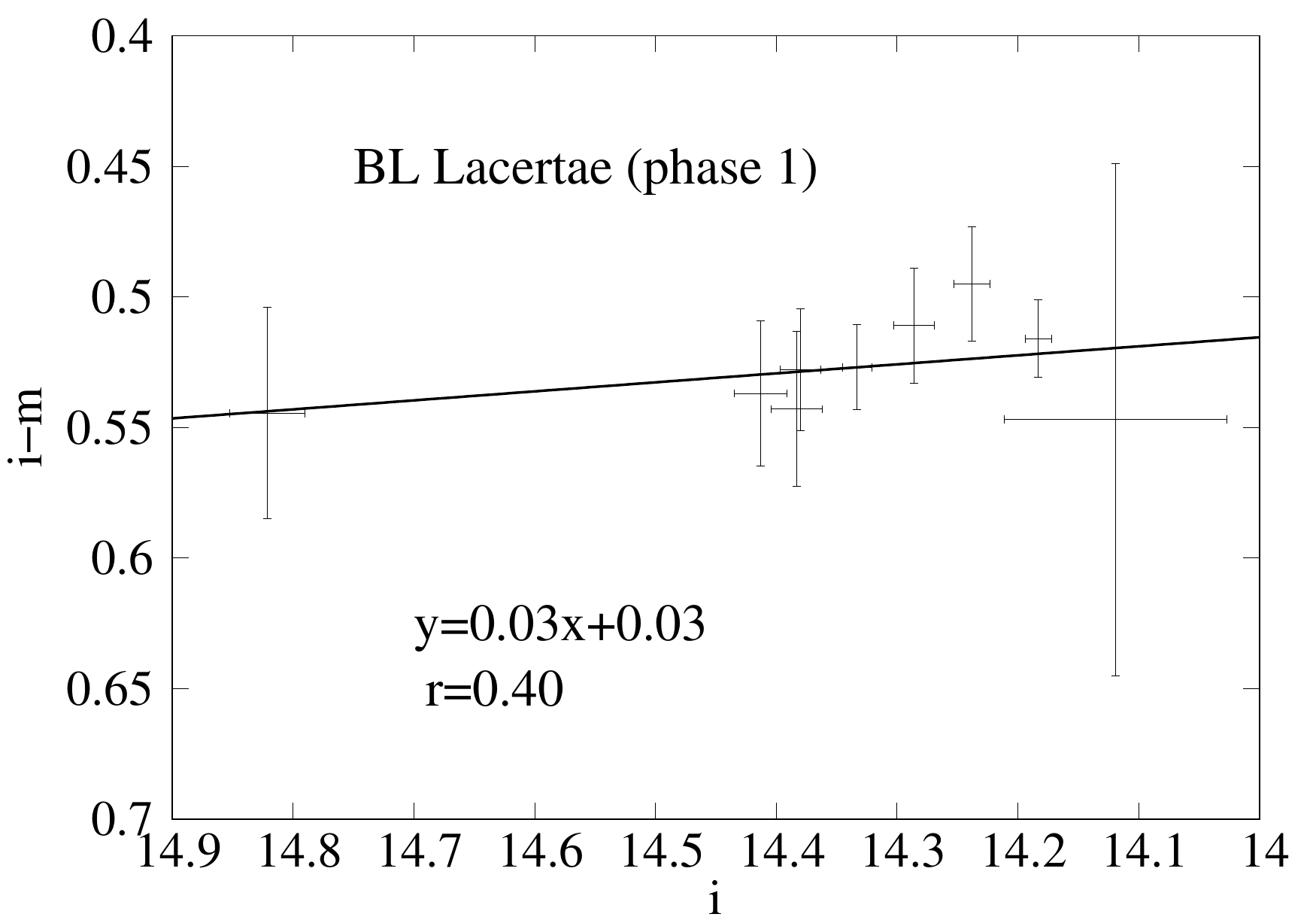}  
   \end{minipage}     
   \begin{minipage}[t]{0.24\textwidth}
	\includegraphics[scale=0.24]{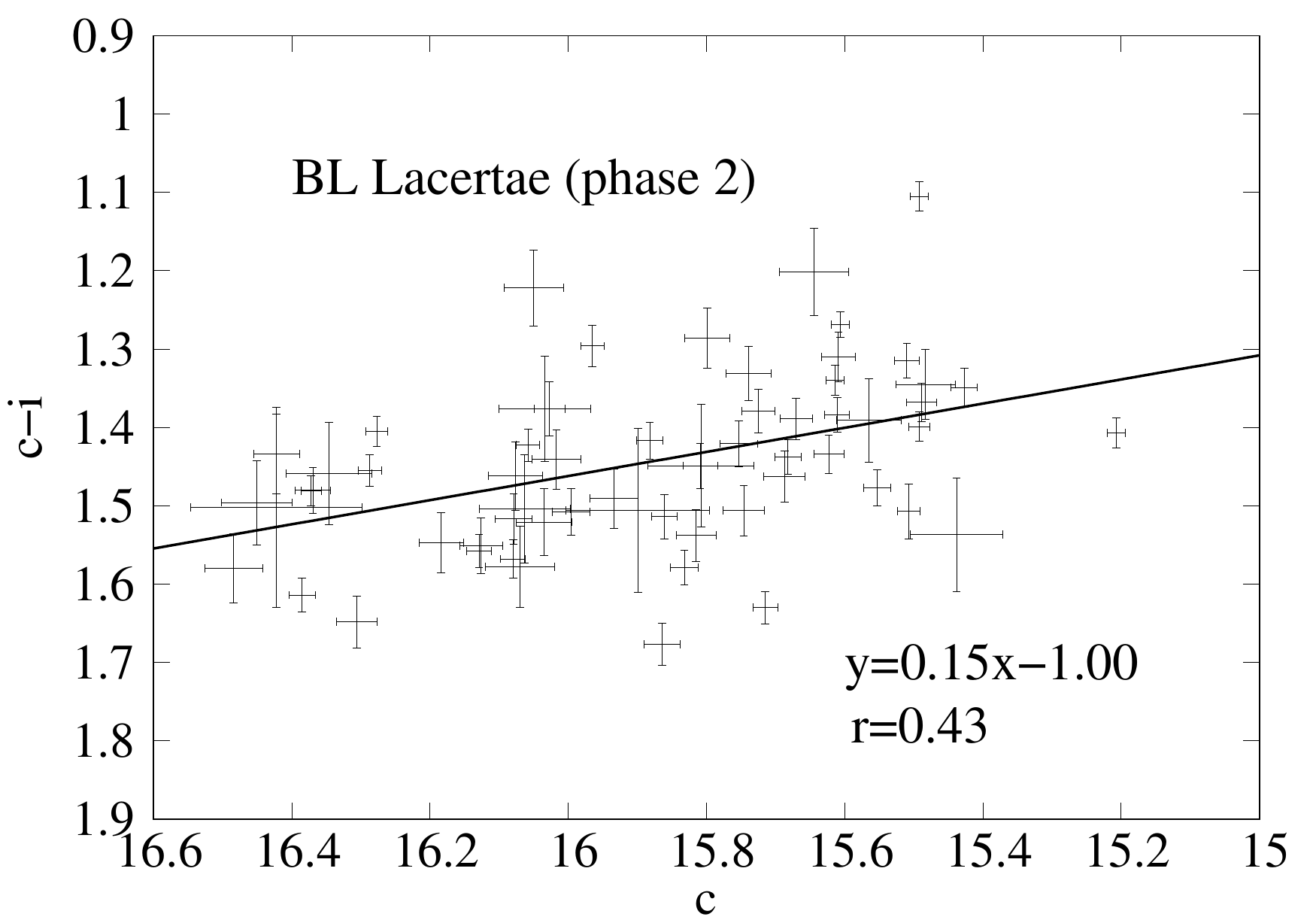}  
   \end{minipage}  
   \begin{minipage}[t]{0.24\textwidth}
	\includegraphics[scale=0.24]{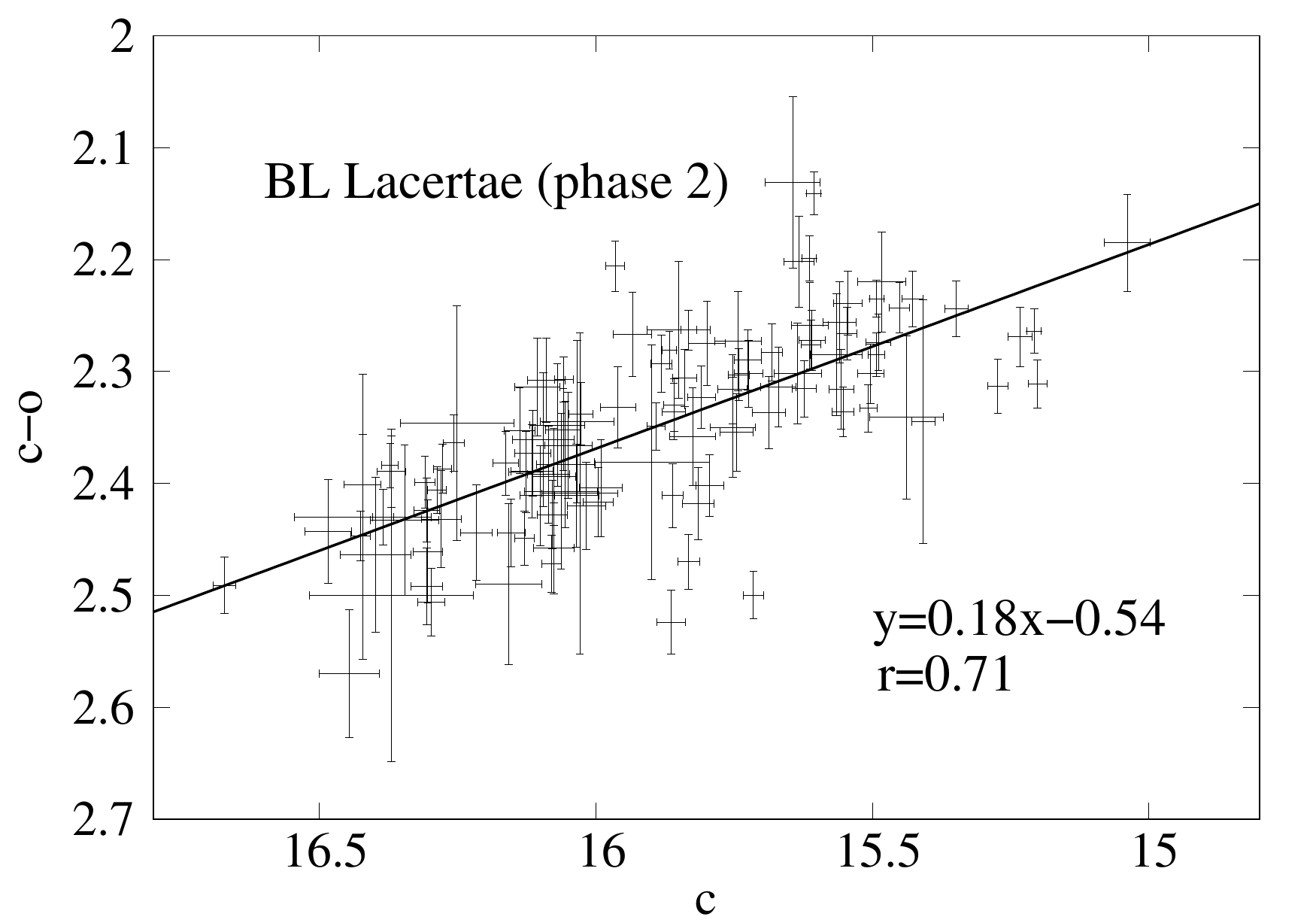}  
   \end{minipage}   
      \begin{minipage}[t]{0.24\textwidth}
	\includegraphics[scale=0.24]{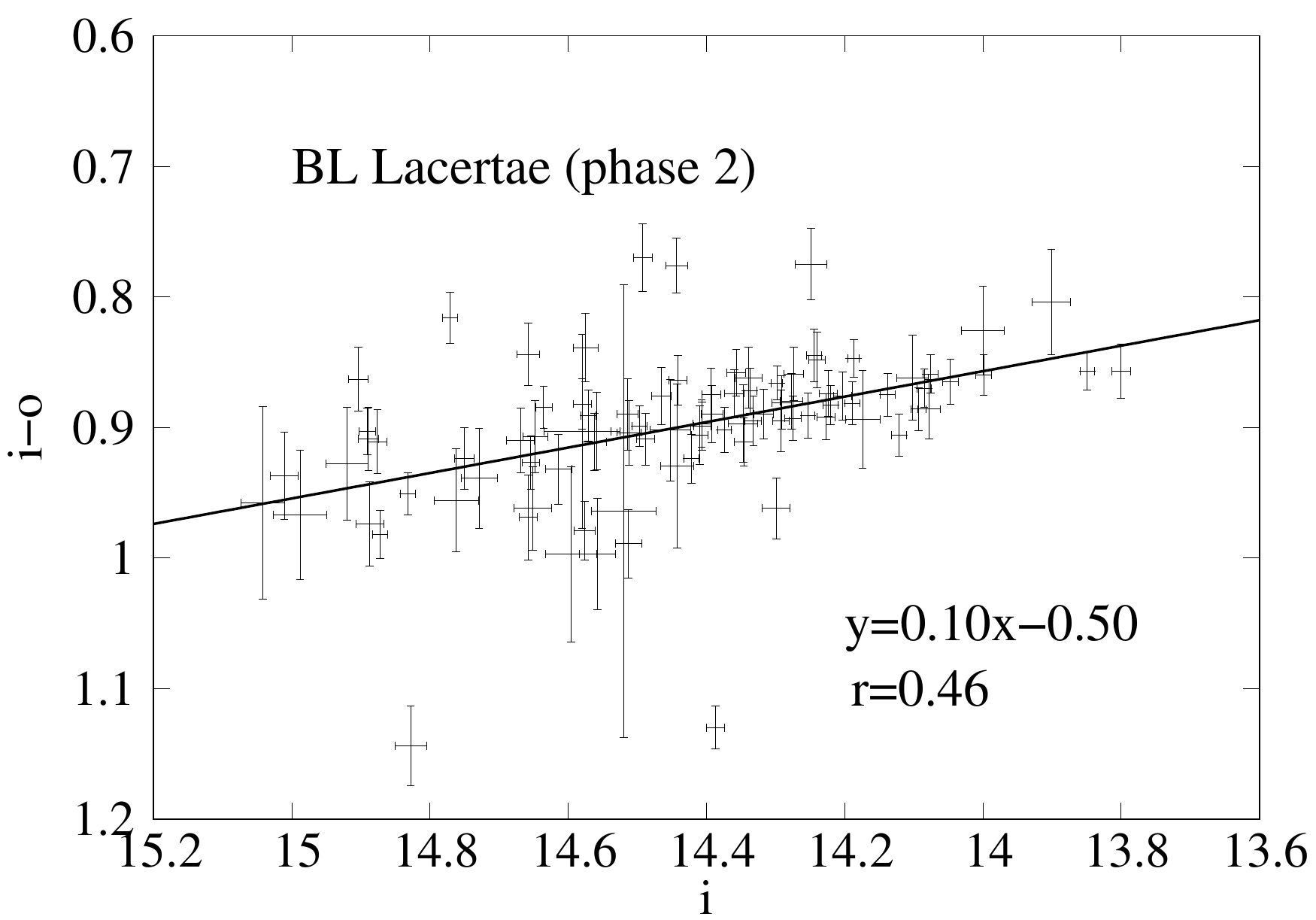}  
   \end{minipage}

      \begin{minipage}[t]{0.24\textwidth}
	\includegraphics[scale=0.24]{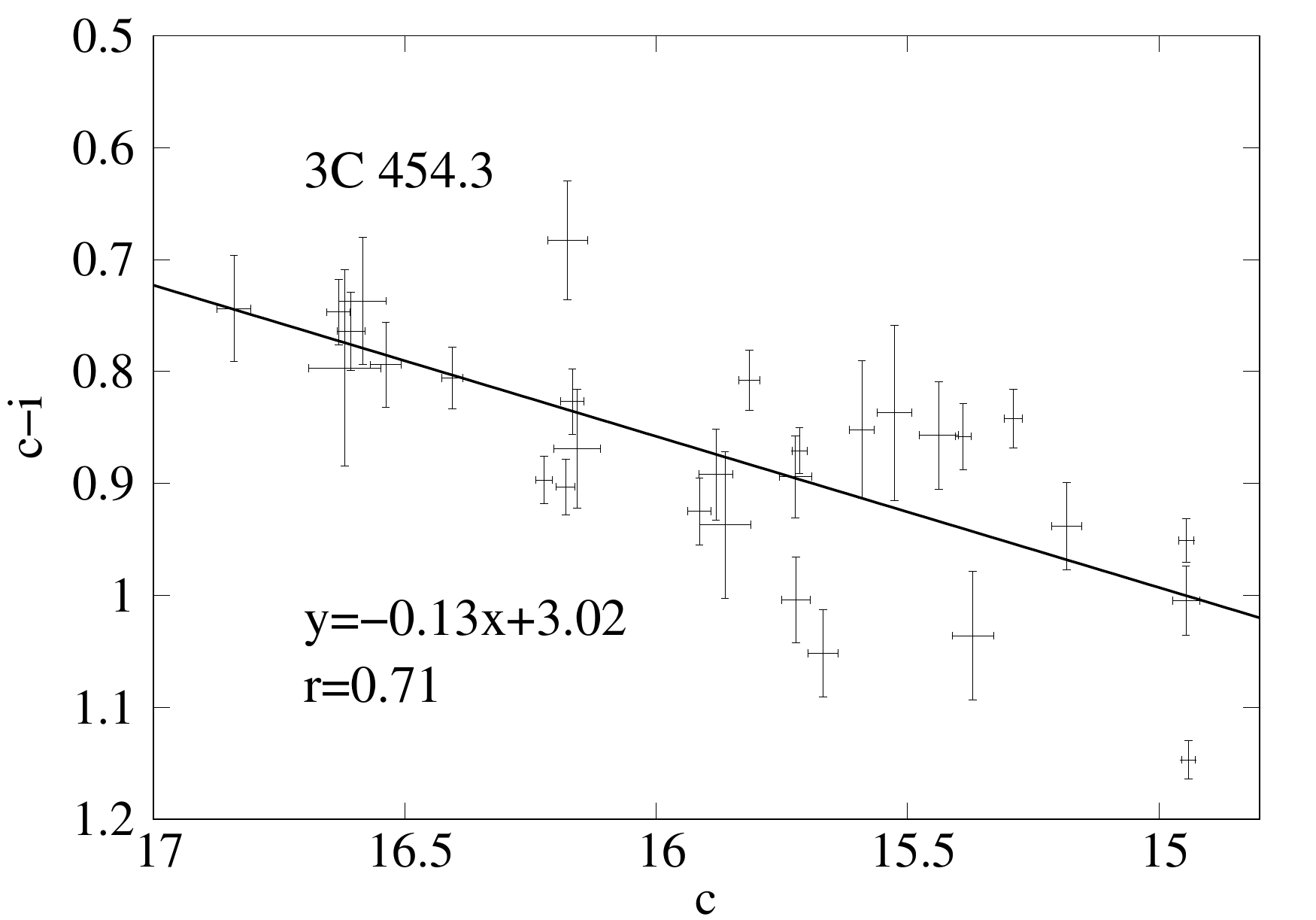}  
   \end{minipage}  
      \begin{minipage}[t]{0.24\textwidth}
	\includegraphics[scale=0.24]{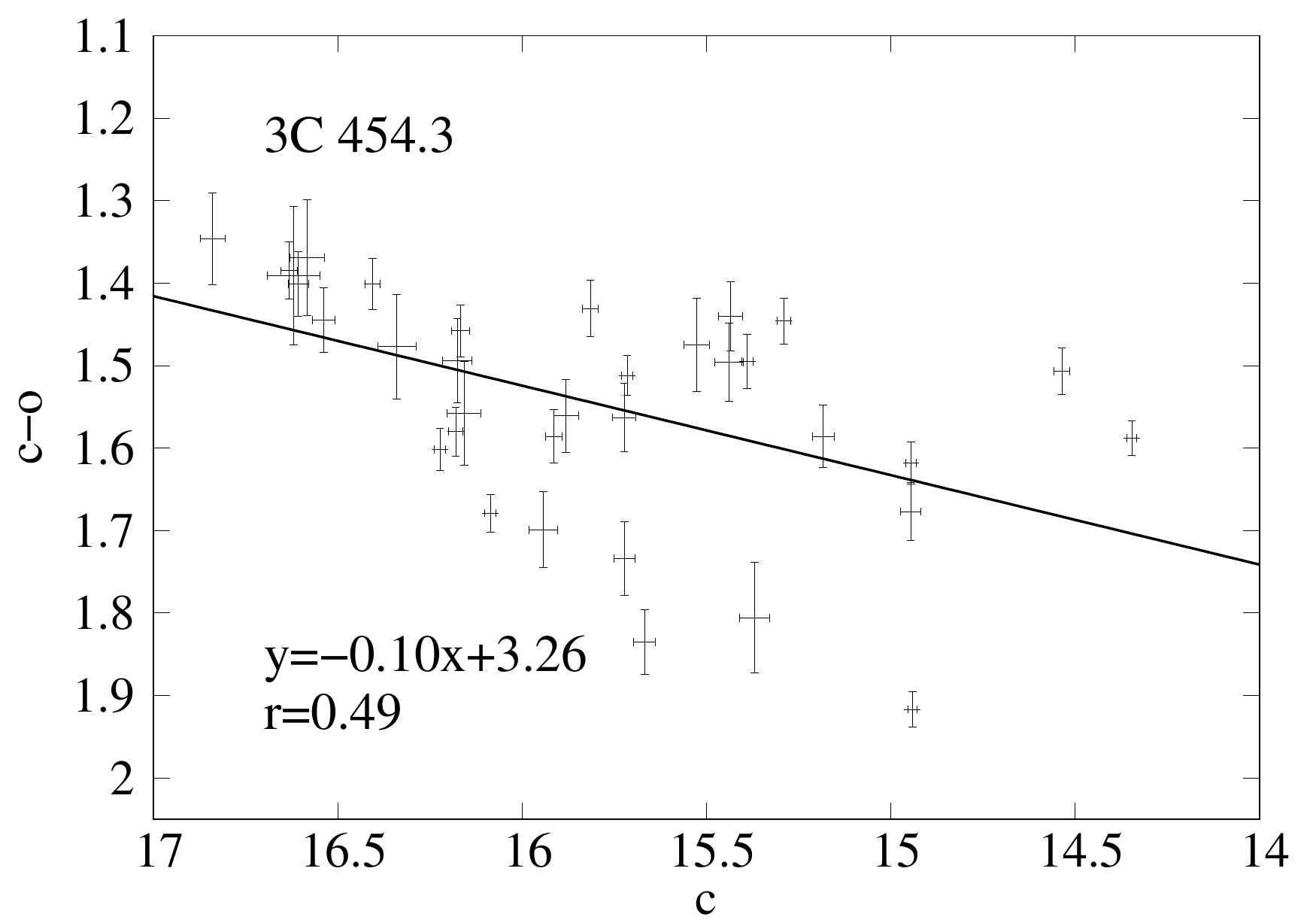}  
   \end{minipage}  
      \begin{minipage}[t]{0.24\textwidth}
	\includegraphics[scale=0.24]{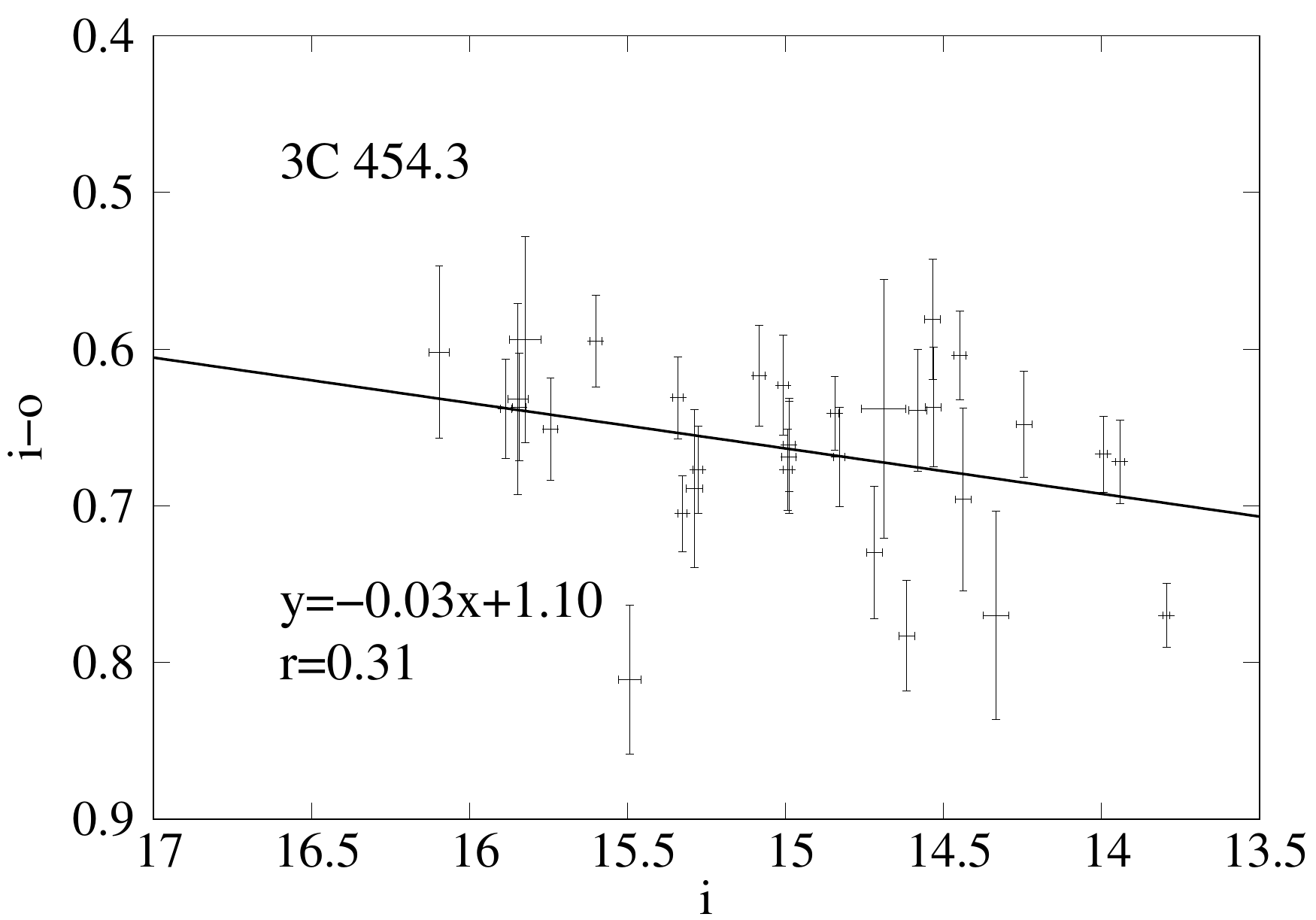}  
   \end{minipage}  
    \caption{Color-magnitude diagrams. The solid lines are the best linear fit.}
    \label{fig:allcolor} 
\end{figure*}

\begin{figure*}
\centering
	\includegraphics[scale=0.6]{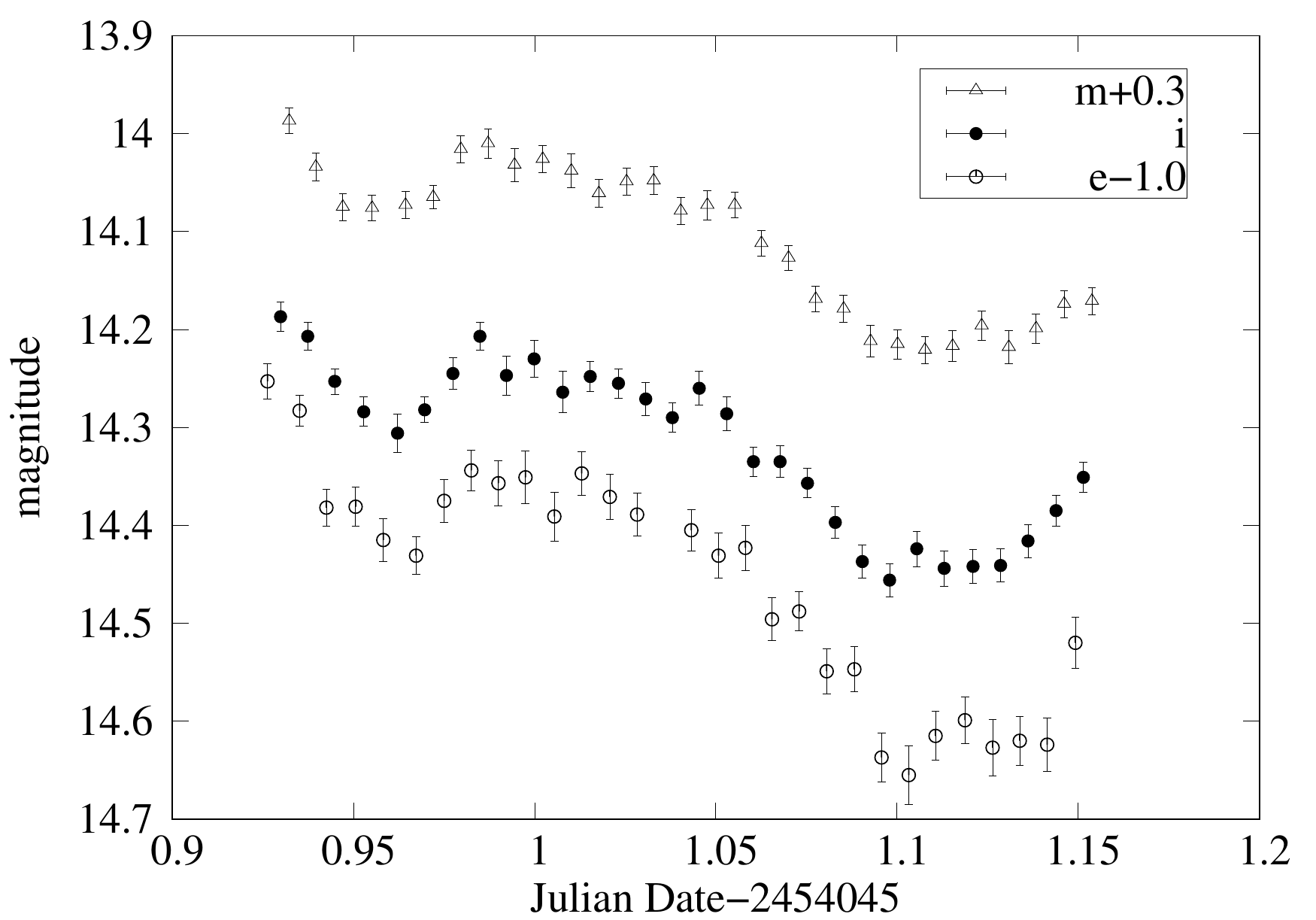}
    \caption{IDV light curves of BL Lacertae in the $e$, $i$, and $m$ bands on 2006 November 6. For clarity, the $m$ and $e$ band magnitudes are shifted by 0.3 and $-1$ magnitudes, respectively.}
    \label{fig:idv}
\end{figure*}

\begin{figure}
	\centering
	\includegraphics[scale=0.6]{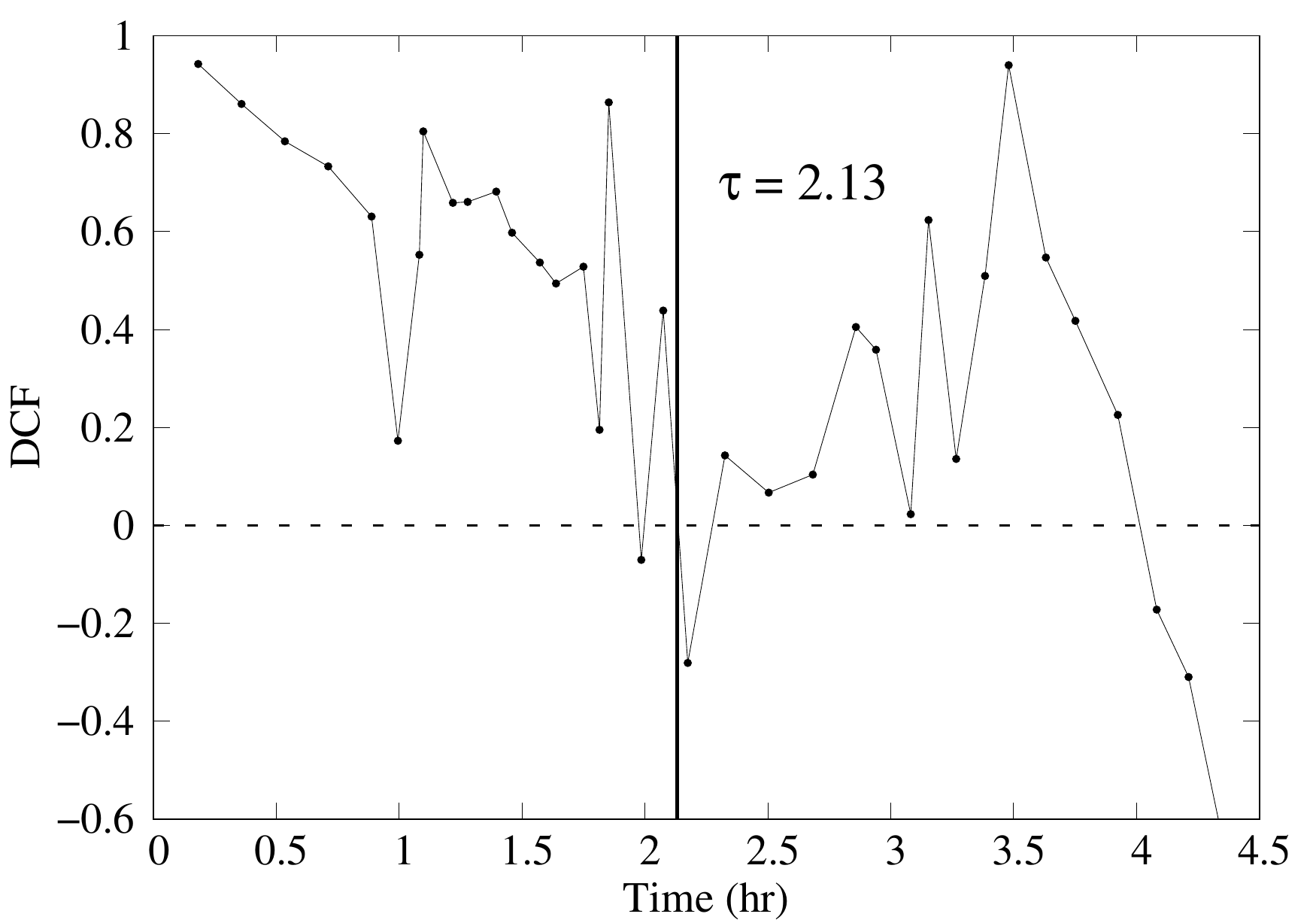}

      \caption{The ZDCF (in autocorrelation mode) results for BL Lacertae on 2006 November 6. $\tau$ gives the calculation of timescale.}
      \label{fig:BHmass}
\end{figure}

\begin{figure*}
\centering
	\includegraphics[scale=0.6]{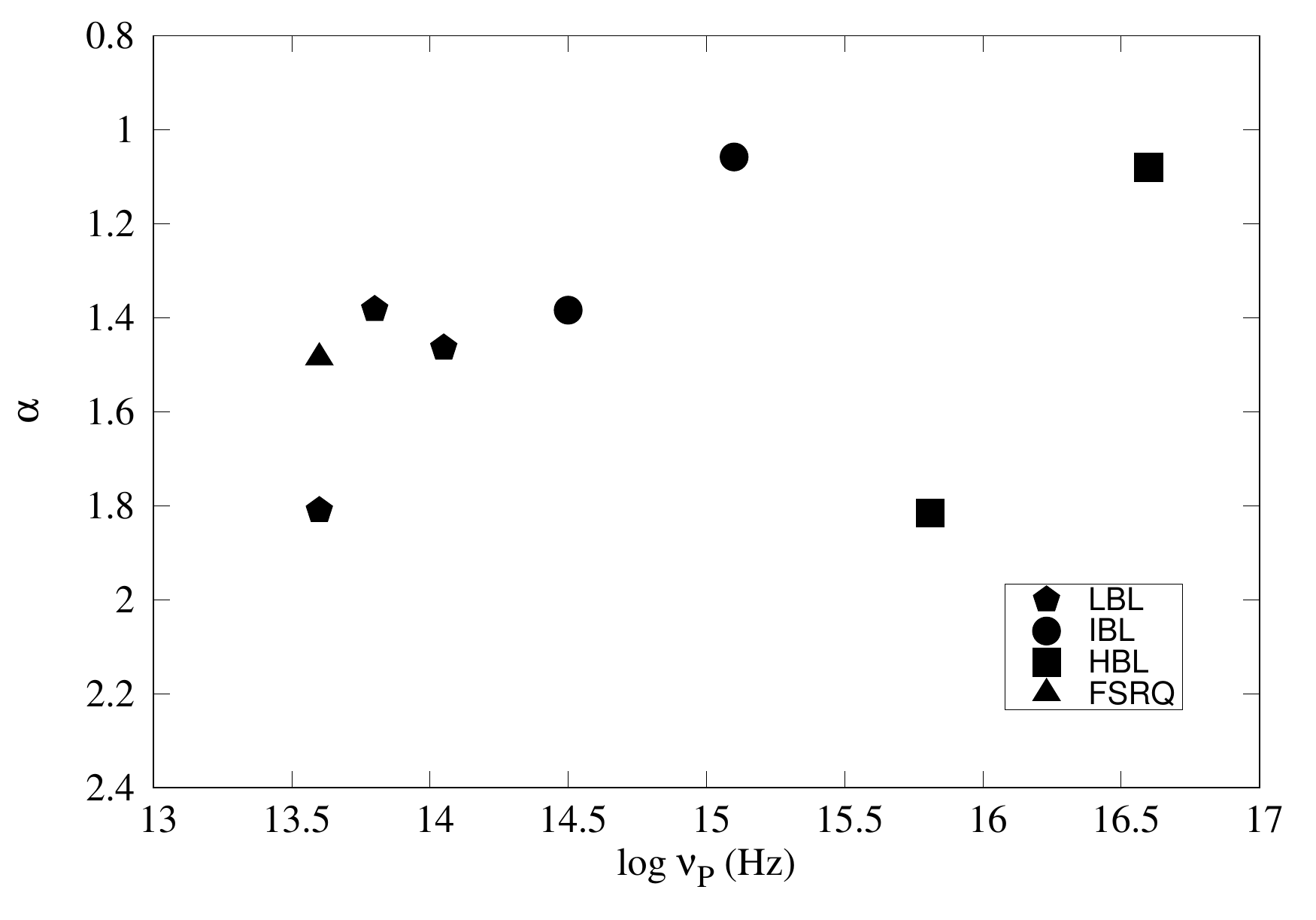}
    \caption{Average spectral index vs synchrotron peak frequency. Mrk 421 and PG 1553+113 are excluded since their spectral indices were not calculated.}
    \label{fig:spectra1}
\end{figure*}

\begin{figure*}
\centering
	\includegraphics[scale=0.8]{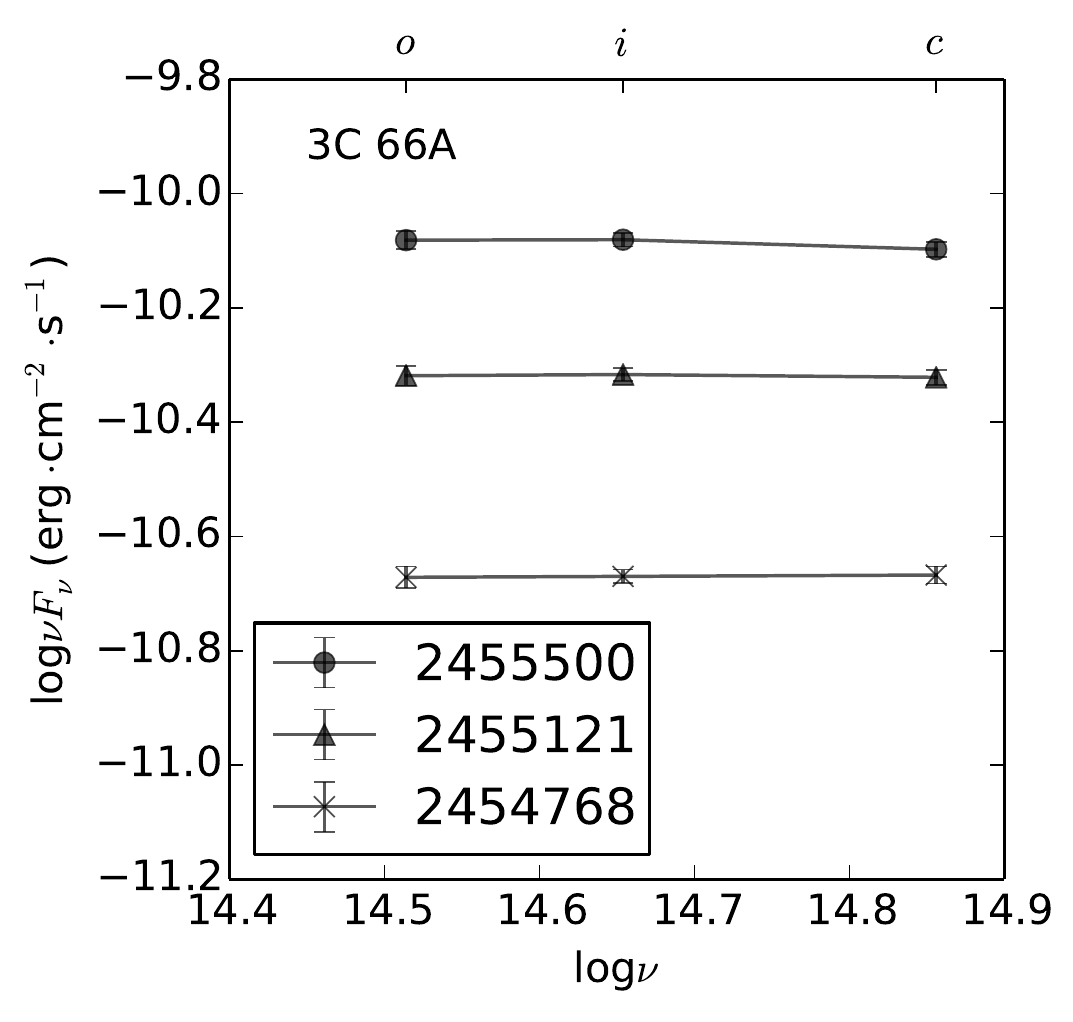}
	\includegraphics[scale=0.8]{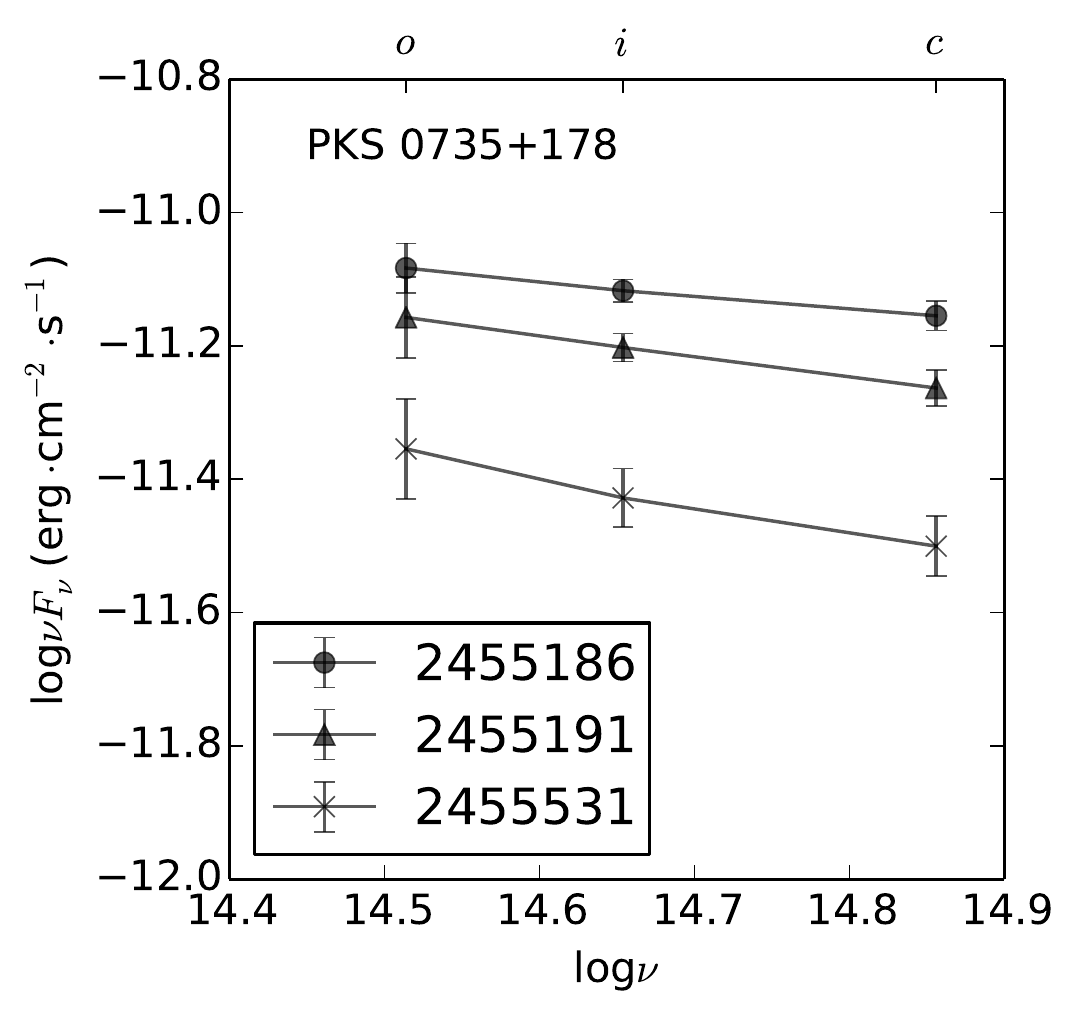}
	\includegraphics[scale=0.8]{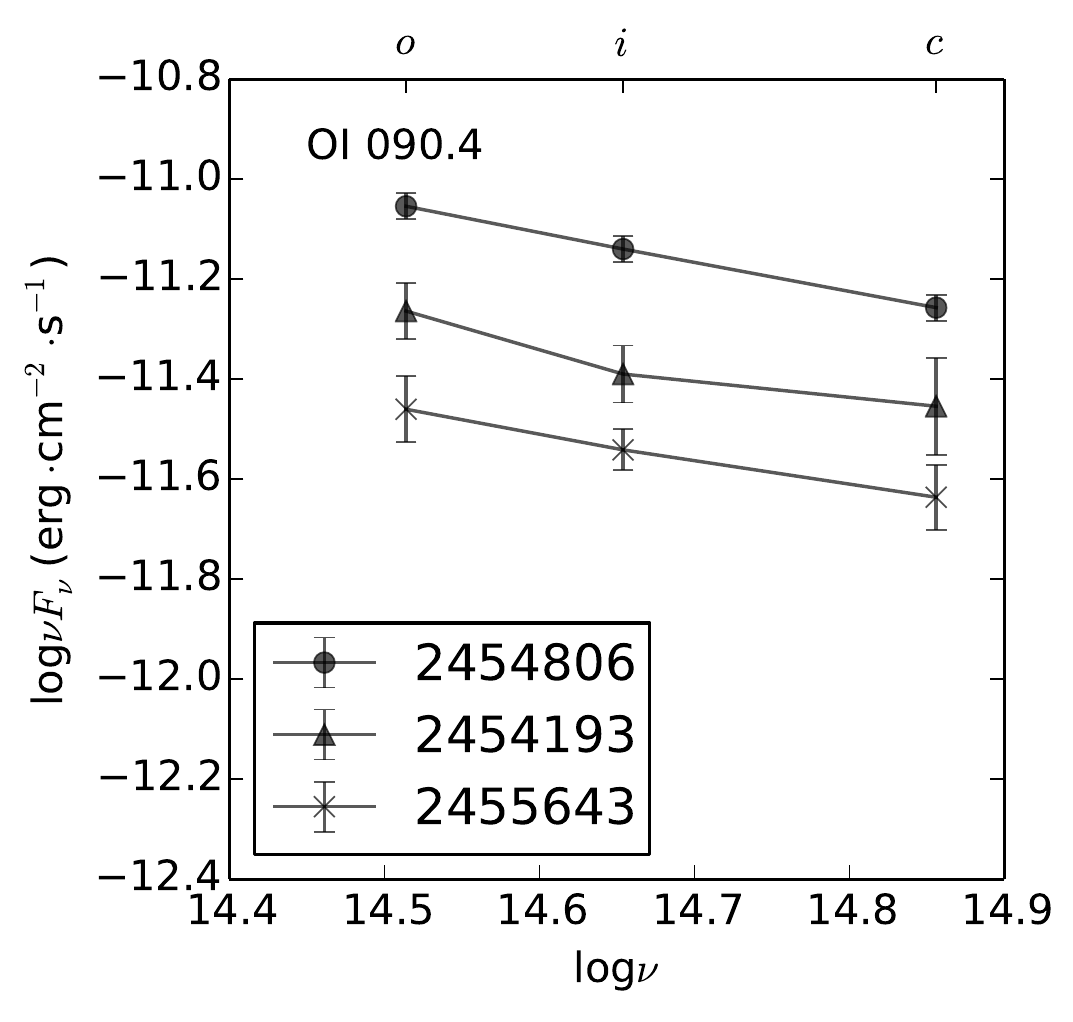}
	\includegraphics[scale=0.8]{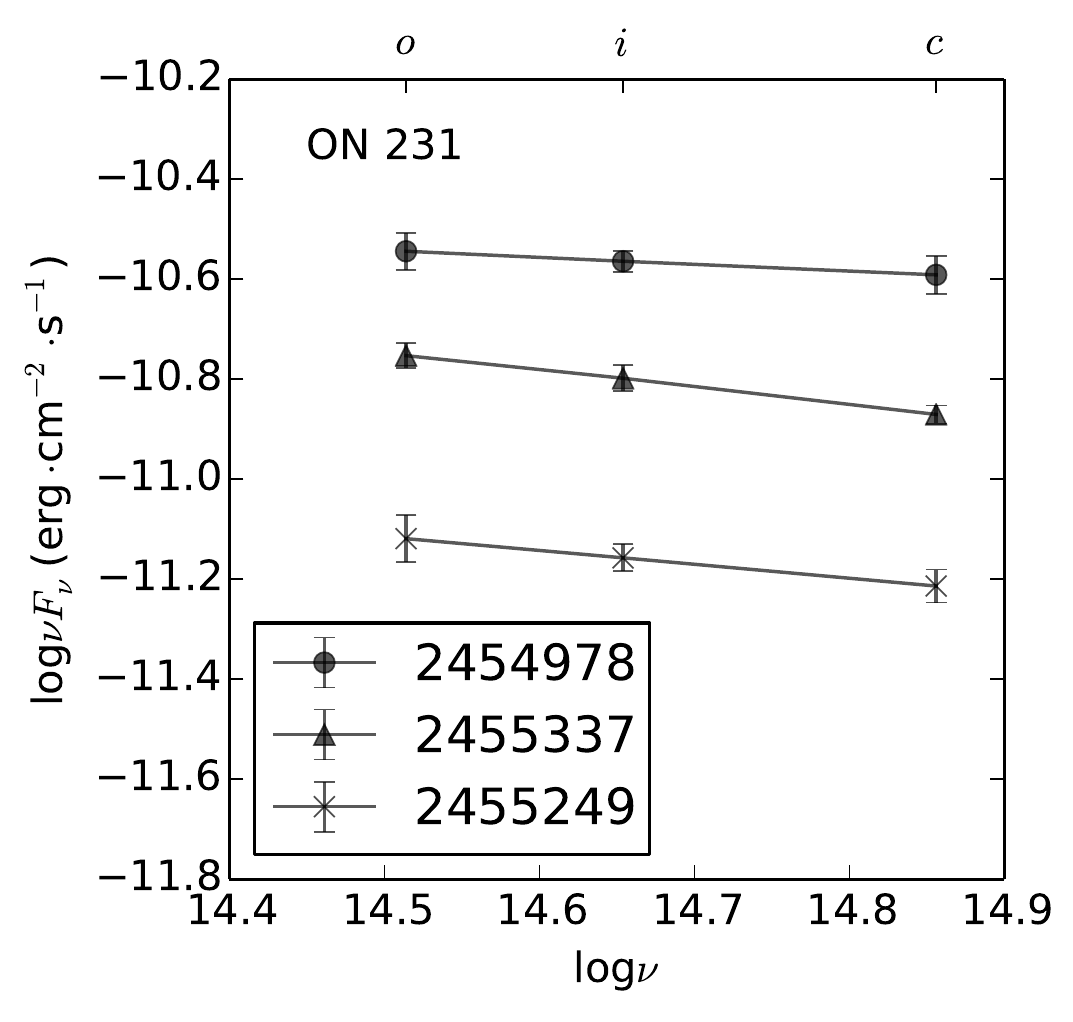}
    \caption{The SEDs in the highest, intermediate, and lowest states of eight blazars. For OT 546, the second highest SED is also presented. See text for details.}
    \label{fig:sed} 
\end{figure*}

\renewcommand{\thefigure}{\arabic{figure} }
\addtocounter{figure}{-1}
\begin{figure*}
\centering
	\includegraphics[scale=0.8]{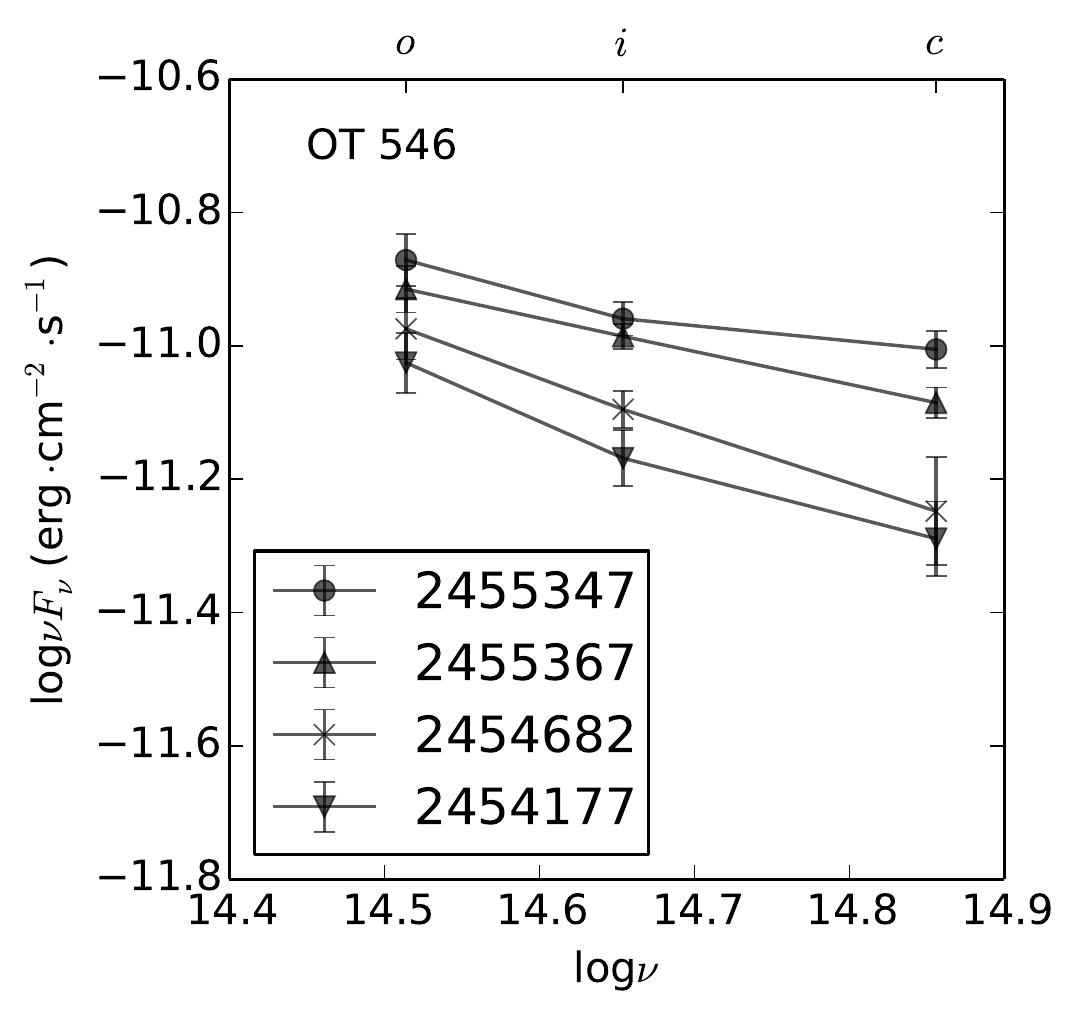}
	\includegraphics[scale=0.8]{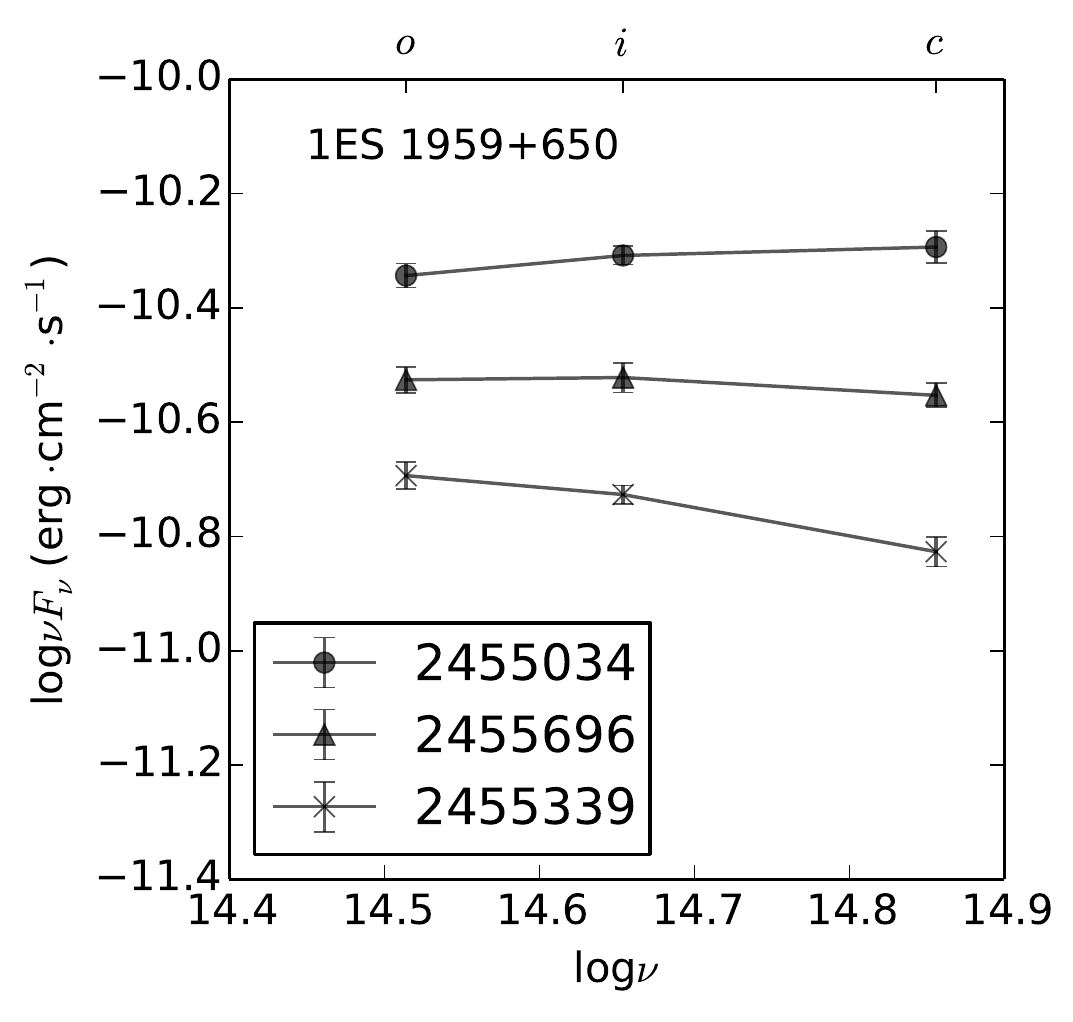}
	\includegraphics[scale=0.8]{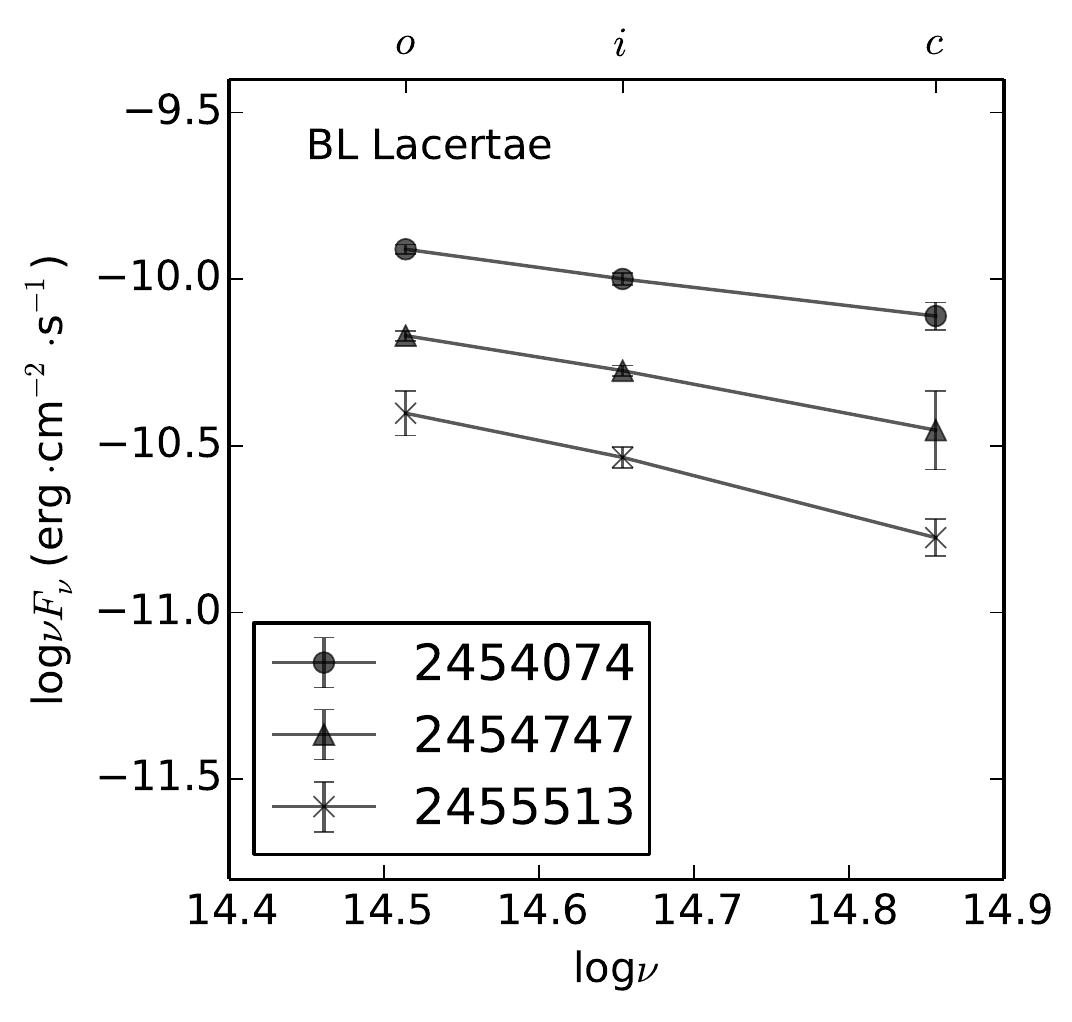}
	\includegraphics[scale=0.8]{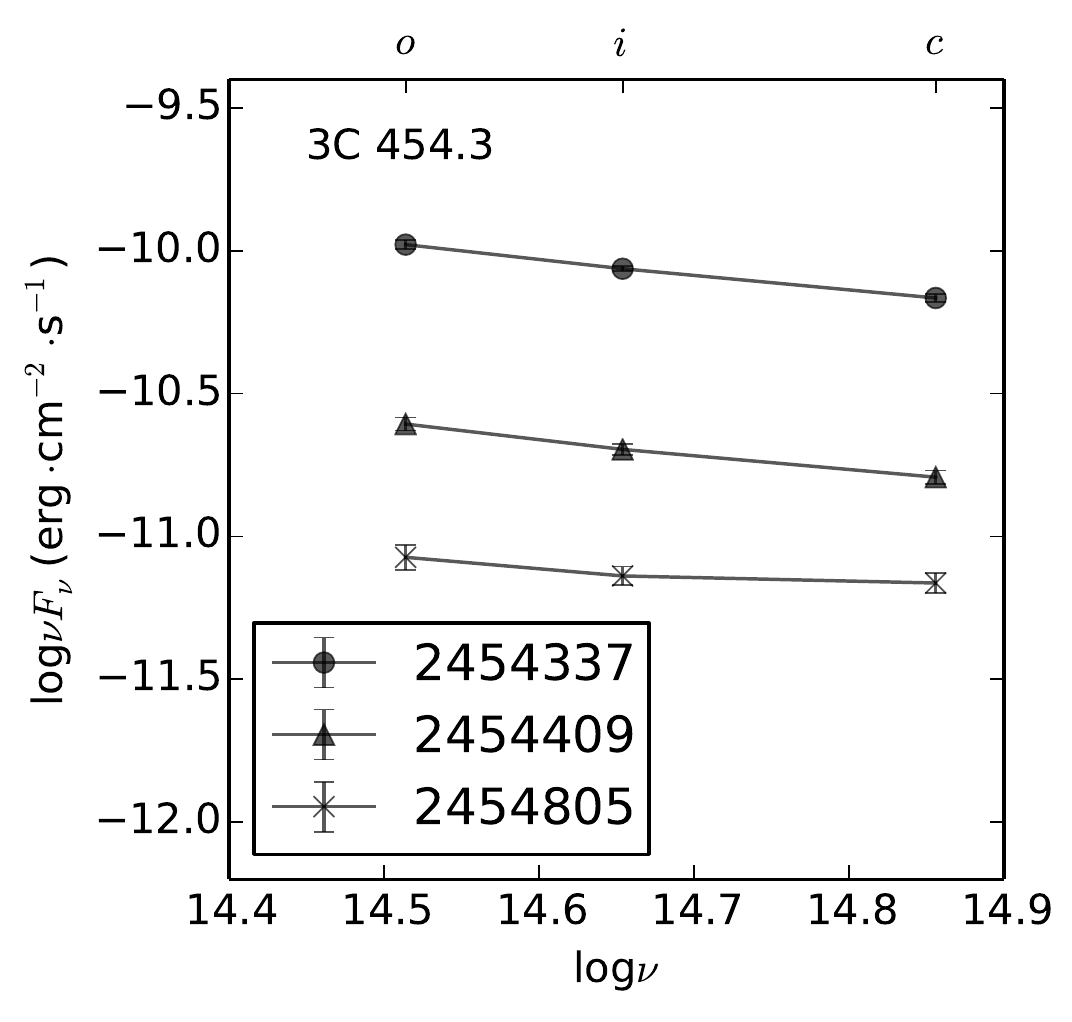}
    \caption{continued}
    \label{fig:sed} 
\end{figure*}

\end{document}